\documentclass[fleqn,usenatbib]{mnras}
\usepackage{newtxtext,newtxmath}
\usepackage[T1]{fontenc}
\DeclareRobustCommand{\VAN}[3]{#2}
\let\VANthebibliography\thebibliography
\def\thebibliography{\DeclareRobustCommand{\VAN}[3]{##3}\VANthebibliography}

\usepackage{graphicx}
\usepackage{xspace}    \usepackage[dvipsnames]{xcolor} 

\DeclareRobustCommand{\sigone}{\ensuremath{\Sigma_\mathrm{1kpc}}\xspace}
\DeclareRobustCommand{\sigseq}{\ensuremath{\Sigma_\mathrm{1kpc,Qu-seq}}\xspace}
\DeclareRobustCommand{\msun}{\ensuremath{\mathcal{M}_{\sun}}\xspace}
\DeclareRobustCommand{\LX}{\ensuremath{L_\mathrm{X}}\xspace}
\DeclareRobustCommand{\mstel}{\ensuremath{\mathcal{M}_*}\xspace}
\DeclareRobustCommand{\mbh}{\ensuremath{\mathcal{M}_\mathrm{BH}}\xspace} 
\DeclareRobustCommand{\Hband}{\ensuremath{H}\xspace}
\DeclareRobustCommand{\ergs}{erg~s$^{-1}$\xspace}

\newcommand{\jaedit}[1]{#1} \newcommand{\redit}[1]{{#1}}

\usepackage{xstring}

\title[AGN across compact and extended galaxy evolution phases]{
AGN accretion and black hole growth across compact and extended galaxy evolution phases
}

\author[Aird, Coil, Kocevski]{
James Aird,$^{1,2}$\thanks{james.aird@ed.ac.uk} 
Alison L. Coil$^{3}$ 
and
Dale D. Kocevski$^4$
\\
$^1$Institute for Astronomy, University of Edinburgh, Royal Observatory, Edinburgh EH9 3HJ, UK\\
$^2$School of Physics \& Astronomy, University of Leicester, University Road, Leicester LE1 7RJ, UK\\
$^3$Center for Astrophysics and Space Sciences, Department of Physics, University of California, 9500 Gilman Dr. MC 0424, La Jolla, CA 92093-0424, USA\\
$^4$Department of Physics and Astronomy, Colby College, Waterville, ME 04961, USA
}

\date{Accepted 2022 July 22. Received 2022 June 24; in original form 2022 January 26.}

\pubyear{2021}

\begin{document}
\label{firstpage}
\pagerange{\pageref{firstpage}--\pageref{lastpage}}
\maketitle

\begin{abstract}
The extent of black hole growth during different galaxy evolution phases and the connection between galaxy compactness and AGN activity remain poorly understood. 
We use Hubble Space Telescope imaging of the CANDELS fields to identify star-forming and quiescent galaxies at $z=$0.5--3 in both compact and extended phases
and
use Chandra X-ray imaging to 
measure the distribution of AGN accretion rates and 
track black hole growth within these galaxies.
Accounting for the impact of AGN light changes $\sim$20\% of the X-ray sources from compact to extended galaxy classifications. We find that $\sim$10--25\% of compact star-forming galaxies host an AGN, a mild enhancement (by a factor $\sim$2) compared to extended star-forming galaxies or compact quiescent galaxies of equivalent stellar mass and redshift.
However, AGN are not ubiquitous in compact star-forming galaxies and this is not the evolutionary phase, given its relatively short timescale, where the bulk of black hole mass growth takes place.
Conversely, we measure the highest AGN fractions ($\sim$10--30\%) within the relatively rare population of extended quiescent galaxies. For massive galaxies that quench at early cosmic epochs, substantial black hole growth in this extended phase is crucial to produce the elevated black hole mass--to--galaxy stellar mass scaling relation observed for quiescent galaxies at $z$$\sim$0. We also show that AGN fraction increases with compactness in star-forming galaxies and decreases in quiescent galaxies \textit{within} both the compact and extended sub-populations, 
demonstrating that AGN activity depends closely on 
the structural properties of galaxies. 
\end{abstract}


\section{Introduction}
\label{sec:intro}

The evolution of galaxies appears closely linked with the growth of the supermassive black holes at their centres, which is dominated by periods of accretion that produce Active Galactic Nuclei (AGN).
Higher galaxy star formation rates (SFRs) are associated with increased AGN activity throughout cosmic time \citep[e.g.][]{chen_correlation_2013,aird_x-rays_2019}. 
In the local Universe, the masses of central black holes
are correlated with tracers of galaxy mass, with quiescent galaxies with ageing stellar populations typically hosting more massive black holes than their actively star-forming counterparts
\citep[e.g.][]{reines_relations_2015,terrazas_quiescence_2016,greene_intermediate-mass_2020}.
Models of galaxy evolution require feedback from AGN to quench star formation and regulate the growth of massive galaxies \citep[e.g.][]{croton_evolution_2006,somerville_semi-analytic_2008,booth_cosmological_2009,weinberger_Supermassive_2018,dave_SIMBA_2019}.
The purported role of AGN feedback in quenching has motivated numerous studies to explore whether AGN are preferentially found in certain galaxy types 
i.e.~those that are already quenched, are in the process of quenching, or are expected to quench imminently \citep[e.g.][]{nandra_aegis_2007,xue_color-magnitude_2010,georgakakis_observational_2011,georgakakis_investigating_2014,rosario_host_2015,kocevski_candels_2017,yang_linking_2018}.
While an overall connection between AGN and galaxy evolution is well established, the mechanisms that trigger AGN at different points of the galaxy lifecycle, the extent of black hole growth during different galaxy evolution phases, and the impact of AGN on the evolutionary pathways of their host galaxies remain poorly understood.

\jaedit{Understanding
how and why certain galaxies undergo quenching and the importance of different physical quenching mechanisms (including AGN feedback) thus remains an area of active study.}
Analysis of the stellar populations and structural properties of galaxy samples at high redshifts indicates that individual galaxies follow a variety of pathways to build up their stellar mass and---for some fraction---quench their star formation and transform into the quiescent galaxies found at late cosmic times \citep[e.g.][]{baldry_quantifying_2004,faber_galaxy_2007,moustakas_primus_2013}.
Of particular recent interest has been the identification of distinct relationships between the effective size and total stellar mass for the star-forming versus the quiescent galaxy populations that provide insight into different quenching mechanisms and the evolution of galaxies during this process.
Quiescent galaxies follow a relatively tight correlation between their stellar mass and size, whereby more massive quiescent galaxies are generally physically larger 
\citep{franx_Structure_2008,barro_candels_2013,van_der_wel_3d-hstcandels_2014}.
This sequence is even more clearly defined if quantified in terms of central stellar mass density \citep[e.g. \sigone, the stellar mass density within the central 1~kpc, see][]
{bezanson_Relation_2009,cheung_dependence_2012,fang_link_2013,barro_structural_2017}.
The quiescent size--mass relation evolves such that quiescent galaxies of a given stellar mass are generally compact at high redshifts and are more extended at later cosmic times \citep[e.g.][]{newman_Can_2012,toft_size-star_2009, trujillo_Size_2006, van_der_wel_3d-hstcandels_2014, williams_Evolving_2010}. 
In contrast, star-forming galaxies tend to be more extended  (i.e. have lower \sigone) than quiescent galaxies of equivalent stellar mass and redshift, although their broad distribution does include a population of compact star-forming galaxies that lie close to the quiescent galaxy size--mass relation \citep{barro_candels_2013,van_dokkum_dense_2014}.

Thus, it is is possible to identify distinct galaxy populations at a given redshift based on both their star formation and structural properties. 
\citet{barro_candels_2013} suggest that these different galaxy populations correspond to distinct phases of galaxy evolution and that individual galaxies follow a range of evolutionary pathways that can move them between the different classifications over time \citep[see also][and references therein]{van_dokkum_forming_2015,barro_structural_2017}. 
Specifically, at high redshifts ($z\gtrsim1.4$), \citet{barro_candels_2013} suggest a ``fast-track'' evolutionary pathway. 
In this scenario galaxies typically form as the ubiquitous extended star-forming galaxies, assembling their stellar mass through star formation throughout the galaxy. 
In a subset of these galaxies internal instabilities, substantial gas inflow from the larger cosmic web, or galaxy merger events trigger substantial additional star formation within the central kiloparsec or may redistribute existing stars to the galactic centre, leading to a galaxy ``compaction'' that rapidly increases \sigone and transforms them into compact star-forming galaxies.
The star formation in these compact star-forming galaxies subsequently quenches \citep[likely due to a combination of gas exhaustion and feedback processes:][]{zolotov_Compaction_2015,kocevski_candels_2017}, transforming them into compact quiescent galaxies. This compact star-forming phase is thought to be comparatively short-lived \citep[$\sim$0.3--1~Gyr,][]{barro_candels_2013,van_dokkum_forming_2015} \redit{given the relatively low number densities of these compact star-forming galaxies.}

Additionally, the redshift evolution of the size--mass relation for quiescent galaxies indicates an overall size growth for the quiescent galaxy population, with the typical central stellar density (i.e. \sigone) becoming lower at later cosmic times.
Such evolution may be due to the size growth of \emph{individual} quiescent galaxies driven by mergers that re-distribute stars \citep[e.g.][]{hopkins_Discriminating_2010, naab_Minor_2009, oser_Two_2010}, or an adiabatic expansion as a result of
mass loss due to either stellar winds \citep[e.g.][]{damjanov_red_2009} or AGN-driven outflows \citep[e.g.][although see also~\citealt{silverman_where_2019}]{fan_cosmic_2010,ishibashi_can_2013}.
Alternatively, the size evolution may be due to star-forming galaxies quenching at lower \sigone at later times to produce new quiescent galaxies and thus shifting the overall \emph{population} rather than reflecting size growth of individual galaxies \redit{i.e.~the apparent size evolution is due to a ``progenitor bias'' \citep{van_dokkum_fundamental_1996, shankar_avoiding_2015}.}
Such a scenario is supported by studies showing that more extended galaxies tend to have younger stellar populations and thus quenched more recently \citep[e.g.][]{van_der_wel_Size_2009, fagioli_Minor_2016, williams_Morphology_2017, wu_fast_2018, hamadouche_combined_2022}. 
 The paucity of compact star-forming galaxies at lower redshifts also suggests that alternative, ``slow-track'' quenching mechanisms may become more important at later cosmic times \citep[e.g.][]{peng_Mass_2012,barro_candels_2013}, which transform extended star-forming galaxies into quiescent galaxies directly without requiring significant compaction.

A number of prior studies explore the incidence of AGN during these putative compact and extended galaxy evolution phases.
These studies measure a high AGN fraction in compact star-forming galaxies ($\sim$20-40\%) that is enhanced (by a factor $\sim$3--30) compared to in either extended star-forming galaxies or compact quiescent galaxies of comparable stellar mass \citep[e.g.][]{barro_candels_2013,kocevski_candels_2017,habouzit_linking_2019}. 
It was proposed by \citet{kocevski_candels_2017} that these high AGN fractions indicate that the compaction process that builds up substantial stellar mass in the centres of compact star-forming galaxies also drives gas into the vicinity of the central black hole and triggers periods of AGN activity.
Additionally, 
\citet{ni_does_2019} find that \sigone may be the strongest predictor of sample-averaged black hole accretion rate in star-forming galaxies, indicating a close association between galaxy compactness and black hole growth \citep[see also][]{ni_revealing_2021}.
Expanding on these findings, \citet{chen_quenching_2020} present a phenomenological model that attributes the quenching of galaxies to the cumulative heating of the host halo gas due to AGN feedback, thus tying the suppression of star formation in galaxies to the overall assembly of central black hole mass that is inferred to occur in the compact star-forming phase. 
However, it is unclear whether such a model is consistent with the observed enhancement of AGN in the short-lived population of compact star-forming galaxies and the fact that the bulk of AGN are found in normal star-forming galaxies \citep[e.g.][]{rosario_mean_2013,azadi_primus_2015,yang_linking_2018}  with only a weak enhancement in AGN fraction for galaxies with suppressed SFRs indicating they are undergoing quenching \citep[e.g.][]{shimizu_decreased_2015,aird_x-rays_2019}.

In this paper, we perform new measurements to determine both the incidence of AGN and their accretion rates across compact and extended galaxy evolution phases.
Rather than measuring X-ray detected fractions \citep[as in e.g.][]{kocevski_candels_2017} or sample-averaged accretion rates \citep[as in e.g.][]{ni_revealing_2021}, we instead follow the approach of \citet{aird_x-rays_2018,aird_x-rays_2019} to robustly measure the full accretion rate probability distributions within the different populations. 
These measurements allow us to quantify both the incidence of AGN within a given galaxy sample and their typical accretion rates, providing insights into both the triggering of AGN and the amount of black hole growth during different phases of galaxy evolution.
We also refine our measurements of the structural properties and stellar populations of our galaxy samples to account for any light from a central AGN, which has an important impact on our results.

Section~\ref{sec:data} below describes the selection of our parent sample of galaxies at $z=0.5-3$ from the deep \emph{Hubble Space Telescope} (\textit{HST}) imaging of the five CANDELS survey fields and the \textit{Chandra} X-ray imaging used to identify AGN.
Our measurements of total stellar mass, SFRs and \sigone, in all cases allowing for potential contamination due to an AGN, are described in Section~\ref{sec:measurements}.
Section~\ref{sec:results} presents our measurements of accretion rate distributions and AGN fractions
\redit{(see also Appendix~\ref{appendix:robustness} that presents careful tests of the robustness of these results).}
In Section~\ref{sec:discuss} we place our results in the context of galaxy evolution, compare with prior studies, assess the extent of black hole growth during different galaxy evolution phases and reconcile these findings with measurements of black hole--galaxy scaling relations in the local Universe. 
Section~\ref{sec:conclusions} summarizes our results and our overall conclusions.
{\emph{Our primary results are shown in Figure~\ref{fig:mainresults} and summarised by Figure~\ref{fig:fagn_vs_classl}, while Figures~\ref{fig:BHgrowth} and \ref{fig:pathways} illustrate our interpretation and key conclusion that significant black hole growth may occur in extended quiescent phases after star formation has quenched.}}
We assume a \citet{chabrier_galactic_2003} stellar initial mass function when deriving galaxy properties and adopt a flat cosmology with $\Omega_\Lambda = 0.7$ and $H_0 = 70$~km~s$^{-1}$~Mpc$^{-1}$ throughout this paper.

\section{Data}
\label{sec:data}

\subsection{CANDELS imaging and parent galaxy sample}
\label{sec:galaxydata}

Our parent sample of galaxies is drawn from HST/WFC3 F160W (\emph{H}-band) selected catalogs in the five CANDELS fields \citep{grogin_candels_2011,koekemoer_candels_2011}.
These fields include the Great Observatories Origins Deep Survey \citep[GOODS:][]{giavalisco_great_2004} North and South, the UKIDSS Ultra Deep Survey \citep[UDS:][]{lawrence_ukirt_2007,cirasuolo_evolution_2007}, the Extended Groth Strip \citep[EGS:][]{davis_all-wavelength_2007}, and the COSMOS \citep{scoville_cosmic_2007} regions.  While point source depths vary among the CANDELS fields (from $H=27$ in the wide fields to $H=27.7$ in the deep fields) in this paper we apply a uniform, conservative magnitude limit of $\Hband<25.5$ across all five fields, which allows us to select a complete sample of galaxies above our stellar mass limits (see Section~\ref{sec:galsample} below).
Multi-wavelength photometry ($U$-band to 8$\mu$m) was measured in each field using the TFIT routine \citep{laidler_tfit_2006} as described in detail in \citet{guo_candels_2013,galametz_candels_2013,stefanon_candels_2017,nayyeri_candels_2017} and \citet{barro_candelsshards_2019} for the GOODS-S, UDS, EGS, COSMOS, and GOODS-N fields, respectively. 

We compile spectroscopic redshifts from a wide variety of campaigns following up galaxy and X-ray selected AGN samples \citep[see][and references therein]{skelton_3d-hst_2014,aird_evolution_2015,nandra_aegis-x_2015,kriek_mosfire_2015,le_fevre_vimos_2015,tasca_evolving_2015,marchesi_chandra_2016}, providing redshifts for $\sim$15\% of our magnitude-limited CANDELS samples. 
When a high resolution spectroscopic redshift is unavailable, we adopt reliable redshifts based on the low-resolution WFC3 grism spectroscopy of the CANDELS fields obtained as part of the 3DHST project \citep{brammer_3d-hst_2012,momcheva_3d-hst_2016}, which are assigned to a further $\sim$25\% of the CANDELS sources.
For the remaining sources, we adopt photometric redshift estimates. 
Photometric redshifts for all objects are provided by \citet{dahlen_critical_2013} and have typical errors of $\Delta z/(1+z)=3\%$.
We also adopt photometric redshifts derived specifically for X-ray identified AGN using appropriate templates \citep{salvato_dissecting_2011,nandra_aegis-x_2015,marchesi_chandra_2016}, preferring these redshifts over the \citet{dahlen_critical_2013} estimates when they are available (for 148 X-ray detected CANDELS sources).

We identify and remove stars based on the spectroscopic classification (if available) as well as any point-like objects with \Hband-band magnitudes $<25.0$ and 
 $   r < -0.01 \Hband +1.6$,
where $r$ is the radius corresponding to the full-width half maximum as provided by \textsc{SExtractor} \citep[FLUX\_RADIUS parameter,][]{bertinl_sextractor_1996}.
Galaxy morphologies and sizes were measured from the \emph{H}-band images using GALFIT \citep{peng_detailed_2002} as described in \citet{van_der_wel_3d-hstcandels_2014}.
The measurements include Sersic indicies and circularized effective (half-light) radii ($r_{e}=a_{\rm eff} \sqrt{b/a}$), where $a_{\rm eff}$ is the half-light radii on the major axis. 
Section~\ref{sec:galfit} below describes both how these measurements are used to estimate structural parameters of galaxies and our updated GALFIT analysis of sources with X-ray detected AGN.

\subsection{X-ray imaging and AGN identification}

We adopt deep \textit{Chandra} ACIS-I imaging that has been obtained across all five of the CANDELS fields with exposure times reaching $\sim$160~ks in COSMOS \citep{elvis_chandra_2009}, $\sim$600~ks in UDS \citep{kocevski_x-uds_2018}, $\sim800$~ks in EGS \citep{nandra_aegis-x_2015}, $\sim$2~Ms in GOODS-N \citep{alexander_chandra_2003}, and $\sim$4~Ms in GOODS-S \citep{xue_chandra_2011}.
All of these \textit{Chandra} surveys have been analysed using a consistent data reduction and source detection procedure described by \citet{laird_aegis-x_2009}, \citet{nandra_aegis-x_2015} and \citet{kocevski_x-uds_2018} 
\citep[see also][]{georgakakis_new_2008,georgakakis_investigating_2014,georgakakis_observational_2017,rangel_x-ray_2013,aird_evolution_2010,aird_evolution_2015}.
X-ray point sources are identified in the full (0.5--7~keV), soft (0.5--2~keV), hard (2-7~keV) and ultrahard (4--7~keV) energy bands and the resulting source lists are merged to produce a combined catalogue.
Our overall sample consists of 1066 significant X-ray detections within the 0.235~deg$^2$ footprint of the CANDELS imaging. 

We cross-match our X-ray detected samples with the \Hband-selected CANDELS catalogues using the likelihood ratio technique \citep[see][and references therein for details]{luo_identifications_2010,aird_evolution_2015}, which allows us to identify robust \Hband counterparts to 961 of our X-ray detections (90.2\%). 
Following \citet{aird_x-rays_2017}, we extract X-ray information (total counts, background counts and effective exposures in the full, soft and hard energy bands) at the positions of all of the remaining \Hband-selected CANDELS sources, excluding any objects that lie close to a significant X-ray source but are not associated with the X-ray source according to our likelihood ratio matching. 
The extracted total counts, background counts, and exposures provide X-ray constraints for galaxies that fall below our nominal X-ray detection threshold as well as provide information on the overall sensitivity of the X-ray imaging. 
All of this information, along with the directly detected X-ray sources, is incorporated into the Bayesian analysis described in \citet{aird_x-rays_2017,aird_x-rays_2018} that we use to perform the measurements of black hole accretion rates presented in Section~\ref{sec:results} below.

\section{Measurement of galaxy stellar population and structural properties}
\label{sec:measurements}

\subsection{Galaxy stellar population properties}
\label{sec:sedfit}

We use spectral energy distribution (SED) fitting to determine stellar masses (\mstel) and star formation rates (SFRs) for all galaxies in our CANDELS \Hband-selected samples. 
 We use an updated version of the \textsc{fast} code to fit SED models to the $U$-band to 8$\mu$m multiwavelength photometry described in Section~\ref{sec:galaxydata} above. 
\textsc{Fast} was originally developed by \citet{kriek_ultra-deep_2009} and was updated by \citet{aird_x-rays_2017,aird_x-rays_2018} to allow for both a galaxy and an AGN component. 
We assume a \citet{chabrier_galactic_2003} initial mass function (IMF), flexible stellar population synthesis (FSPS) models \citep{conroy_connecting_2009,conroy_propagation_2010}, fixed solar metallicity, dust reddening of $A_V$ in the range  0--4 mag \citep[assuming the][dust attenuation curves]{kriek_dust_2013}, and ``delayed-$\tau$'' star formation histories with $\tau$ in the range 0.1--10~Gyr \citep[e.g.][]{maraston_star_2010}. We allow for galaxy ages in the range $\sim$100~Myr to 13~Gyr but exclude ages that are older than the observable universe at a given $z$ and apply a redshift-dependent minimum age to exclude the youngest templates at lower redshifts \citep[see appendix~A of][]{aird_x-rays_2017}. 

We allow for an AGN component in the SED for galaxies with significant X-ray detections, adopting a library of eight empirically determined AGN templates \citep[see appendix~A of][]{aird_x-rays_2018}. 
Ultimately, we decide whether to adopt the two-component galaxy + AGN fits depending on whether a point source component is required in our analysis of the two-dimensional \Hband\ images, as described in Section \ref{sec:galfit} below.

\subsection{Galaxy structural properties}
\label{sec:galfit}

Following \citet{barro_structural_2017}, we calculate $\Sigma_{\rm 1 kpc}$, the central surface mass density within 1 kpc,
\begin{equation}
\Sigma_{\rm 1 kpc} = \frac{M_{*}(<1~{\rm kpc})}{\pi (1~{\rm kpc})^{2}}
\end{equation}
by integrating the best-fit surface brightness profile of each galaxy out to a radius of 1 kpc. 
\redit{We assume a constant mass-to-light ratio for a given galaxy based on the best fitting stellar population parameters from the overall SED fit described in Section~\ref{sec:sedfit} above
and that mass follows the \Hband-band light with a S\'ersic profile (the assumption of a constant mass-to-light is examined in Appendix~\ref{sec:mass_vs_light} and shown to have a negligible impact on our final results).} The projected mass profile is thus given by
\begin{equation}
M(r) = M_{\rm e}~{\rm exp}(-b_{\rm n}[(r/r_{\rm e})^{(1/n)} - 1]),
\end{equation}
where $n$ is the S\'ersic index and $M_{\rm e}$ is the projected mass at the effective radius, $r_{\rm e}$.  Integrating to $r=1$ kpc  results in the relationship
\begin{equation}
\log \Sigma_{\mathrm{1 kpc}} = \log \mstel - \log \pi + \log P(2n, b_{n} r_{\rm e}^{-1/n}).
\label{eq:sigone}
\end{equation}
Here \mstel is the total stellar mass of the system, $P(s,x)$ is the \redit{regularised lower incomplete gamma function}\footnote{\redit{The regularised lower incomplete gamma function is given by $P(s,x)=\gamma(s,x)/\Gamma(s)$, where $\gamma(s,x)= \int_0^x t^{s-1} e^{-t} dt$ and $\Gamma(s) = \int_0^\infty t^{s-1} e^{-t} dt$.}}
and the constant $b_{n}$, which depends on $n$, is estimated using the asymptotic approximation of \citet{ciotti_analytical_1999}:
\begin{equation}
b_{n} \approx 2n - \frac{1}{3} + \frac{4}{405n} + \frac{46}{25515 n^{2}}.
\label{eq:bn}
\end{equation}

An initial determination of $\Sigma_{\rm 1 kpc}$ is made for all galaxies using the best-fit S\'ersic profile parameters from \citet{van_der_wel_structural_2012} and \mstel from our SED fitting.  Following \citet{van_dokkum_dense_2014}, we apply a small correction (typically about 10\%) to the resulting $\Sigma_{\rm 1 kpc}$ values to account for any difference between the total magnitude implied by the S\'ersic fit and the $H$-band magnitude in the CANDELS photometry catalogs. 

For galaxies that host an X-ray detected AGN, we take additional steps to account for potential unresolved nuclear light that may artificially increase our $\Sigma_{\rm 1 kpc}$ measurements.  
\begin{figure*}
    \centering
    \includegraphics[height=14cm]{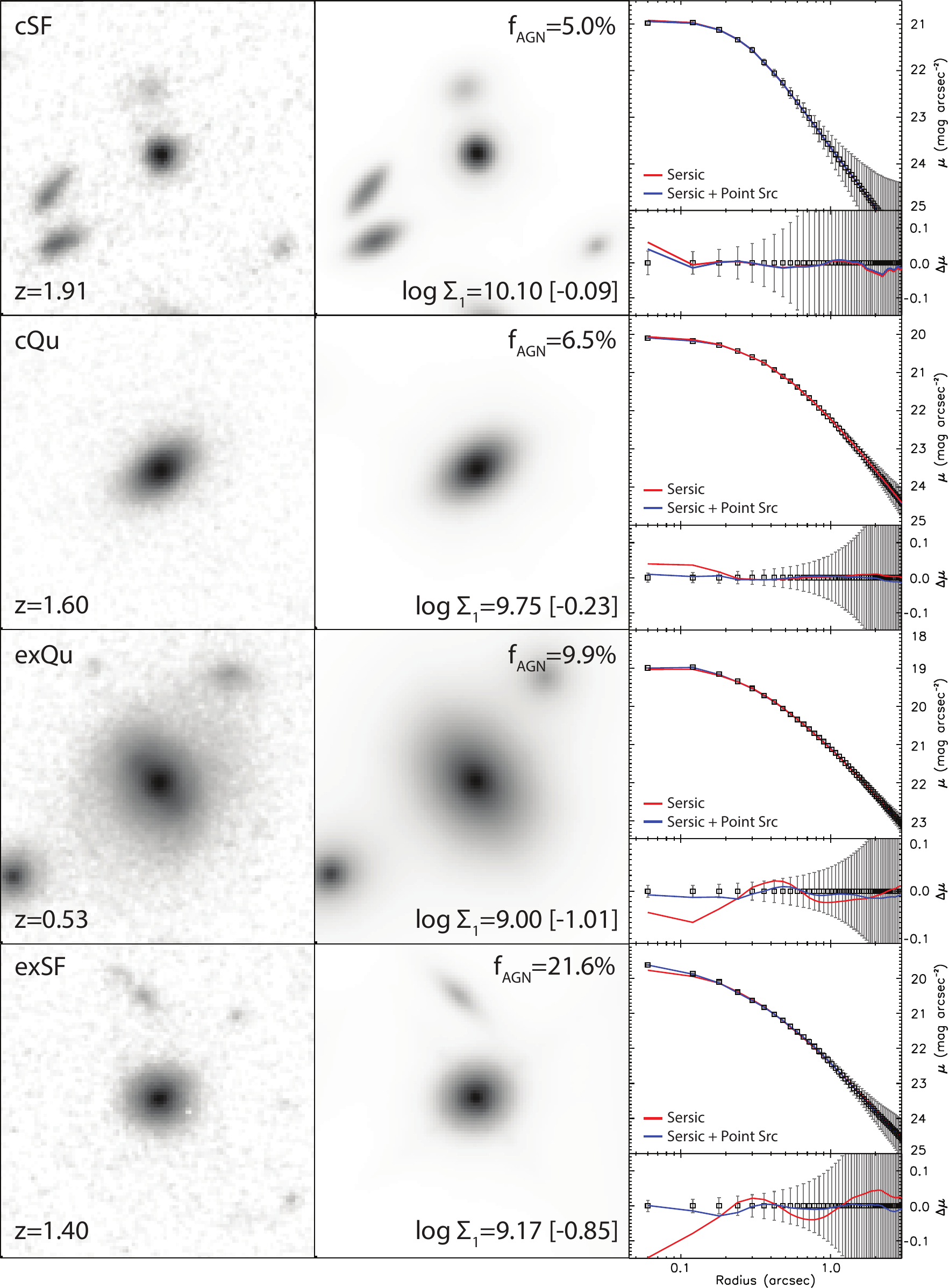}
        \caption{Example \Hband-band (HST/WFC3 F160W) images of galaxies in each of our four populations defined by structural and star formation properties (left), showing compact star forming (cSF), compact quiescent (cQu), extended quiescent (exQu), and extended star forming (exSF) galaxies from top to bottom, along with their best-fit GALFIT models (middle) and corresponding surface brightness profile fits (right) with and without including a central point source.   The fractional contribution of the AGN to the light at $1.6\mu m$ is listed in the middle panel.  Also shown are the calculated best-fit $\Sigma_{\rm 1 kpc}$ values for each galaxy, followed by (in brackets) how these values changed as a result of adding point-like emission from the AGN to our models.  Negative values in the brackets indicates a decrease in $\Sigma_{\rm 1 kpc}$ when a two component (galaxy+AGN) model is used.   These galaxies were chosen to demonstrate the wide range of AGN contributions observed in our sample. 
        }
    \label{fig:galfit_examples}
\end{figure*}
When our two component SED fit indicates that more than 2\% of a host galaxy's $H$-band light originates from the AGN, we carry out a series of surface brightness profile fits with GALFIT to test for the presence of nuclear contamination from a central point source.  We perform GALFIT runs on each galaxy using the following three models:

\begin{itemize}
\item[1.] A S\'ersic profile with an additional central point source component whose magnitude is fixed to the $H$-band magnitude predicted by our two component SED modeling.
\item[2.] A S\'ersic profile with an additional central point source component whose magnitude is allowed to float.
\item[3.] A S\'ersic profile only, with no additional point source component.
\end{itemize}

For these fits we provide GALFIT with custom PSF models appropriate for each CANDELS field \citep[see][for details]{van_der_wel_structural_2012} and custom-made noise images that account for both the intrinsic image noise (e.g., background noise and readout noise) as well as added Poisson noise due to the objects themselves.  Instead of allowing GALFIT to determine the local background flux level, we also pass along an independent background estimate specific to each galaxy from \citet{van_der_wel_3d-hstcandels_2014}.  Neighboring objects are fit simultaneously using single S\'ersic fits.  Examples of our GALFIT modeling are shown in Figure \ref{fig:galfit_examples}.

To determine if an additional point source component is required over that of a single S\'ersic profile, we use the residual flux fraction \citep[RFF;][]{hoyos_hstacs_2011} as a goodness-of-fit indicator.  The RFF is a measure of the fraction of the signal contained in the residual image that cannot be explained by fluctuations of the background.  We compute the RFF for each fit using the relationship
\begin{equation}
\mathrm{RFF} = \frac{ \left( \sum{ \mathrm{|RES|}} - 0.8\times\sigma_{\mathrm{F160W}}\right)}{\mathrm{FLUX}}.
\end{equation}
Here |RES| is the absolute value of the residual image, defined as the difference between the $H$-band image and the model fit, $\sigma_{\rm F160W}$ is the background noise image, and the summation is conducted within the Kron ellipse defined by SExtractor (Bertin \& Arnouts 1996). FLUX is the total flux of the galaxy in the $H$-band image and the constant of 0.8 is chosen to ensure that for a Gaussian noise error image, the expected value of the RFF is 0 \citep[see][for details]{hoyos_hstacs_2011}.

We calculate the RFF for all three of the GALFIT models described above and the one that results in the lowest RFF value is chosen as our best-fitting model.  A large fraction (43\%) of our host galaxies are best-fit by a S\'ersic profile with no additional point source component, as might be expected for the moderate luminosity AGN that dominate our sample.  
\jaedit{In these cases, we retain the galaxy-only SED fits for our \mstel and SFR measurements.}
For galaxies where the point source component improved our surface brightness fit, we find the median AGN contribution to be 10\% at $1.6\mu m$. 
\jaedit{When our GALFIT model requires a different AGN fraction to our SED fits (i.e.~case 2 above), we re-scale the estimated \mstel so that the AGN-to-host ratio is consistent with the GALFIT result.}
Accounting for this unresolved component works to increase $r_{\rm e}$ and decrease $n$ for the underlying hosts, both of which decrease $\Sigma_{\rm 1 kpc}$.  
We find that our $\Sigma_{\rm 1 kpc}$ measurements for these galaxies decreases by a median 0.4 dex when we allow for point source contamination in our surface brightness fits.  
The change in $\Sigma_{\rm 1 kpc}$ as a function of our best-fitting AGN fraction at $1.6\mu m$ is shown in Figure 2.

\begin{figure*}
    \centering
    \includegraphics[height=8cm]{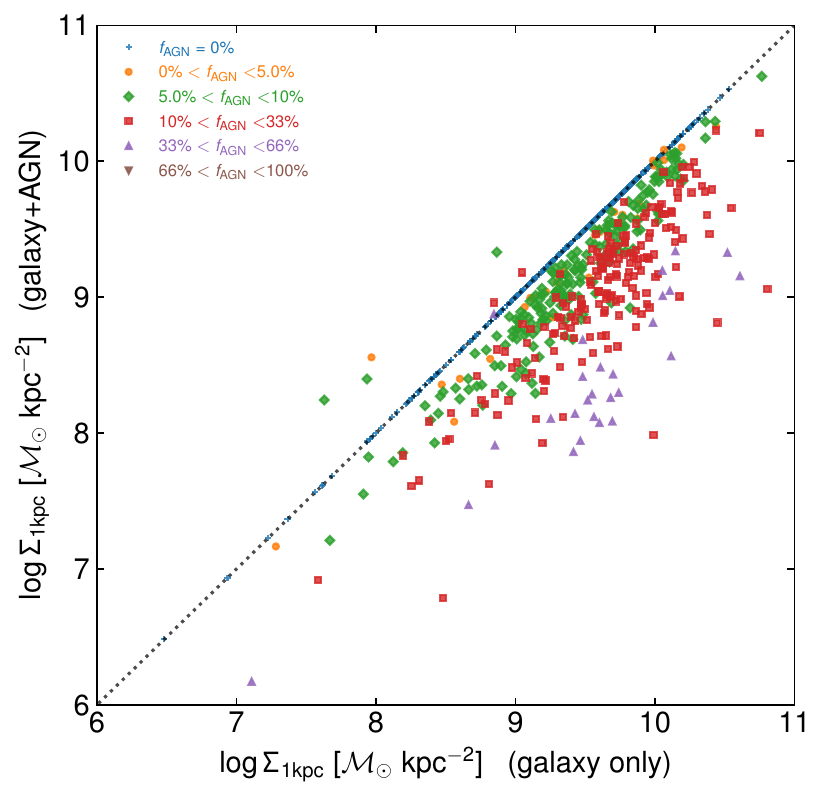}
     \includegraphics[height=8cm]{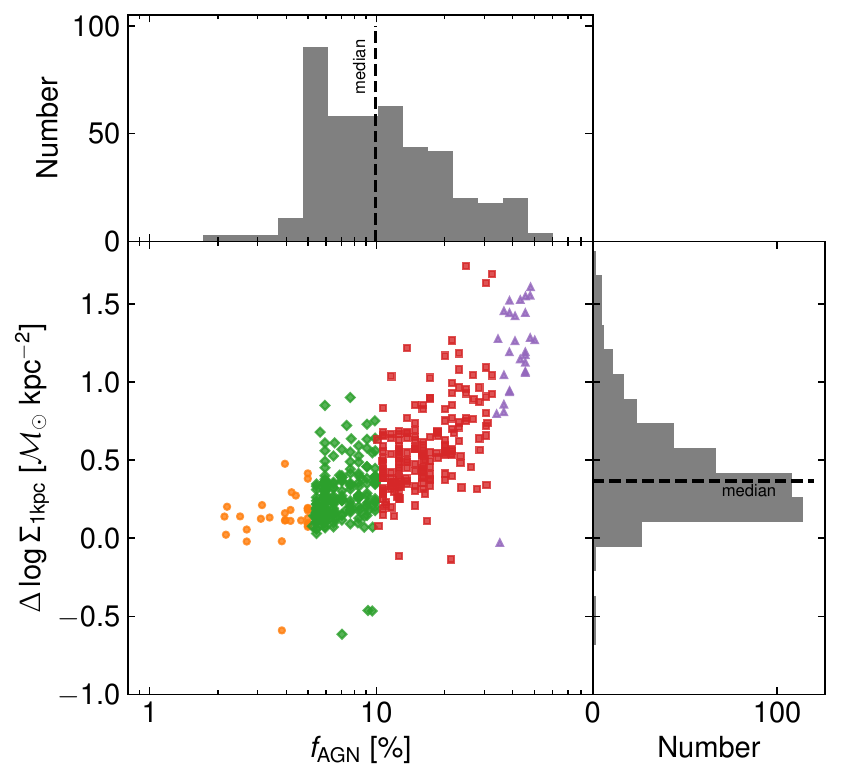}
    \caption{
    \textit{Left:} 
   Measurements of \sigone\ of X-ray sources for the combined galaxy$+$AGN analysis compared to the galaxy-only estimates. Colours/symbols indicate our best estimate of $f_\mathrm{AGN}$, the fractional contribution of the AGN to the optical light for each source.
    We note the small number of sources ($<1$\% of our X-ray detected sample) where our estimate of \sigone \emph{increases} when we allow for AGN light; this is due to changes in the best-fitting stellar population that alters the mass-to-light ratio.  In the vast majority of cases allowing for an AGN contribution results in a lower \sigone estimate.
    \textit{Right:} 
    The change in \sigone\ when allowing for a central point source contribution versus $f_\mathrm{AGN}$ (colours/symbol types as in the left panel). 
    Only sources where an AGN component is required (i.e.~$f_\mathrm{AGN}>0$) are shown on this plot, corresponding to 57\% of the X-ray detections in our sample.  
    The histograms indicate the overall distributions, with black dashed lines indicating the median values. 
    For the sources where we require an AGN component, we typically assign $\sim$10\% of the light to the AGN and reduce our \sigone values by $\sim$0.4~dex. As expected, the change in  \sigone\ increases as $f_\mathrm{AGN}$ increases.
    }
    \label{fig:sigma1_vs_fagn}
\end{figure*}

\subsection{Selection of galaxy populations with different structural and star formation properties}
\label{sec:galsample}

\begin{figure*}
    \centering
    \includegraphics[width=\textwidth,trim=0 0.5cm 0 0]{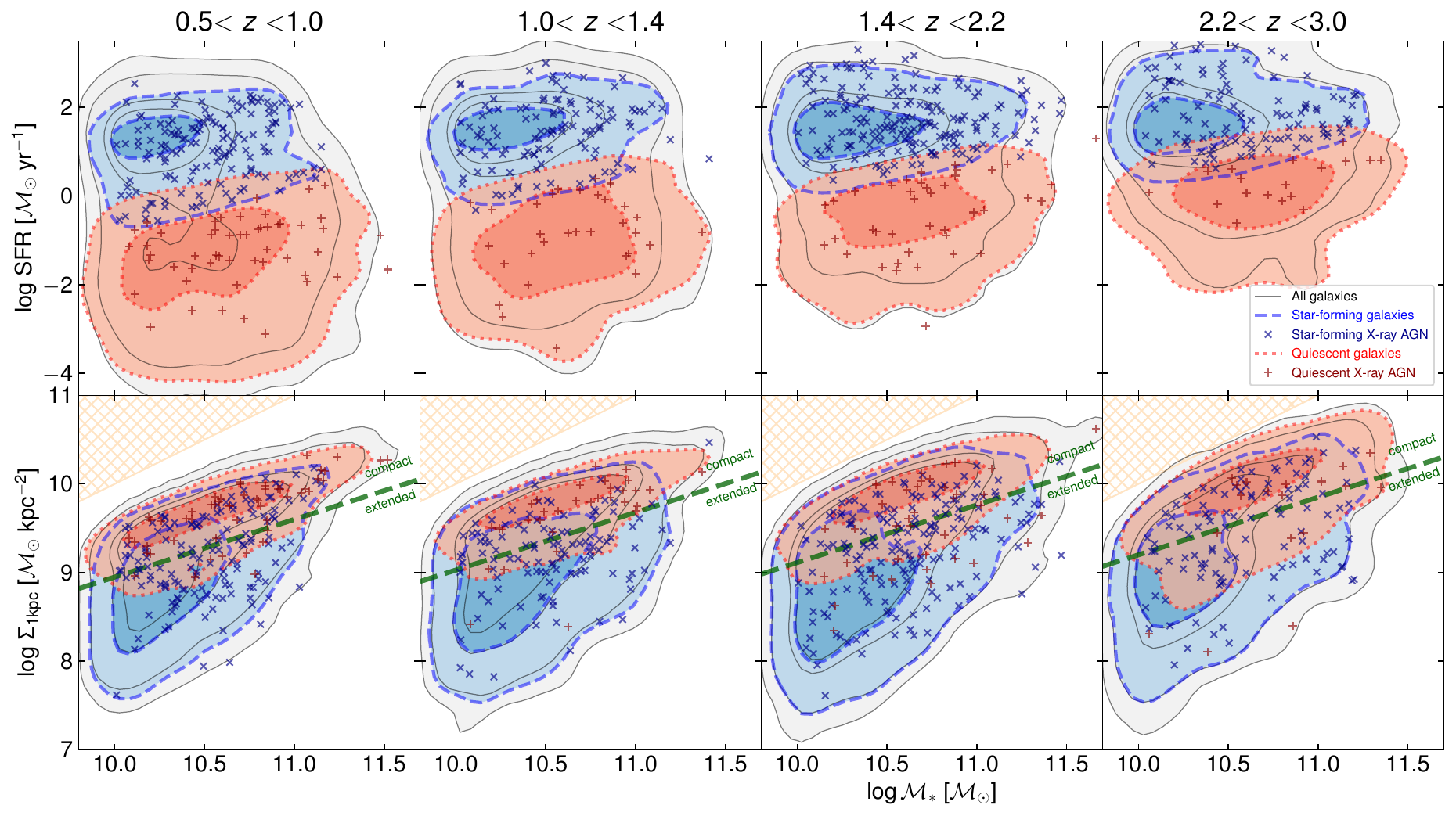}
    \caption{
    SFR (top) and \sigone\ (bottom) as a function of \mstel\ for the parent galaxy population in each of our redshift bins. The underlying grey contours enclose 50, 68, 95 and 99\% of galaxies (above our $\mstel>10^{10}\msun$ limit), whereas the dashed blue and dotted red contours enclose star-forming and quiescent galaxies, respectively (50 and 95\% levels). 
    \redit{We note the overlap between the star-forming and quiescent galaxy contours in the top panels, which is due to the cut in terms of SFR evolving within a given redshift bin (see Equation~\ref{eq:mainseq_cut}).}
    The dark blue crosses indicate hard (2--10~keV) X-ray detected star-forming galaxies whereas dark red pluses indicate hard (2--10~keV) X-ray detected quiescent galaxies. The green dashed line in the lower panels shows our evolving division between compact and extended galaxies. 
    }
    \label{fig:sfrandsigma1_vs_mstel} 
\end{figure*}

With our estimates of \mstel, SFR and \sigone in hand, we now define the galaxy samples used for the analysis in this paper. 
We first place galaxies into redshift bins spanning the range $z=0.5-3.0$ where sufficient cosmological volume is probed by the deep CANDELS fields.
We also apply a uniform stellar mass limit of $\mstel>10^{10}\msun$ above which all five CANDELS fields are complete throughout our redshift range \citep{tal_observations_2014}. This also ensures that we are able to identify AGN down to relatively low specific black hole accretion rates without being substantially impacted by stellar-mass-dependent selection biases \citep{aird_primus_2012}. 
Applying these limits results in a final sample of 7477 galaxies, of which 678 are significant X-ray detections.  
The top panels of  Figure~\ref{fig:sfrandsigma1_vs_mstel} show the distribution of our galaxy samples (and X-ray detections) in each of our redshift bins in the \mstel-SFR plane. 
We further divide our sample into star-forming galaxies, which lie along an evolving ``main sequence of star formation'' in the \mstel-SFR plane, and quiescent galaxies that exhibit suppressed SFRs and lie significantly below the main sequence at a given redshift \citep[e.g.][]{noeske_star_2007,karim_star_2011,schreiber_herschel_2015}, using the cut given in \citet{aird_x-rays_2018}:
\begin{equation}
    \log \mathrm{SFR_{cut}} [\msun \mathrm{yr}^{-1}]
    = -8.9 + 0.76\log \frac{\mstel}{\msun} + 2.95\log (1+z).
    \label{eq:mainseq_cut}
\end{equation}

The lower panels of Figure~\ref{fig:sfrandsigma1_vs_mstel} show where our galaxy samples lie in the \sigone\ versus \mstel\ plane. 
At these redshifts, the majority of quiescent galaxies follow a well-defined sequence in this plane, with relatively high values of \sigone. 
This sequence is well described by 
\begin{equation}
    \log \Sigma_\mathrm{1kpc,Qu-seq} = 0.65 \left(\log \frac{\mstel}{\msun} - 10.5\right)+ 0.80 \log(1+z) + 9.5
    \label{eq:sigone_qu_seq}
\end{equation}
which is in good agreement with prior measurements by \citet{barro_structural_2017}.\footnote{We have applied a small systematic shift of the quisecent galaxy \sigone\ sequence to 0.14~dex higher values compared to \citet{barro_structural_2017} to better match our own analysis of the data.} 
We identify galaxies that lie on this sequence, with $\log \sigone/\Sigma_\mathrm{1kpc,Qu-seq}>-0.4$, as ``compact'' galaxies, whereas galaxies lying below this cut are described as ``extended''.
The majority of star-forming galaxies are ``extended'' in this plane at all redshifts---with a lower fraction of their total mass isolated within the central kpc---but we are able to identify a population of compact star-forming galaxies that lie within 0.4~dex of the quiescent \sigone\ sequence.  
Such galaxies are expected to be short-lived ($\lesssim$1~Gyr) and will subsequently quench their star formation and thus build up the population of compact quiescent galaxies
\citep{barro_candels_2013,van_dokkum_forming_2015}.

We note that our definition of compact and extended changes with redshift due to the redshift dependence of Equation~\ref{eq:sigone_qu_seq}, reflecting the ``size growth'' of the quiescent galaxy population with time whereby the typical quiescent galaxy (at a given \mstel) is larger (i.e.~has a lower \sigone) at lower redshifts \citep{van_dokkum_growth_2010,damjanov_quiescent_2019}. 
This size growth proceeds alongside the overall increase in the number densities of quiescent galaxies due to the quenching of star forming galaxies.
The relatively rare population of extended quiescent galaxies at a given redshift may correspond to galaxies that quenched early via the ``fast-track'' mechanism and have already grown in size, as well as any galaxies that have undergone ``slow-track'' quenching (directly transforming from extended star-forming galaxies without a compact phase). 

We are thus able to identify four distinct galaxy populations: extended star-forming galaxies, compact star-forming galaxies, compact quiescent galaxies, and extended quiescent galaxies (hereafter exSF, cSF, cQu and exQu galaxies) within each of our redshift bins.
The slow- and fast-track quenching pathways suggest evolutionary links between these galaxy populations, 
\redit{although we note that galaxies may not be transforming between these phases \textit{within the same redshift bin}, depending on the width of the redshift bin and how long galaxies spend in a certain evolutionary phase.}
In Section~\ref{sec:results} below we explore the incidence of AGN and their accretion rates within each galaxy population defined here, before returning to a discussion of the evolutionary pathways followed by galaxies as they assemble their stellar mass and their black holes in Section~\ref{sec:galpathways}.

\subsection{Impact of AGN emission on galaxy classification}
\label{sec:agnclass}

\begin{figure*}
    \centering
    \includegraphics[width=\textwidth,trim=0 0.8cm 0 0]{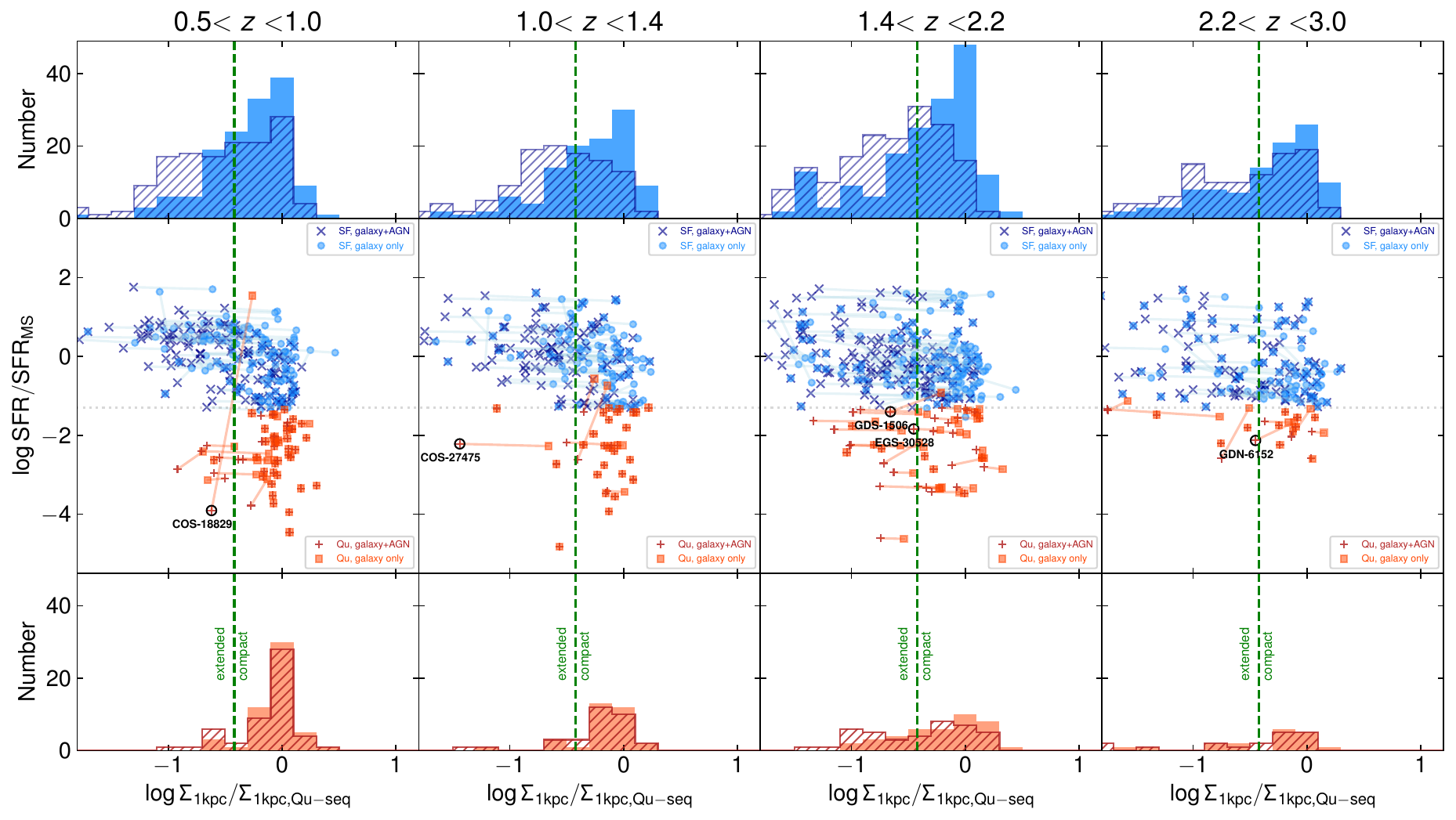}
    \caption{
    Change in classification of X-ray sources when accounting for AGN light in the SED fits and the potential point source contamination of the structural measurements. 
    The central panels show the measurements of the SFR relative to the main sequence ($\log \mathrm{SFR/SFR_{MS}}$) and the measurements of \sigone\ relative to the quiescent galaxy sequence ($\log \sigone/\Sigma_\mathrm{1kpc,Qu-seq}$) i.e. the parameters that we use to divide star-forming vs. quiescent galaxies and compact vs. extended galaxies. 
    The light blue circles and light red squares show our original measurements that do not account for any AGN contribution, which are linked to the corresponding final measurements (dark blue crosses and dark red pluses), with the colour scheme indicating the final star-forming vs. quiescent classification. 
    The upper and lower panels show histograms of the \sigone measurements for the star-forming and quiescent galaxies, respectively, where the light filled histograms are the original ``galaxy-only'' measurements and the darker hatched histograms are the results with our updated ``galaxy$+$AGN'' fits. 
    The black circles highlight five sources that undergo substantial changes in their measured values with our two-component fitting, which are examined in Figure~\ref{fig:color_thumbnails} below.
    }
    \label{fig:change_in_class}
\end{figure*}

Having classified our full galaxy sample into four different populations, 
we now re-visit the classification of the X-ray detected sources
to determine and correct for the impact of the central AGN emission on these classifications. 
In the central row of Figure~\ref{fig:change_in_class} we show where detected sources lie in terms of their SFRs relative to the main sequence of star formation ($\mathrm{SFR/SFR_{MS}}$) and their \sigone compared to the quiescent galaxy sequence ($\sigone / \Sigma_\mathrm{1kpc,Qu-seq}$), i.e.~the space that we use to separate star-forming versus quiescent and compact versus extended galaxies. 
The light blue circles and light red squares in the central panels show the measurements for each X-ray source \emph{before} accounting for the AGN light, whereas the dark blue crosses and dark red pluses show our measurements \emph{after} allowing for two components in the SED fitting (AGN and galaxy components) and in our structural measurements (Sersic and point source components), if deemed necessary by our process described in Section~\ref{sec:galfit} above. 
Lines link these two sets of measurements to illustrate the change for individual sources. 
In the top and bottom rows of Figure~\ref{fig:change_in_class} we also show the distributions of $\log (\sigone / \Sigma_\mathrm{1kpc,Qu-seq})$ for X-ray sources (for star-forming and quiescent classifications, respectively) before and after correcting for an AGN component (solid versus hatched histograms).

\begin{figure*}
    \centering
    \includegraphics[height=5cm]{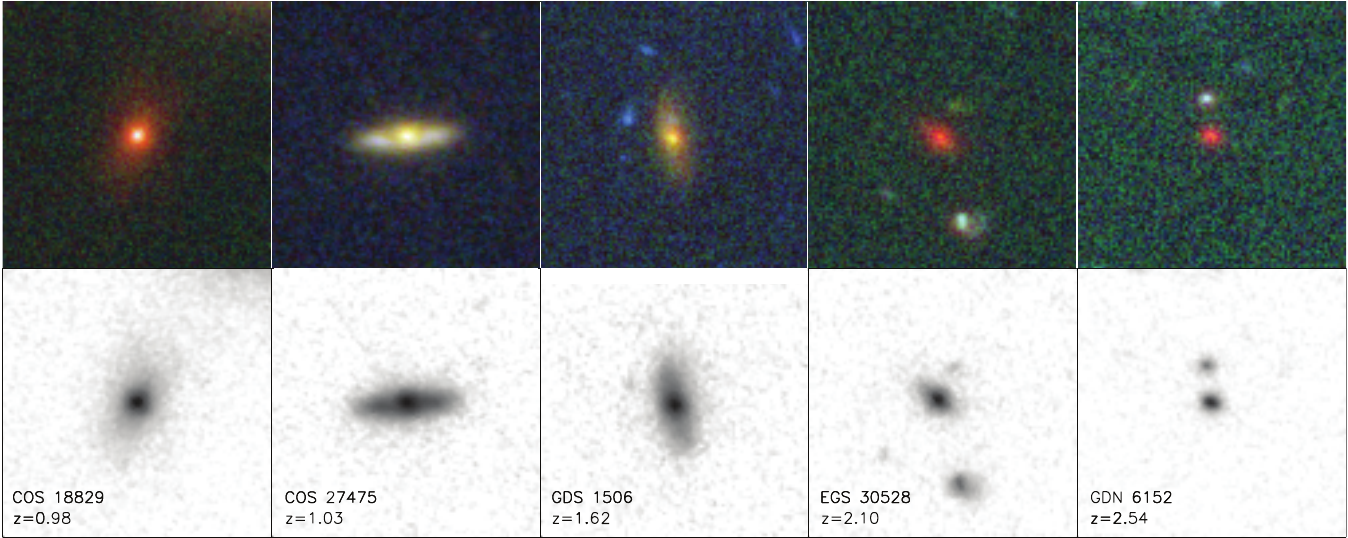}
    \caption{Colour composite and F160W grey-scale images of five galaxies that undergo significant changes to their classification when we include AGN emission in our SED and GALFIT modeling.  These five galaxies are highlighted in Figure \ref{fig:change_in_class}.  In most cases, point-like emission embedded in a more extended structure is clearly visible.  Taking this emission into account reduces $\Sigma_{\rm 1 kpc}$ and, occasionally, decreases the galaxy's calculated SFR. This is most dramatically exemplified by COS 18829, where we see a point source \redit{with relatively bluer colours (appearing yellow in the image)} embedded in an otherwise red extended galaxy. The classification of COS 18829 changes from cSF to exQu as a result of our two-component modeling.}
    \label{fig:color_thumbnails}
\end{figure*}

For star-forming galaxies, allowing for the emission from a blue, central point source (i.e. an AGN) has a tendency to decrease our estimates of \sigone\ for the X-ray detected sources and there is thus an overall shift of the sample from the cSF to the exSF class.
In contrast, there is much less change in the 
structural distribution of quiescent galaxies 
even if the estimates for \emph{individual} sources can change significantly. 
\redit{Our quiescent galaxy sample tends to be higher stellar mass and to host relatively lower X-ray luminosity AGN, particularly at lower redshifts, and thus the host galaxy generally dominates the optical/NIR light and requires little correction for the AGN contribution (at higher redshifts, our X-ray flux limits mean we only detect the most X-ray luminous sources in quiescent galaxies and thus a higher proportion require corrections for AGN contamination).}
Overall, 142 of the 678 X-ray detected sources (21\%) are changed from a compact to an extended classification as a result of the corrections for AGN light.
While there are a few notable individual exceptions, very few X-ray sources move between the star-forming and quiescent classifications; in general, the overall light is dominated by the galaxy component for these relatively faint AGN.

In Figure~\ref{fig:color_thumbnails} we present images for a selection of sources that undergo the most extreme changes in their classification. These sources are marked in Figure~\ref{fig:change_in_class}, with increasing redshift from left to right. 
COSMOS 18829 changes classification from cSF to exQu when accounting for AGN light; indeed the extended emission can be seen in the F160W image of this source. COSMOS 27475 does not change classification, but the \sigone\ value for this source decreases substantially.  
GOODS-S 1506 changes classification from cSF to exQu, while EGS 30528 changes from cQu to exQu.  Here again the extended nature of the source can be seen in the F160W image. GOODS-N 6152 also changes classification from cQu to exQu, just crossing the threshold for our definition of an extended galaxy; we note that the typical quiescent galaxy is very compact at these redshifts (i.e.~the normalisation of \sigseq is high) and thus---while accurately identified as exQu---this source appears relatively compact in the image shown in Figure~\ref{fig:color_thumbnails}, albeit with a weak extended component visible in the grey-scale image.
The impact of these corrections on our results is explored in Section~\ref{sec:robustness} below.

\section{The distribution of black hole accretion rates for galaxies with different structural and star formation properties}
\label{sec:results}

In this section, we use the Bayesian methodology of \citet{aird_x-rays_2017,aird_x-rays_2018} to measure the probability distributions of AGN accretion rates within our different galaxy samples, i.e. the probability density (per dex in accretion rate) of hosting an AGN of a given accretion rate.  
Our measurements are based on the hard X-ray imaging and fully incorporate the information from both detected sources and non-detections. 
The expected contribution to the X-ray emission from galactic processes (i.e. high- and low-mass X-ray binary populations within a galaxy, rather than a central AGN) is accounted for as an additional background component, as described in appendix~B of \citet{aird_x-rays_2018}. 
Based on these accretion rate probability distributions, we are able to provide robust estimates of the X-ray AGN fraction \emph{to well-defined limits} that are corrected for incompleteness due to the varying sensitivity limits of the \textit{Chandra} imaging, both within and between fields. 
Accounting for the effects of incompleteness, galactic processes, and non-detections is vital to determine meaningful and robust measurements of the true AGN fraction within galaxy populations.
Section~\ref{sec:mainresults} presents our primary results for the exSF, cSF, cQu and exQu galaxy populations.
\redit{A number of tests to check the robustness of these results are described in Appendix~\ref{appendix:robustness}.}
In Section~\ref{sec:refined} we repeat our analysis in more refined bins in central stellar mass density to further explore the dependence of AGN activity on \sigone \emph{within} each of the four galaxy populations.

\subsection{Incidence of AGN as a function of structural properties using various definitions of AGN accretion rates}
\label{sec:mainresults}

\begin{figure*}
    \centering
    \includegraphics[width=0.82\textwidth,trim=0 0.5cm 0 0.1cm]{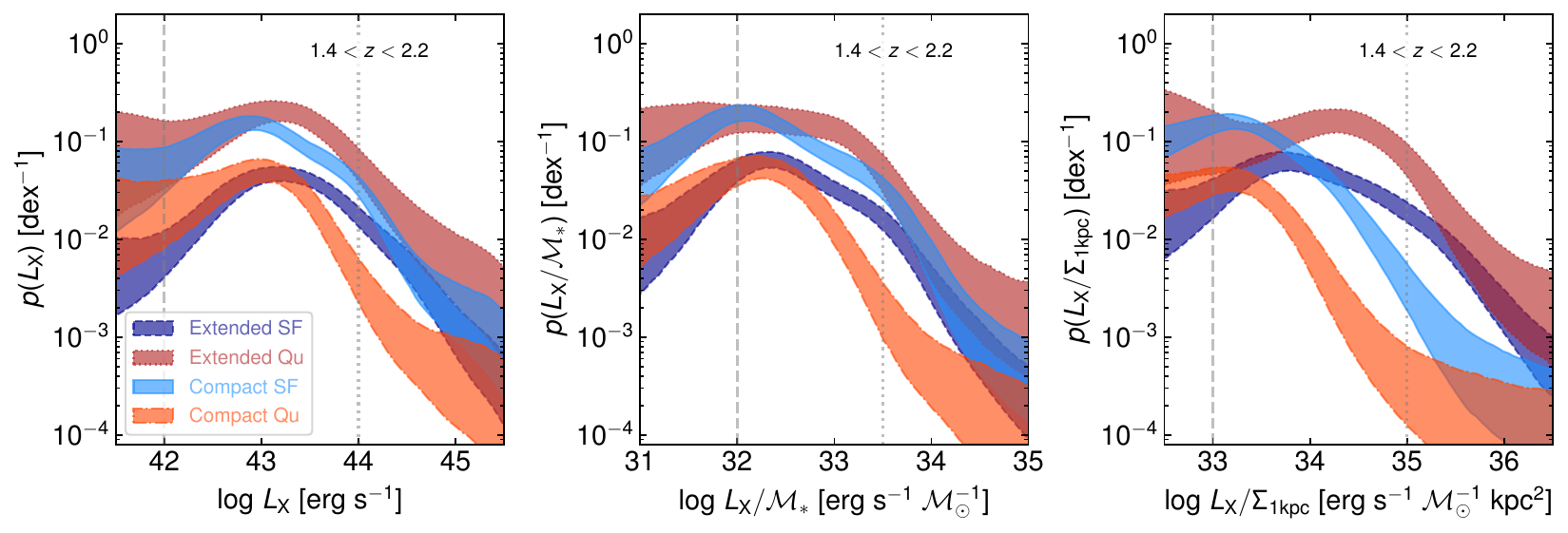}
    \caption{ 
    Measurements of accretion rate probability distributions in one of our redshift bins ($1.4<z<2.2$), comparing the distributions in exSF, cSF, cQu and exQu galaxies (as indicated by the colours) for our three different accretion rate tracers: \LX (left), \LX/\mstel or ``specific accretion rate'' (centre) and \LX/\sigone (right). 
     The normalisation of $p(\LX)$ and $p(\LX/\mstel)$ is enhanced in cSF galaxies compared to exSF galaxies, whereas cQu galaxies have both a lower normalisation and the distribution is shifted to typically lower accretion rates. The distributions are highest within the rare population of exQu galaxies, showing that such galaxies have a high incidence of AGN, with a wide range of accretion rates. 
    The differences are exaggerated when quantifying AGN activity relative to the central stellar mass (i.e.~using \LX/\mstel, right panel) indicating that both exSF and exQu galaxies are assembling their black holes at higher rates relative to their central stellar densities compared to cSF or cQu galaxies.
        }
    \label{fig:mainresults_p}
    \includegraphics[width=0.82\textwidth,trim=0 0.5cm 0 0.2cm]{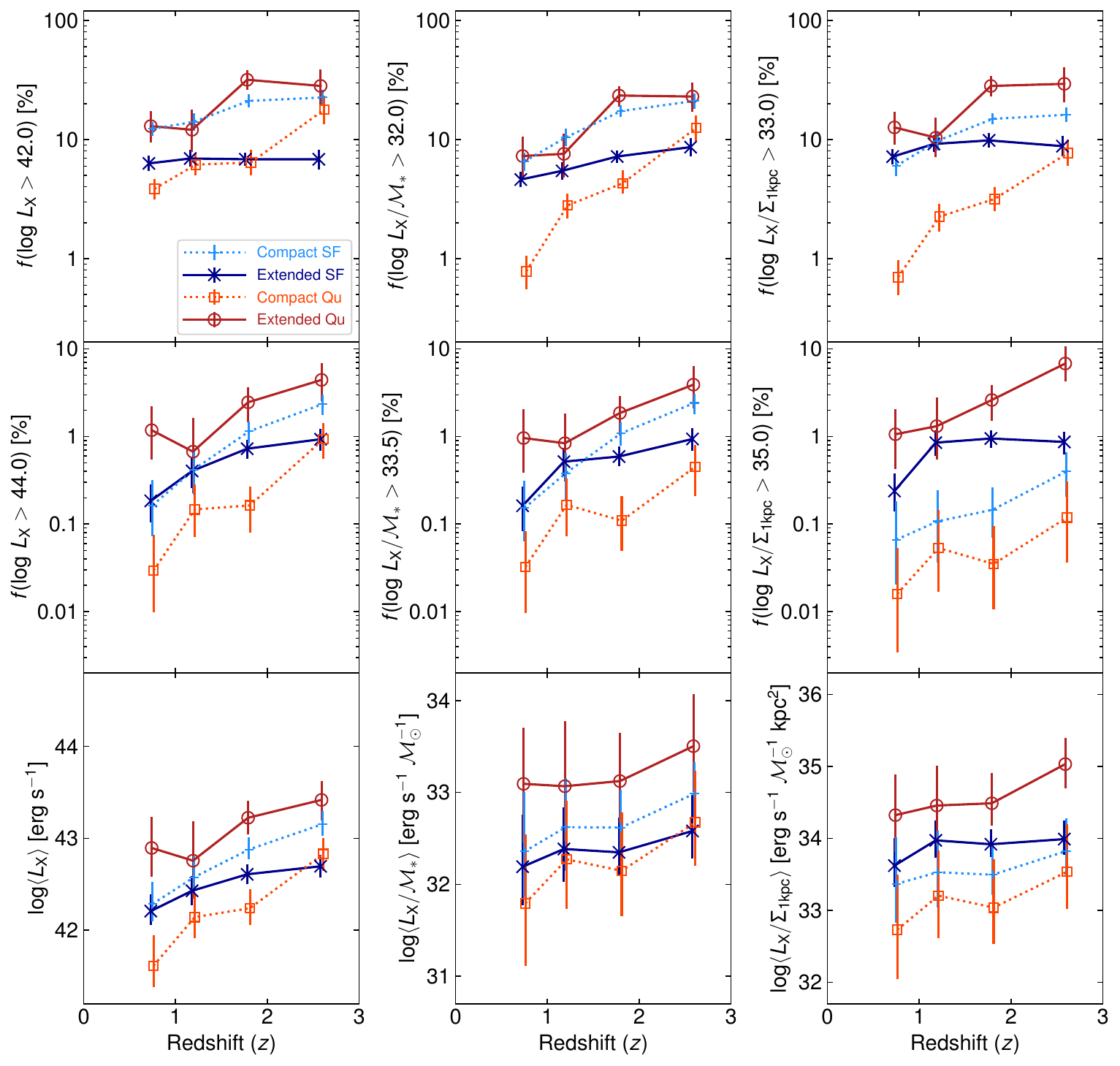}
   \caption{
    Measurements of AGN fractions to different limits (top and middle rows, as indicated by the y-axis labels) and average accretion rates (bottom row) as a function of redshift, which summarise our full accretion rate probability distributions (e.g.~Figure~\ref{fig:mainresults_p}), comparing the \redit{cSF (light blue pluses), exSF (dark blue crosses), cQu (light red squares) and exQu (dark red circles)} galaxy populations. 
    Each column adopts a different accretion rate tracer: \LX (left), \LX/\mstel or ``specific accretion rate'' (centre), and \LX/\sigone (right) that tracks black hole growth rates relative to the central stellar density.
    In general we measure the highest AGN fractions in exQu galaxies at all redshifts. 
    Based on \LX or \LX/\mstel, the fraction is enhanced in cSF galaxies compared to exSF galaxies, especially when deriving AGN fractions to lower limits (top row), while the fraction in cQu galaxies is suppressed. 
    In contrast, when tracking accretion relative to the central stellar mass (i.e. using \LX/\sigone, right column), we find that AGN activity is enhanced in both exSF and exQu  compared to compact galaxies.
    We note that the average accretion rates shown in the bottom row tend to be noisier than the AGN fractions and do not fully capture the underlying differences in the probability distributions. 
    }
    \label{fig:mainresults}
\end{figure*}

Figure~\ref{fig:mainresults_p} presents our measurements of the probability distribution of AGN accretion rates within our four different galaxy populations (shown for the $1.4<z<2.2$ redshift bin) while Figure~\ref{fig:mainresults} shows quantities derived therefrom across all of our redshift bins from $z=0.5$ to 3:~robust AGN fractions to specified limits and the sample averaged accretion rates. 
As described in \citet[see figure~1 therein for an illustration]{aird_x-rays_2019}, we derive AGN fractions by integrating our probability distributions to fixed limits,
\begin{equation}
    f(\log \lambda> X_\mathrm{lim}) = 
    \int_{X_\mathrm{lim}}^\infty p(\lambda) \; \mathrm{d} \log \lambda
\end{equation}
where $\lambda$ indicates our chosen accretion rate tracer (\LX, \LX/\mstel, or \LX/\sigone, as described in turn below) and $X_\mathrm{lim}$ is a chosen limit in this quantity (in logarithmic units) by which we define a robust AGN fraction.
Sample averaged accretion rates are determined by averaging $\lambda$ over the probability distributions,
\begin{equation}
    \langle \lambda \rangle = 
     \int_{-\infty}^\infty \lambda \; p(\lambda) \; \mathrm{d} \log \lambda
\end{equation}
and provides an indicator of the ``typical'' accretion rate for all galaxies in a given sample.

In the left-hand column of Figures~\ref{fig:mainresults_p} and \ref{fig:mainresults} we use the observed X-ray luminosity\footnote{Based on the observed 2--7~keV fluxes and converted to rest-frame 2--10~keV luminosities assuming an X-ray spectrum with photon index $\Gamma=1.9$, hereafter \LX\ and given in units of \ergs.}  directly and obtain robust measurements of the fraction of galaxies with AGN to luminosity limits of $\log \LX>42$ and $\log \LX>{44}$ as well the sample-averaged X-ray luminosity, $\langle \LX \rangle$, as a function of redshift.
We emphasise that these fractions are not limited to directly detected X-ray AGN and instead represent a statistically robust result that is inferred using our Bayesian analysis and accounts for (and allows us to probe below) the sensitivity limits of the X-ray data.

We find that AGN span a broad range of X-ray luminosities ($42\lesssim \log \LX\lesssim45$) in all four galaxy populations. 
Our measurements of $p(\LX)$ at $1.4<z<2.2$ (Figure~\ref{fig:mainresults_p}) show a clear enhancement in the incidence of low-to-moderate \LX AGN in cSF galaxies compared to exSF galaxies at that same redshift, which is reflected in the AGN fractions (Figure~\ref{fig:mainresults}), in particular the AGN fraction above our lower luminosity limit, $f(\log\LX>42$).
This enhancement in cSF galaxies becomes weaker at lower redshifts
and is not significant for $z<1.4$ for the higher luminosity limit ($\log\LX>44$). 
The sample averaged luminosity, $\log \langle \LX \rangle$, also shows a mild enhancement in cSF galaxies compared to the exSF population, which increases with increasing redshift.

We find that the AGN fractions and average luminosities are consistently a factor $\sim2-10$ lower in cQu galaxies compared to cSF galaxies at the same redshift, which appears to be driven by both a lower normalization in $p(\LX)$ and a shift in $p(\LX)$ to lower luminosities.
In contrast, we find a surprisingly high AGN fraction in the exQu galaxy samples that in most luminosity and redshift ranges exceeds the incidence in cSF galaxies. 
We note that exQu galaxies are relatively rare at any redshift (given our definition relative to the evolving quiescent galaxy size--mass sequence) and thus while the fraction of exQu galaxies with an X-ray AGN is high, such sources remain a minority of the overall AGN population. The vast majority of X-ray AGN are still found within the (much more numerous) star-forming galaxy populations. 

The identification of X-ray AGN is known to be strongly affected by stellar-mass-dependent selection biases \citep[][]{aird_primus_2012,aird_primus_2013,mendez_primus_2016}.
The observed fraction of X-ray AGN (to a given luminosity limit) is found to rise significantly with increasing stellar mass. 
Broadly, more massive galaxies are expected to host more massive central black holes and will produce a higher observable luminosity even if they are only weakly accreting, relative to the mass of the galaxy.
To account for this selection bias over our broad stellar mass bin ($10<\log \mstel/\msun<11.5$), in the central columns of Figures~\ref{fig:mainresults_p} and \ref{fig:mainresults} we present measurements of the distribution of specific black hole accretion rates, $\LX/\mstel$, normalising the observed X-ray luminosity by the stellar mass.\footnote{For clarity, in this paper we do not translate $\LX/\mstel$ into ``Eddington-rate-equivalent'' units, in contrast to our prior work  \citep{aird_x-rays_2018,aird_x-rays_2019} as this requires assumptions on the AGN bolometric correction and---most crucially---the highly uncertain scaling between total stellar mass and black hole mass. Dividing $\LX/\mstel$ by $10^{34}$ converts our values to the roughly ``Eddington-rate-equivalent'' specific black hole accretion rate, $\lambda_\mathrm{sBHAR}$, using the assumed scale factors from \citet{aird_x-rays_2018}, and thus our limit of $\log \LX/\mstel>32.0$ roughly corresponds to Eddington ratios of $\gtrsim$1\%.}
We also present AGN fractions to the specified limits ($\log \LX/\mstel>32.0$ and $>33.5$, where \LX\ is in \ergs\ and \mstel is in units of \msun) and the sample averaged $\log \langle \LX/\mstel \rangle$ as a function of redshift.
Our Bayesian methodology naturally allows for this conversion and the impact on our measurements due to the varying X-ray sensitivity limits across our fields, the differences in redshift, and the differences in stellar masses.
Accounting for this bias improves the robustness and interpretation of our results, and the overall patterns for our $\LX/\mstel$ results are consistent with the \LX-based results: an enhancement in the incidence of moderate accretion rate AGN in cSF versus exSF galaxies at $z\gtrsim1.4$; a lower fraction in cQu galaxies;
and a high incidence of AGN in exQu galaxies.

While stellar-mass-dependent selection biases are known to have a significant and important effect on AGN samples, the total stellar mass may be a somewhat poor tracer of the central black hole mass, especially when considering galaxies with a broad range of star formation and structural properties (\citealt{kormendy_coevolution_2013}; but see also \citealt{reines_relations_2015}, \citealt{bentz_black_2018}).
Black hole mass assembly may be more directly related to the build up of the central stellar bulge of galaxies \redit{\citep[e.g.][]{caplar_agn_2018}.} 
Thus, in the right-hand columns of Figures~\ref{fig:mainresults_p} and \ref{fig:mainresults}, we present measurements of the incidence of X-ray AGN across our four galaxy populations in terms of $\LX/\sigone$, normalising the X-ray luminosity by the central stellar mass density.
These measurements allow us to assess the extent of central black hole growth \emph{relative to stellar mass built up in the central regions of galaxies}. 
This translation has a substantial impact on our results.\footnote{\citet{fang_link_2013} suggest that \sigone may trace black hole mass as $\mbh\propto\sigone^{2.0}$. Allowing for this additional exponent in our accretion rate tracer (i.e. measuring \LX/$\sigone^2$) would further exaggerate the differences that we see between the compact and extended populations.}
Both of the compact galaxy populations, cSF and cQu galaxies, have relatively \emph{high} \sigone values (compared to their total \mstel), leading to typically \emph{low} values of $\LX/\sigone$, whereas exSF and exSF have lower \sigone\ and are thus enhanced in terms of \LX/\sigone. 
When considering consistent limits in \LX/\sigone, the extended galaxy populations generally have higher AGN fractions at all redshifts, particularly when considering the higher threshold AGN fractions, $f(\log \LX/\sigone>35.0$) (middle right panel of Figure~\ref{fig:mainresults}).
In particular, cSF galaxies, which are found to have a reasonably high incidence of AGN in terms of \LX or \LX/\mstel, are actually accreting at \emph{lower} rates in terms of \LX/\sigone compared to their extended counterparts. 
Thus, while cSF galaxies may host relatively luminous X-ray AGN in absolute terms, exSF galaxies may be growing their central black holes at higher rates relative to the stellar mass already assembled within the central regions.
Such differences suggest that the assembly of a central black hole could \emph{precede} the assembly of the central stellar bulge.
We also note that it is the exQu galaxy population that shows the highest AGN fractions and average accretion rates in terms of \LX/\sigone out of our four galaxy populations: these galaxies are not assembling more stars (they have low SFRs) and they are relatively large in physical size (whether they formed that way initially or have undergone a growth in size over their recent history), yet they are continuing to produce AGN emission and grow their central black holes. 
We investigate the implications of these findings on our understanding of the growth of black hole mass during the different phases of galaxy evolution in Sections~\ref{sec:bhgrowth} and~\ref{sec:galpathways} below.

\begin{figure*}
\includegraphics[width=0.9\textwidth]{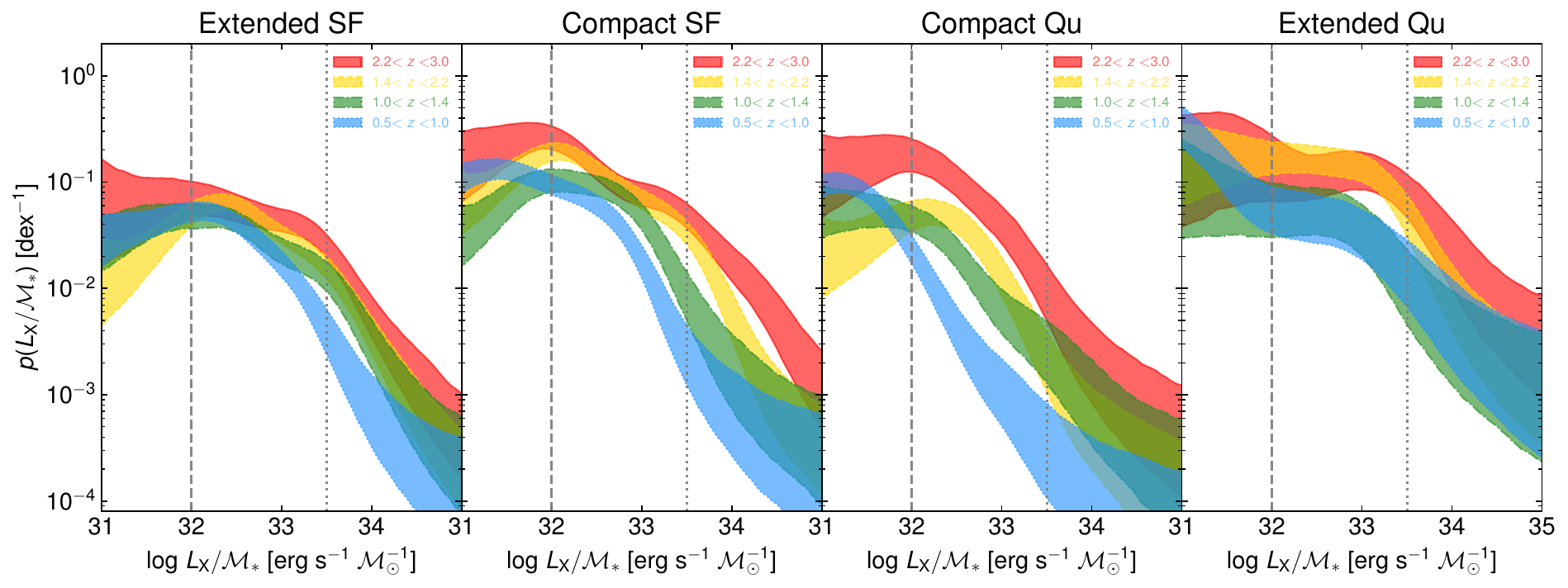}
\caption{
Measurements of $p(\LX/\mstel)$ for each of the four galaxy populations as a function of redshift. Significant evolution is seen in all four populations. 
In exSF galaxies, there is a small shift in $p(\LX/\mstel)$ toward higher accretion rates at higher redshift, resulting in moderate evolution in the AGN fraction above the higher accretion rate limit, $\log\LX/\mstel>33.5$, but negligible evolution in the overall AGN fraction above the lower limit of $\log\LX/\mstel>32.0$ (see Figure~\ref{fig:mainresults}). In cSF, cQu and exQu galaxies there is an evolution in both the normalisation and typical accretion rate of $p(\LX/\mstel)$ with increasing redshift.
}
\label{fig:pledd_vs_z}
\end{figure*}

The AGN fractions and average accretion rates shown in Figure~\ref{fig:mainresults} tend to increase with increasing redshift, consistent with the overall increase in the prevalence of AGN at earlier cosmic times for galaxies with $\mstel\gtrsim10^{10}$~\msun  \citep[see e.g.][]{aird_x-rays_2018}. 
The exception is in exSF galaxies, where the AGN fractions above our lower adopted limits (dark blue \redit{crosses} in the top row of Figure~\ref{fig:mainresults}) are approximately constant or show only a mild increase with redshift. In Figure~\ref{fig:pledd_vs_z} we show the measurements of $p(\LX/\mstel)$ (our preferred accretion rate tracer) as a function of redshift in each of the four galaxy populations.
The evolution of accretion rates in exSF galaxies is driven by a shift toward higher \LX/\mstel at higher redshifts. However, the overall \emph{normalisation} remains approximately constant. The AGN fractions to lower limits, below the overall peak in the distribution, thus remain approximately constant with redshift. The bulk of AGN, found in $\sim$8--10\% of galaxies at a given time, are captured by these limits and thus the overall AGN fraction does not increase, even though those AGN tend to be slightly more luminous at higher redshifts (see bottom row of Figure~\ref{fig:mainresults}). 
In contrast, the accretion rate probability distributions for the cSF, cQu and exQu populations in Figure~\ref{fig:pledd_vs_z} all show a shift in \LX/\mstel toward higher accretion rates at higher $z$ \emph{and} an increased normalisation, representing an increase in the triggering rate of AGN at higher $z$.

We note that the sample average accretion rates alone (whether in terms of \LX, \LX/\mstel or \LX/\sigone) are insufficient to fully characterise the differences in the AGN content of the four galaxy populations. 
While the sample averages follow similar overall trends to the AGN fractions, it is only with a characterisation of the \emph{full probability distributions} that shifts in the normalisation or shifts in the typical accretion rates can be distinguished from each other. 
Furthermore, the sample averaged accretion rates have higher statistical uncertainties and can be severely skewed by the precise shape of the underlying probability distributions at high and low accretion rates.
AGN fractions to well-defined limits---provided these are corrected for the sensitivity limits of the X-ray imaging---are more robust. 
Measuring the full distributions of accretion rates, and summarising these differences in terms of AGN fractions (indicating how often galaxies host an AGN) as well as the sample average accretion rates (indicating how rapidly those black holes are growing) is required to obtain a complete and accurate picture. 

\redit{In Appendix~\ref{appendix:robustness} we describe a variety of tests of the robustness of our results.
Specifically, we show that neglecting the impact of AGN light at optical/IR wavelengths on our measurements of galaxy structural and star formation properties results in a significant overestimate of the AGN fraction in cSF galaxies (by up to a factor $\sim$2) and a correspondingly lower AGN fraction in exSF galaxies, although our results in cQu and exQu galaxies are not significantly changed (see Section~\ref{sec:galonly_vs_best}). 
In contrast, changing the method to classify compact galaxies from our preferred evolving cut in $\sigone/\sigseq$ to a simpler fixed cut in \sigone (see Section~\ref{sec:strictsigone}) or
or accounting for differences in the mass versus the light profile of our galaxies (see Section~\ref{sec:mass_vs_light})
has a minor impact on our AGN fractions for all four galaxy populations, showing that our results are robust to the precise definition of compact or extended galaxy populations  
Using soft (0.5--2~keV) X-ray data to select AGN---rather than the hard (2--7~keV) band with careful corrections for the sensitivity limits---leads to systematically lower estimates of the AGN fraction (by a factor $\sim2$) across all four galaxy populations, indicating that soft X-ray selection fails to identify the dominant, moderately absorbed AGN populations across our entire redshift range (see Section~\ref{sec:hard_vs_soft}).
Finally, we show that there are large field-to-field variations in our measurements, especially for the rarer cQu and exQu galaxy populations (albeit with large statistical uncertainties), demonstrating the need to combine the power of all five CANDELS fields to accurately probe the AGN incidence across compact and extended galaxy populations out to high redshift. 
}

\subsection{Incidence of AGN as a function of central stellar mass density within each galaxy population}
\label{sec:refined}

In this section we explore the dependence of AGN activity on the central stellar mass density of galaxies in more detail, to reveal the dependence on \sigone \emph{within}, as well as between, the four galaxy populations.
We repeat our measurements of AGN fractions within galaxies using bins in central stellar mass density, separating galaxies 
in terms of 
$\log \sigone/\Sigma_\mathrm{1kpc,Qu-seq}$, which measures ``compactness'' {\it relative} to the evolving, stellar-mass-dependent quiescent galaxy sequence, $\Sigma_\mathrm{1kpc,Qu-seq}$, as given by Equation~\ref{eq:sigone_qu_seq}. 
We are thus able to track any dependence on compactness within both the star-forming and quiescent galaxy populations, rather than simply separating them into compact and extended.
To increase the size of the galaxy samples, we merge our two higher and two lower redshift bins for this analysis. 

\begin{figure*}
    \centering
    \includegraphics[width=0.7\textwidth,trim=0 0.5cm 0cm 0]{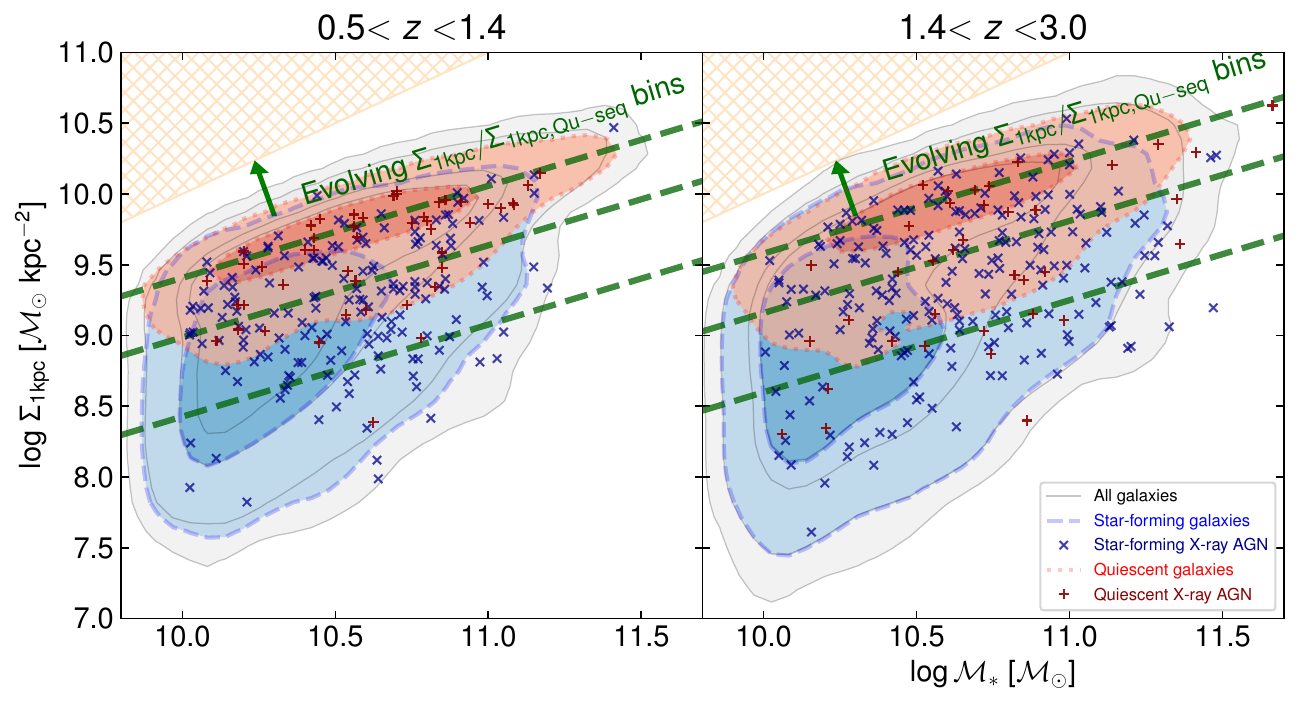}
    \caption{
    Illustration of our additional bins in central stellar mass density that are adopted for the analysis of the AGN incidence \emph{within} the four galaxy populations in Section~\ref{sec:refined}.
        We define ``tilted'' bins defined relative to the evolving $\sigone-\mstel$ sequence for quiescent galaxies, using the fixed bins (as indicated by the green dashed lines) as well as adopting a flexible binning scheme with a minimum bin size of 0.2~dex and at least 100 galaxies in a bin (see Figure~\ref{fig:fagn_vs_Sigma1relative}).
    The underlying contours show the distribution of star-forming (blue) and quiescent (red) galaxies (enclosing 50\% and 95\% of galaxies), while the blue crosses and dark red pluses indicate hard X-ray detected star-forming and quiescent galaxies, respectively.
    }
    \label{fig:refined_sigone_bins}
    \centering
    \includegraphics[width=\textwidth,trim=0 0.8cm 0cm 0]{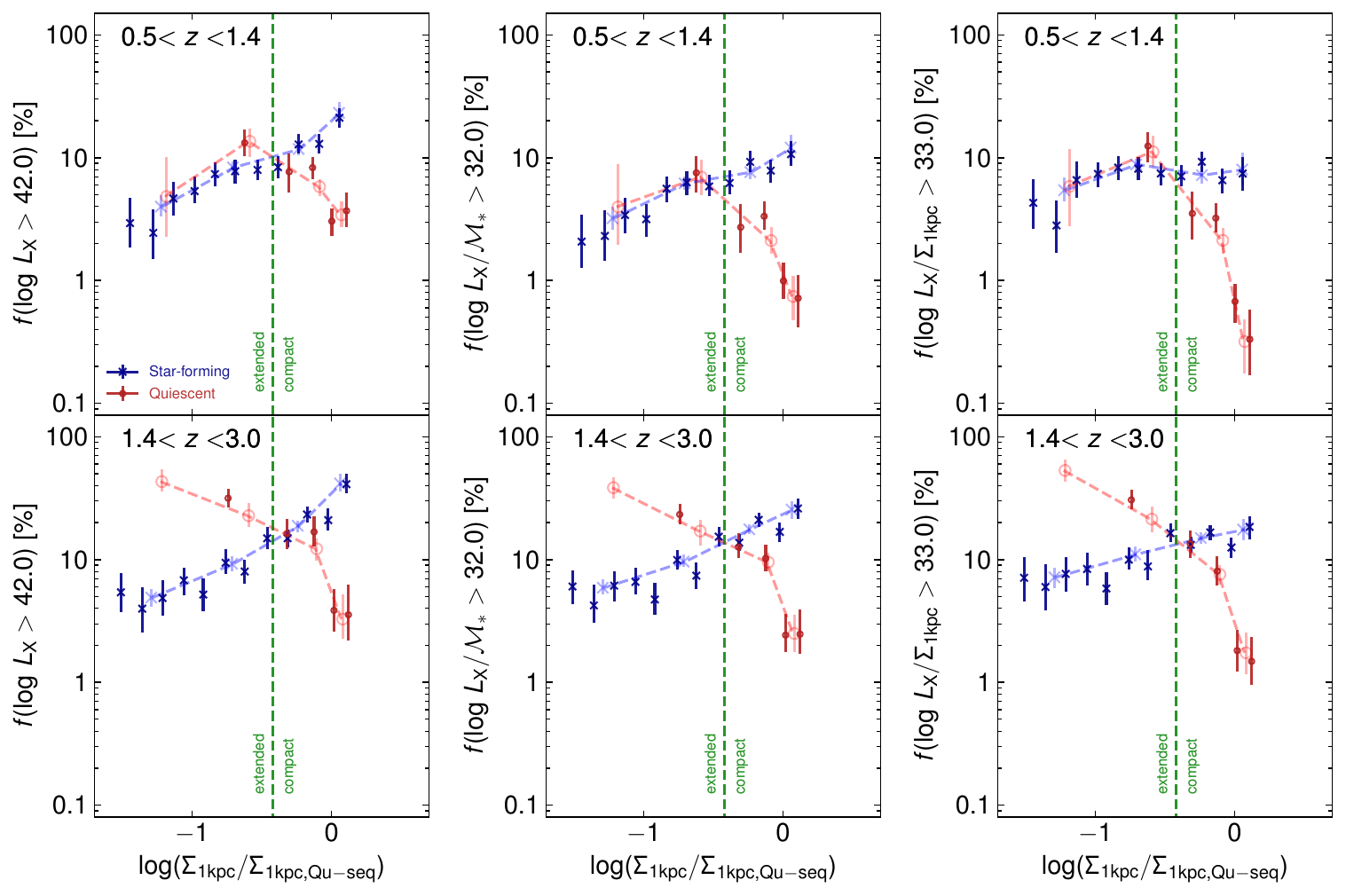}
    \caption{
    Measurements of AGN fractions  as a function of $\sigone/\sigseq$ \emph{within} the galaxy populations.
    We adopt limits in terms of \LX (left panels) \LX/\mstel (middle) and \LX/\sigone (right) for two broad redshift bins (as indicated). 
    The light blue crosses and light red circles show measurements for star-forming and quiescent galaxies, respectively, in fixed $\log \sigone/\sigseq$ bins (as shown by the green dashed lines in Figure~\ref{fig:refined_sigone_bins}), with the data point placed at the median value within a bin. 
    The smaller, darker points instead adopt a flexible binning scheme, requiring at least 100 galaxies in a bin and setting a minimum bin width of 0.2~dex in $\log\sigone/\sigseq$. 
    The vertical green dashed line shows the division between extended and compact galaxies used for our main results (Section~\ref{sec:mainresults}). 
    The results in this figure show that there is a dependence on the AGN fraction on \sigone/\sigseq \emph{within} either compact or extended galaxy populations. 
    In star-forming galaxies, the AGN fraction (to \LX or \LX/\mstel limits) increases as galaxies become relatively more compact; the converse is seen within quiescent galaxies. 
    In the higher redshift bin, the AGN fraction in the most extended quiescent galaxies exceeds the AGN fraction within star-forming with comparable \sigone/\sigseq. 
    }
    \label{fig:fagn_vs_Sigma1relative}
\end{figure*}

Figure~\ref{fig:refined_sigone_bins} shows the initial four bins in $\sigone/\Sigma_\mathrm{1kpc,Qu-seq}$ that we adopt in each of our redshift ranges to further divide each population.
We also adopt a flexible binning scheme where we require a minimum of 100 galaxies in a bin or a minimum bin width of 0.2~dex in \sigone/\sigseq, 
providing us with additional bins to closely track any dependence of \sigone/\sigseq while ensuring there are sufficient galaxies in a bin to obtain an accurate measurement of the AGN accretion rate probability distribution. 
The resulting measurements of AGN fractions are shown by the blue (star-forming) and red (quiescent) points 
in Figure~\ref{fig:fagn_vs_Sigma1relative}. 
We present AGN fractions to defined limits in terms of \LX (left), \LX/\mstel (centre) and \LX/\sigone (right), in all cases adopting the lower thresholds (matching the top row of Figure~\ref{fig:mainresults}).
These refined measurements confirm that the trends identified in our initial analysis of the four galaxy populations (exSF, cSF, cQu and exQu) in Section~\ref{sec:mainresults} are also seen \emph{within} the individual galaxy populations.

For star-forming galaxies, we see a rise in the AGN fraction---whether defined to \LX, \LX/\mstel, or \LX/\sigone limits---with increasing 
values of \sigone/\sigseq.
A higher central stellar mass density is associated with higher levels of AGN activity \emph{within} the cSF and exSF galaxy populations. 
Measuring AGN activity relative to the mass of the central bulge (i.e. $\LX/\sigone$, right column of Figure~\ref{fig:fagn_vs_Sigma1relative}) tends to flatten this overall trend, especially in our lower redshift bin.

Within quiescent galaxies, the trends are substantially different;
we find a \emph{decrease} in the AGN fraction with increasing \sigone/\sigseq. This decrease is found using both \LX and \LX/\mstel as our accretion rate tracers and is exaggerated further when using \LX/\sigone. 
In our higher redshift bin, the AGN fractions in exQu galaxies exceed the AGN fraction in star-forming galaxies. At lower redshifts, AGN fractions in the more extended quiescent populations are comparable to AGN fractions in star-forming galaxies of the same \sigone/\sigseq.
Crucially, we see a dependence of the AGN fraction on \sigone/\sigseq \emph{within} the cQu or exQu populations. The refined analysis presented here confirms that the size of quiescent galaxies (or equivalently their central densities) is correlated with the incidence of AGN, with larger quiescent galaxies more likely to host an X-ray AGN.

\section{Discussion and interpretation} 
\label{sec:discuss}

In this section we first summarize our measurements of AGN activity in different galaxy population and interpret them in terms of different, sequential phases of galaxy evolution (Section~\ref{sec:agn_gal_stages}). 
We then compare our results with prior measurements of AGN fractions in exSF, cSF, cQu and exQu galaxy samples  (Section~\ref{sec:dicuss_prior}) and with prior measurements showing a correlation between compactness and sample-averaged black growth (Section~\ref{sec:discuss_predictor}).
Section~\ref{sec:compactness_vs_SFRMS} explores the relation between our measurements and the increased incidence of X-ray AGN in star-forming galaxies with SFRs that place them below the main sequence, as found by \citet{aird_x-rays_2019}. 
In sections~\ref{sec:bhgrowth} and \ref{sec:galpathways} we explore the implications for how and when black hole assembly takes place within the evolving galaxy population. 
We first (Section \ref{sec:bhgrowth}) combine our measurements with estimates of the timescales of different evolutionary phases to estimate the black hole mass that is assembled by a typical galaxy in each phase at a given epoch. 
Then in Section~\ref{sec:galpathways} we consider the different evolutionary pathways that may be followed by individual galaxies as they transform between the four galaxy populations. 
We also use our measurements to track black hole mass assembly during these different pathways and show how and when different types of galaxies assemble their central black holes to reach the scaling relations between black hole mass and galaxy properties observed in the local Universe.

\subsection{AGN activity across different galaxy evolution phases} 
\label{sec:agn_gal_stages}

The results presented in Section~\ref{sec:results} above show that both the incidence of X-ray AGN and their distribution of accretion rates vary between the four different galaxy populations we consider, classified according to their structural properties (compact versus extended) and star formation properties (star-forming versus quiescent). 
The AGN content of a given galaxy population also changes with redshift, which can be characterised by a shift in the accretion rate probability distributions toward lower accretion rates as redshift decreases (corresponding to a reduction in the typical \emph{fuelling} rate at later cosmic times), a reduction in the normalisation of the probability distribution (corresponding to a reduction in the \emph{triggering} rate of AGN), or---in most cases---a combination of the two effects.

As described in Section~\ref{sec:intro}, these different galaxy populations may correspond to distinct phases of galaxy evolution.
In particular a high-redshift galaxy may follow a ``fast-track'' quenching pathway: transforming from an exSF into a cSF galaxy (due to a compaction event that rapidly increases \sigone), rapidly quenching due to gas exhaustion or feedback processes to form a cQu galaxy, and potentially growing in size over cosmic time to become an exQu galaxy \citep{barro_candels_2013,barro_structural_2017,van_dokkum_forming_2015}.
We adopt this evolutionary sequence here as a framework to interpret our results.
Our results enable us to quantify the amount of AGN activity that is occurring in each of these potential phases of galaxy evolution,
although we caution that a broader range of pathways may be followed by different galaxies and that individual galaxies will not pass sequentially between all four phases within a given redshift interval.

\begin{figure*}
    \centering
    \includegraphics[width=0.9\textwidth,trim=0 0 0cm 0]{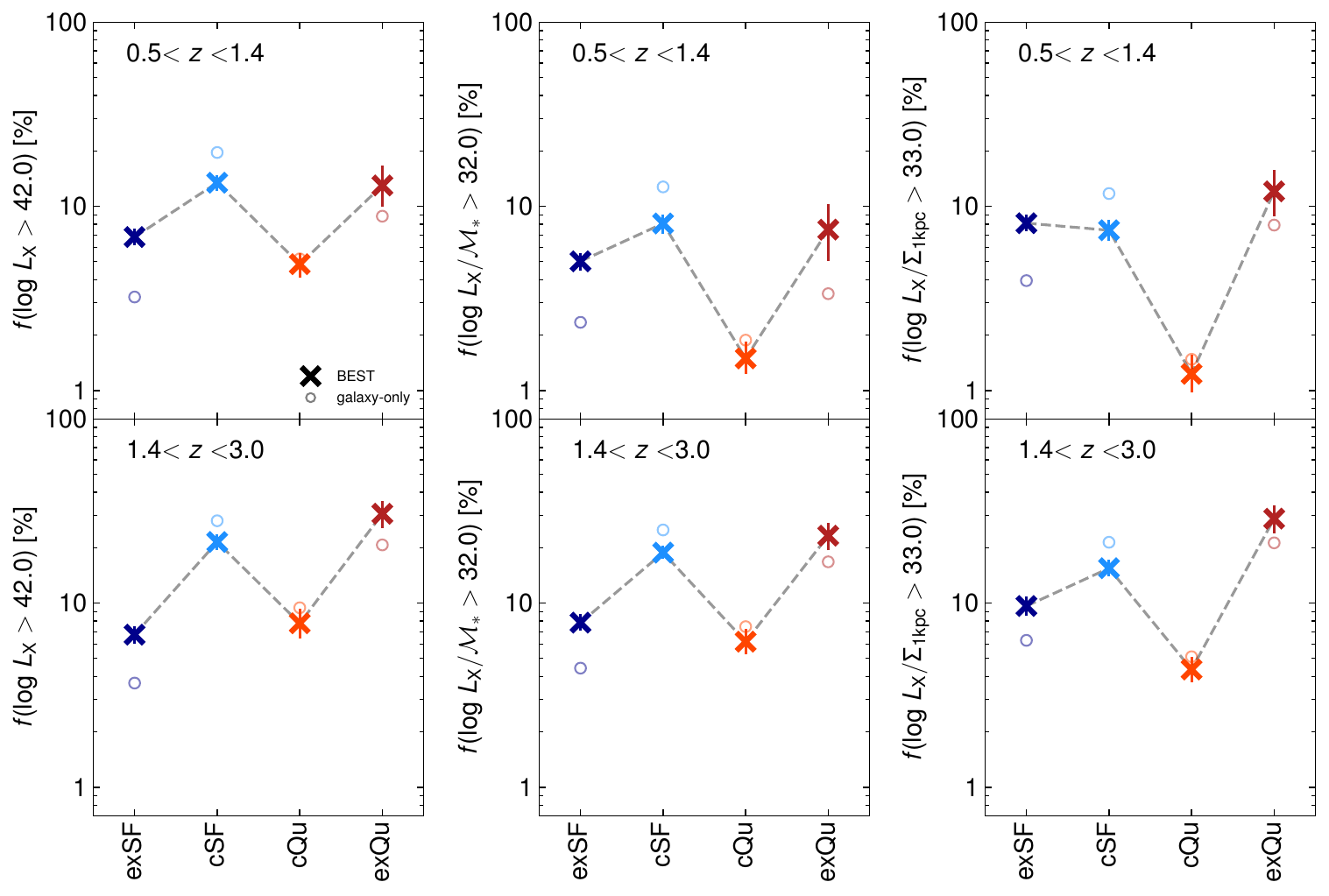}
    \caption{
        Comparing AGN fractions across our four galaxy populations in our two broader redshift ranges (\emph{top}: $0.5<z<1.4$; \emph{bottom}: $1.4<z<3.0$), adopting our three different accretion rate tracers (as indicated).
        Moving left to right within a given panel corresponds to the putative ``fast-quenching'' sequence of \citet{barro_candels_2013}, although we caution that most \emph{individual} galaxies will not move sequentially between these four populations \emph{within a given redshift interval}.
        We find a mild rise in the AGN fraction between exSF and cSF, associated with galaxy compaction (with the exception of the top-right panel where we see a slight decrease indicating a \emph{reduction} in AGN activity relative to the central stellar mass that has been assembled, see Section~\ref{sec:agn_gal_stages} for discussion). 
        The AGN fractions all drop again in cQu, where star formation has quenched, but rise again in exQu galaxies, indicating significant triggering and fuelling of AGN in these comparatively rare galaxies that have already quenched at early cosmic times and may have undergone significant growth in size.
        While the thick crosses indicate our best measurements, the open circles also show our measurements when only the galaxy contribution is considered in the optical/IR light (i.e. AGN light is neglected, see Section~\ref{sec:galonly_vs_best}) which has a tendency to exaggerate the differences between exSF and cSF galaxies.
    }
    \label{fig:fagn_vs_classl}
\end{figure*}

In Figure~\ref{fig:fagn_vs_classl} we present our measurements of AGN fractions (defined based on each of our three accretion rate tracers) in two broader redshift bins as a function of the galaxy classification, moving sequentially through the putative fast-track sequence from the exSF to cSF population (i.e.~following galaxy compaction) onto cQu (quenching) and finally to exQu galaxies (size growth).
Moving from exSF to cSF galaxies, we find an enhancement in the AGN fraction in most cases, indicating the galaxy compaction process---corresponding to substantial build up in the stellar mass of the central kiloparsec---is associated with an increased incidence of X-ray AGN. When using \LX or \LX/\mstel as our accretion rate tracer we generally find an enhancement of the AGN fraction by a factor $\sim$2--3. 
However, for \LX/\sigone (right column of Figure~\ref{fig:fagn_vs_classl}) we find a milder increase at $1.4<z<3.0$ and a \emph{decrease} at $0.5<z<1.4$. 
While galaxy compaction may be associated with both higher luminosity AGN and a higher AGN incidence in absolute terms, any enhancement may occur in-step with the build up of the stellar mass in the galactic centre. 
Indeed, our main results from Section \ref{sec:mainresults} (see right column of Figures~\ref{fig:mainresults_p} and \ref{fig:mainresults}) suggest that it is exSF galaxies that generally host more rapidly accreting AGN relative to their current central stellar mass (\sigone). 
Thus, while cSF galaxies may be associated with increased AGN activity, this increase may simply reflect the fact that such galaxies have \emph{already} assembled a substantial central black hole before galaxy compaction (or early in the process), which continues to accrete gas from the surroundings to produce a luminous AGN.

We note that the observed increase in the AGN fraction between exSF and cSF galaxies in terms of \LX and \LX/\mstel from our best analysis (crosses in Figure~\ref{fig:fagn_vs_classl}) is significantly weaker than when we  consider only the galaxy contribution to the optical/IR light and ignore the AGN contribution (open circles in Figure~\ref{fig:fagn_vs_classl}). 
Correctly accounting for the contribution from a central, point-like AGN and the impact this has on the measured star formation rates and---most crucially---the measured \sigone indicates a less extreme association between the cSF population and AGN activity. 
Many of the X-ray detected sources have significant central, blue light due to the AGN itself but the underlying galaxy is actually extended (see Figure~\ref{fig:change_in_class}). 

Moving from cSF to cQu galaxies we see a significant drop in the AGN fractions, although the level in terms of absolute \LX remains comparable to the AGN fraction in exSF galaxies ($\sim$5--8\% of galaxies hosting an AGN with $\LX>10^{42}$~\ergs). 
The quenching of star formation that transforms cSF into cQu galaxies is associated with a decrease in AGN activity, but the process is not shut off completely.
Given that the duty cycle of a single accretion event is expected to be $\sim$0.1--1~Myr (i.e. much shorter than the lifetime of a cQu galaxy), there must be ongoing fuelling of AGN activity in such galaxies, albeit with a lower triggering rate and prompting lower levels of activity (as demonstrated by the shifts in the distributions see in Figure~\ref{fig:mainresults_p} in both normalisation and toward lower accretion rates). 

Finally, moving from cQu to exQu galaxies, we find an increase in the AGN fraction, similar to or even exceeding our measurements in cSF galaxies. 
Such exQu galaxies have low levels of star formation--indicating little cold gas is present---and are not centrally concentrated, so represent the polar opposite of the cSF population. Thus, it is may be surprising to find such a high fraction hosting X-ray luminous AGN at their centres. 
However, stellar mass loss associated with the ongoing evolution of the stellar population in such galaxies can provide a substantial source of low angular momentum gas that, while not triggering significant levels of star formation, can accumulate within the central regions and provide sufficient fuel to drive substantial black hole growth and periods of bright AGN activity \citep[e.g.][]{ciotti_radiative_2007,kauffmann_feast_2009}. 
We note that such exQu galaxies are rare, with the lowest number densities of the four populations over our entire redshift range. 
In terms of total number density, \emph{most} AGN are hosted by star-forming galaxies. AGN in exQu are a small minority of the overall AGN population, but our measurements show that the \emph{fraction} of exQu galaxies that host AGN is very high, indicating that ongoing AGN activity may be especially important at this stage of galaxy evolution.

While the sequence presented in Figure~\ref{fig:fagn_vs_classl} broadly corresponds to the putative ``fast-track'' identified by \citet{barro_candels_2013}, we caution that most individual galaxies would not pass sequentially between these populations \emph{within the indicated redshift range}. 
Indeed, the majority of the exSF galaxies would remain classified as such throughout both redshift intervals -- only a subset undergo the galaxy compaction process that forms the cSF population. In contrast, the cSF galaxies are thought to be short-lived ($\sim$0.3--1~Gyr) and are expected to quench within this timescale to produce cQu galaxies, either within the redshift interval or within the subsequent bin. 
The cQu galaxies themselves will thus be a mix of these recently-quenched galaxies and those that quenched at an earlier epoch. 
Most notably, the exQu galaxy population at a given redshift will consist of galaxies that have already quenched---at an earlier epoch---and may have \emph{already} undergone significant size growth (decreasing their relative \sigone). 
Such size growth may be due to an adiabatic expansion caused by the ongoing loss of mass from the galaxy in stellar winds \citep[e.g.][]{damjanov_red_2009}, the same process that we suggest can fuel ongoing AGN activity at their centres, providing a connection between the process of size growth of galaxies and the accretion activity of their central black holes. 
\redit{We also note that a proportion of the exQu population may consist of galaxies that have quenched \emph{recently} at lower \sigone and thus without requiring significant further size growth to have occurred.}
In Section~\ref{sec:galpathways} below we explore further the evolutionary pathways that may be followed by individual galaxies and use our measurements to estimate the amount of black hole growth that occurs at the different stages of their lifecycle.

\subsection{Comparison to prior measurements of AGN fractions}
\label{sec:dicuss_prior}

Measurements of AGN fractions within the exSF, cSF, cQu and exQu galaxy populations have been presented in prior work, although different methods were
adopted to classify the galaxies into these different categories. 
Furthermore, most prior studies do not model the contribution of both AGN and galaxy light to the optical/IR, which we have shown has a significant impact on the resulting measurements. 
Additionally,
no prior study of AGN fractions has corrected for the impact of X-ray incompleteness, which we have achieved using our Bayesian modelling that combines both detections and non-detections. 
As such, our work provides robust measurements of AGN activity (to well-defined limits) as a function of the structural and star formation properties of $z\sim0.5-3$ galaxies.\footnote{We acknowledge the limitation of our work that identifies AGN based purely on X-ray emission and the potential for a more sophisticated modelling of obscuration properties that remains beyond our scope here.} 
Here, we compare our measurements with prior studies and examine the overall conclusions drawn from these works. 

In an early study, \citet{barro_candels_2013} found that $\sim$30\% of their sample of cSF galaxies\footnote{We note that \citet{barro_candels_2013} define compact and extended based on \redit{stellar mass densities within the effective radius of the galaxy, $\Sigma_e$}, which differs from our approach based on the central stellar mass density within 1~kpc relative to the quiescent galaxy size--mass sequence, \sigone/\sigseq.}
at $z\sim2$ were X-ray detected (compared to $<1$\% of exSF at the same stellar mass), indicating a high prevalence of AGN that they note may be associated with the compaction process. 
While this result is broadly in line with our best measurements, it is likely that their estimate of the AGN fraction in cSF galaxies is biased high, and conversely the estimate in exSF galaxies is biased low, due to AGN light contaminating their galaxy structural measurements. Their AGN fraction in exSF may also be further biased due to the omission of X-ray completeness corrections. 
While our measurements still show an enhancement in cSF versus exSF galaxies at a comparable redshifts ($1.4<z<3.0$), the increase is reduced from a factor $\sim$30 to $<4$, showing that AGN activity remains prevalent in exSF galaxies and indicating a less direct association between AGN activity and the galaxy compaction (and subsequent quenching) process.

\begin{figure*}
    \centering
    \includegraphics[width=0.7\textwidth,trim=0cm 0.5cm 0cm 0]{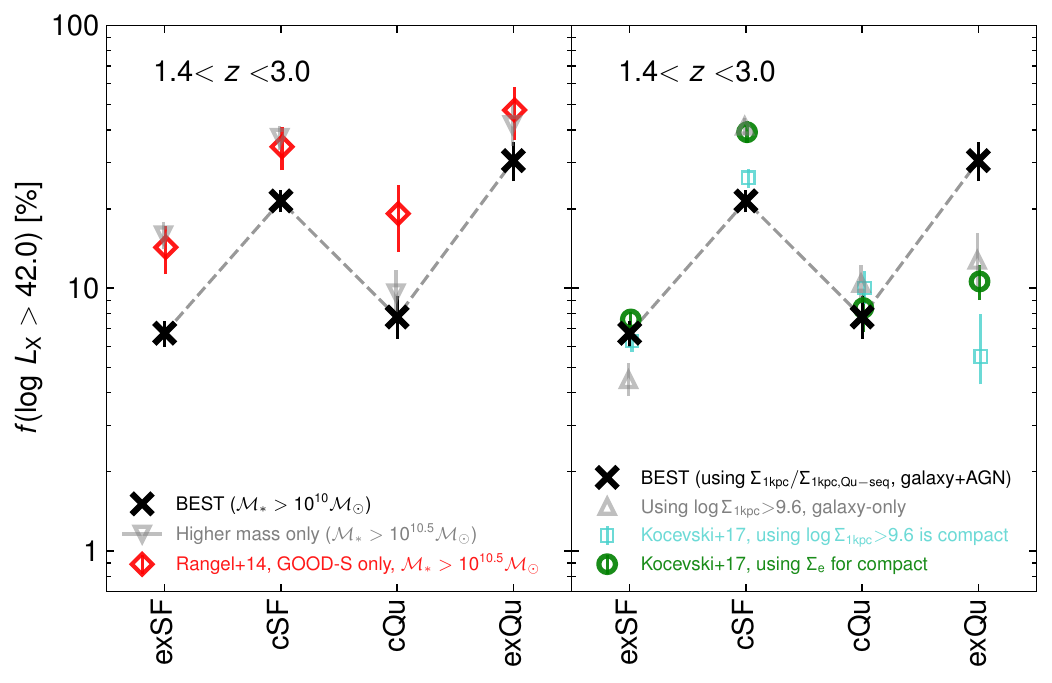}
             \caption{
          Our best measurements of $f(\log\LX>42.0)$ across the four galaxy populations at $1.4<z<3.0$ (black) compared to prior measurements of X-ray detected fractions by \citet[red diamonds, left panel]{rangel_evidence_2014} and \citet[green circles, right panel]{kocevski_candels_2017}. 
          \citet{rangel_evidence_2014} adopted a sample with a higher stellar mass threshold ($\mstel>10^{10.5}\msun$) which leads to a systematically higher AGN fraction in all four galaxy populations; 
          adopting the same \mstel threshold for our measurements (grey inverted triangles) brings our measurements into better agreement. We note the high AGN fraction in exQu galaxies, consistent with our estimates.
          In contrast, later work by \citet{kocevski_candels_2017} found a more moderate AGN fraction in exQu galaxies and measured a very strong (factor $\sim$3) enhancement of the AGN fraction in cSF galaxies. 
       Altering our measurements to use a strict cut of $\log \sigone>9.6$ to define compact galaxies \citep[as used in the appendix of][light blue squares]{kocevski_candels_2017} and only considering the galaxy component in the optical/infrared measurements of galaxy structural and star formation properties (neglecting the AGN contribution) brings our measurements to within reasonable agreement with \citet{kocevski_candels_2017}, most notably biasing the measured fraction high in cSF galaxies and low in exQu galaxies.
    }
    \label{fig:fagn_vs_kocevski}
\end{figure*}

\citet{rangel_evidence_2014} classified a sample of 268 massive ($\mstel>10^{10.5}\msun$) galaxies in the GOODS-S field as exSF, cSF, cQu and exQu based on $\Sigma_e$ \citep[\redit{stellar mass density within the effective radius, as in}][]{barro_candels_2013}.
X-ray detected fractions (using a combination of energy bands) were then measured using the 4~Ms \textit{Chandra} imaging. 
In the left panel of Figure~\ref{fig:fagn_vs_kocevski} we compare these results with our robust AGN fractions to a fixed X-ray luminosity limit, $f(\log \LX>42.0)$. 
The \citet{rangel_evidence_2014} results all lie systematically above our measurements, which is likely due to the higher stellar mass threshold of $\mstel>10^{10.5}\msun$ adopted by \citet{rangel_evidence_2014} compared to our analysis using $\mstel>10^{10}\msun$. 
Repeating our measurements of  $f(\log \LX>42.0)$ using the same stellar mass threshold (inverted grey triangles in the left panel Figure~\ref{fig:fagn_vs_kocevski}) brings our results into better agreement with \citet{rangel_evidence_2014}. 
The increase in the AGN fraction to fixed, absolute luminosity limits with increasing stellar mass is well established and is due to lower Eddington rate AGN, which are more common, being easier to detect in higher mass galaxies (which generally host more massive black holes) as a higher absolute luminosity is produced \citep{aird_primus_2012}. The effect is clearly present across all four of the galaxy populations. 
Recent work by \citet{ji_agn_2022} also discussed the important impact of stellar mass and its effect on X-ray detected fractions and advocated a relative, \mstel-dependent definition of galaxy compactness, similar to our \sigone/\sigseq approach but not accounting for the non-unity slope of the \sigone-\mstel sequence.
Adopting limits in terms of \LX/\mstel \citep[as advocated by][]{aird_x-rays_2018,aird_x-rays_2019} and applying appropriate corrections for X-ray incompleteness \citep[not considered by][]{ji_agn_2022} will also mitigate the impact of stellar-mass-dependent biases when determining AGN fractions. 

We note that \citet{rangel_evidence_2014} found an especially high AGN fraction in exQu galaxies ($\gtrsim$40\%), as found in our work.
\citet{rangel_evidence_2014} also found that compact galaxies (whether star-forming of quiescent) contain a higher proportion of obscured or Compton-thick AGN than the extended galaxy populations,  based on X-ray spectral analysis.
They suggested that this indicates distinct accretion modes in compact versus extended galaxies, although these conclusions are reliant on small samples (in just the GOODS-S field) without corrections for X-ray incompleteness, which may limit the conclusions that can be drawn from direct comparison of the detected sources. 
While we have not performed detail X-ray spectral fitting, our comparison of soft versus hard X-ray selection (see Section~\ref{sec:hard_vs_soft} and row 3 of Figure~\ref{fig:fagn_checks}) suggests that obscuration has a similar impact across all four galaxy populations. 
In general, we find similar, broad distributions of accretion rates in all  galaxy populations, indicating X-ray luminous AGN may be produced in all galaxies.  This suggests similarities in the underlying accretion processes powering the AGN, although differences in the precise shapes and normalisations of our distributions may indicate different \emph{fuelling} mechanisms are responsible, broadly consistent with the conclusions of \citet{rangel_evidence_2014}.

In the right panel of Figure~\ref{fig:fagn_vs_kocevski} we compare our results to the measurements by  \citet{kocevski_candels_2017}. 
This work used four of the five CANDELS fields (excluding COSMOS due to the shallower X-ray coverage), and provided overall X-ray detected fractions (in a combination of bands) without corrections for X-ray completeness. 
The primary results (green circles in Figure~\ref{fig:fagn_vs_kocevski}) divided the galaxy sample into exSF, cSF, cQu and exQu populations based on cuts in terms of $\Sigma_e = \mathcal{M}_*/r_e$ and found an especially high AGN fraction in cSF galaxies ($\sim$40\%)---a factor $\sim$2 higher than our best estimate---compared to a fraction of $\sim$10\% in each of the three other galaxy populations. 
Notably, no enhancement was seen in the exQu galaxy population. 
In their appendix \citet{kocevski_candels_2017} also provided AGN fractions based on adopting a threshold of $\log \sigone$ [\msun kpc$^{-2}$] $>9.6$ to identify compact galaxies, which we show as the light blue squares in Figure~\ref{fig:fagn_vs_kocevski}; this leads to a significantly lower AGN fraction in cSF galaxies, in much better agreement with our best estimate (although the estimate in exQu galaxies becomes even lower). 
To provide a more direct comparison to \citet{kocevski_candels_2017}, we also show our measurements if we adopt a strict $\log \sigone>9.6$ (see Section~\ref{sec:strictsigone}) and neglect the AGN contribution to the optical/IR (i.e.~our galaxy-only results, see Section~\ref{sec:galonly_vs_best}) as grey triangles. 
These changes result in a higher AGN fraction in cSF galaxies and a lower fraction in exQu, in agreement with the fiducial \citet{kocevski_candels_2017} results (although still inconsistent with the measurements using $\log \sigone>9.6$ to define compact galaxies).
Ultimately, we conclude that \citet{kocevski_candels_2017} may have over-estimated the AGN fraction in cSF galaxies (and thus the enhancement relative to exSF) due to AGN light in the optical/IR leading to contamination of the cSF sample, whereas they likely underestimated the AGN fraction in exQu galaxies due to applying a cut based on \sigone or $\Sigma_e$ (rather than the relative definition we adopt) as well as neglecting X-ray incompleteness corrections.
Our measurements presented here are more robust and present a somewhat different picture, where the enhancement of AGN activity in cSF galaxies is less pronounced and the highest AGN fraction is found in the exQu population.

More recently  \citet{habouzit_linking_2019} compared AGN fractions in simulated galaxies from the large-scale cosmological hydrodynamic simulation IllustrisTNG \citep{marinacci_First_2018, naiman_First_2018, nelson_First_2018, pillepich_First_2018, springel_First_2018} and observed samples from CANDELS, applying consistent redshift- and mass-dependent criteria to classify by compactness (traced by $\Sigma_e$) and star formation properties and 
using an empirical model to include the impact of AGN obscuration on the simulated estimates.
They found consistent AGN fractions of $\sim$16--20\% in cSF and $\sim$6--10\% in cQu in both the observed and simulated samples, in agreement with our measurements.
In contrast, for exSF galaxies the simulation predicts an AGN fraction of $\sim$15--17\%, which is significantly higher than the $\sim$4--5\% AGN fraction that \citet{habouzit_linking_2019} measure in the corresponding observed sample and our estimate of $f(\log \LX>42.0)\approx7$\%.
For exQu galaxies the simulations predict AGN fractions of $\sim$8--12\%, which is close to the measurement in the observed samples by \citet{habouzit_linking_2019} of $\sim$5--10\% but is significantly lower than our measurements which reach $\sim$30\% at high redshift ($z=1.4-3$). 
Thus the Illustris simulation may not fully capture the physical mechanisms that can trigger AGN activity in both exSF and exQu galaxy populations, although a detailed comparison is beyond the scope of the present paper.

\subsection{Galaxy compactness as a predictor for AGN activity}
\label{sec:discuss_predictor}

The results presented in Section~\ref{sec:results} above show that both the incidence of X-ray AGN and their typical accretion rates depend on the compactness of the potential host galaxies. 
In star-forming galaxies, we measure a rise in the AGN fraction with increasing compactness (as measured using either an absolute \sigone threshold or relative to the \mstel--\sigone sequence for quiescent galaxies) that is found \emph{within} both the exSF and cSF populations (see Figure~\ref{fig:fagn_vs_Sigma1relative}). 
Such a trend suggests a commonality between the processes that build up the central regions of star-forming galaxies, increasing their central stellar mass, and the processes that lead to the triggering of an X-ray luminous AGN. 

Recent work by \citet{ni_revealing_2021} suggested that galaxy compactness (traced by \sigone) is a stronger predictor of the overall level of AGN activity in star-forming galaxies (traced by sample-averaged black hole accretion rates) than other galaxy properties such as the total stellar mass, \mstel, or the overall SFR \citep[see also][]{ni_does_2019,yang_linking_2018}.
\citet{ni_revealing_2021} concluded that this relation suggests a link between the gas density within the central kpc of star-forming galaxies and the gas that is accreted onto the central black hole.
Our measurements allow us to quantify not only the sample-averaged accretion rate (traced by, e.g. $\langle \LX \rangle$) but also the fraction of star-forming galaxies with an AGN as a function of compactness, showing that the \emph{incidence} of AGN activity increases with host galaxy compactness in star-forming galaxies and indicating an increase in the rate at which AGN are \emph{triggered} in galaxies with denser central regions.

\begin{figure*}
    \centering
    \includegraphics[width=0.7\textwidth,trim=0 0.4cm 0 0]{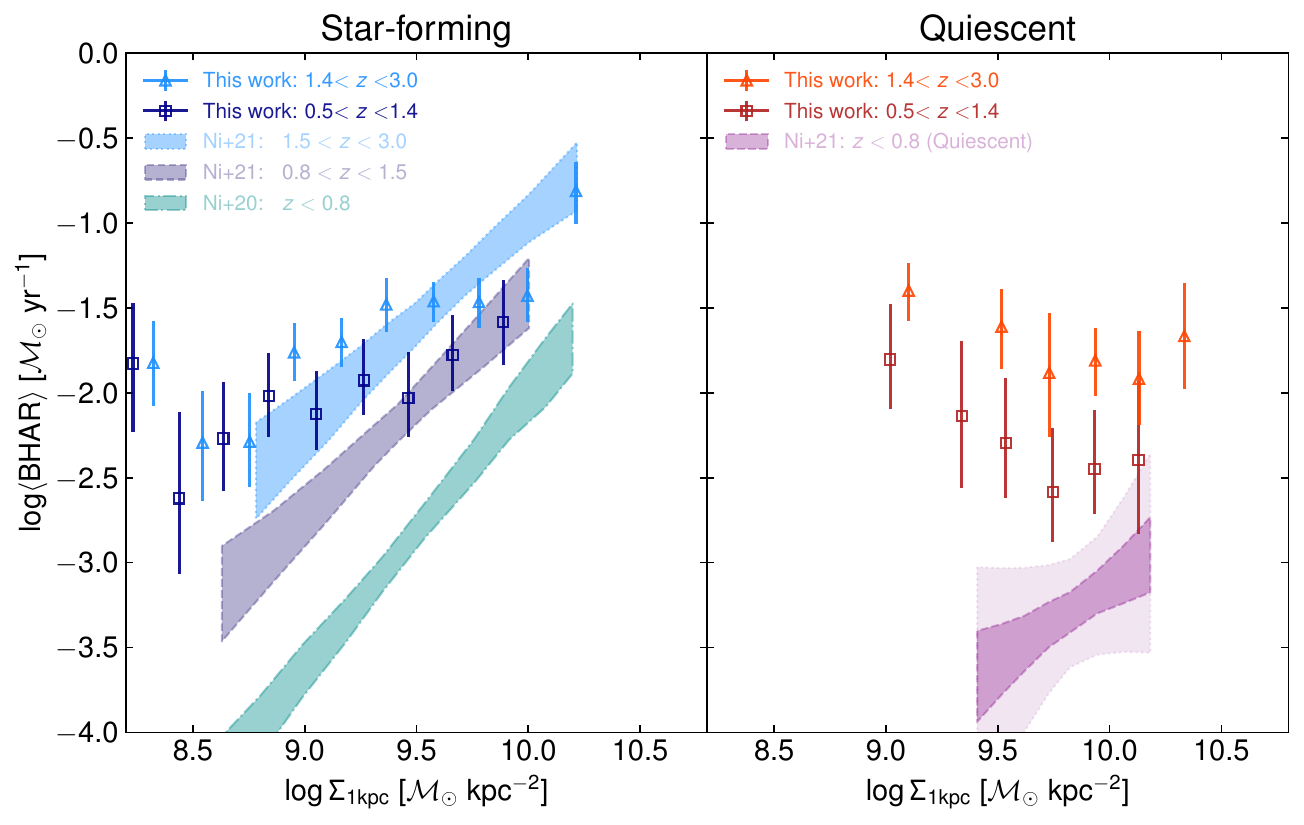}
    \caption{
    Sample-averaged black hole accretion rate, $\langle \mathrm{BHAR}\rangle$, (derived from $\langle \LX \rangle$ using Equation~\ref{eq:bhar}) for star-forming galaxies (left) and quiescent galaxies (right) as a function of \sigone, compared to prior measurements from \citet{ni_revealing_2021}. 
    In star-forming galaxies, we find a positive correlation at both $0.5<z<1.4$ (dark blue squares) and $1.4<z<3.0$ (light blue triangles), although the slope is slightly shallower than measured by \citet[light blue, dark blue and green regions showing the 1$\sigma$ uncertainties on the measured correlations at the indicated redshifts]{ni_revealing_2021}.
    In quiescent galaxies, we find only a mild, negative trend of  $\langle \mathrm{BHAR}\rangle$ with increasing \sigone, with large uncertainties in the individual measurements (cf. the stronger relation found between AGN fraction and relative \sigone/\sigseq in Figure~\ref{fig:fagn_vs_Sigma1relative}). 
    For a lower redshift sample ($z<0.8$), \citet{ni_revealing_2021} find a shallow positive correlation, albeit with large uncertainties and consistent with a flat relationship (dark and light purple regions indicating the 1$\sigma$ and 3$\sigma$ uncertainties, respectively).
    }
    \label{fig:avBHAR}
\end{figure*}

In Figure~\ref{fig:avBHAR} we compare with the \citet{ni_revealing_2021} results. 
To enable a direct comparison, we measure $p(\LX)$ for star-forming galaxies at a given redshift divided into bins of absolute \sigone (cf. our relative measure of compactness adopted in Section~\ref{sec:refined}).  
We extract mean X-ray luminosities, $\langle \LX \rangle$, by averaging over these distributions and, following \citet{ni_revealing_2021}, convert these to sample-averaged black hole accretion rates, 
\begin{align}
    \langle \mathrm{BHAR} \rangle 
    &=\frac{1-\epsilon}{\epsilon} \frac{k_\mathrm{bol} \langle \LX \rangle}{c^2}\nonumber\\
    &=\frac{1.58}{10^{46} \mathrm{erg\; s^{-1}}} \times 
              20 \times   \langle \LX \rangle \; [\mathcal{M}_\odot\;\mathrm{yr^{-1}}]
              \label{eq:bhar}
\end{align}
where we assume a typical X-ray to bolometric correction factor of $k_\mathrm{bol}=20$ and a radiative efficiency of $\epsilon=0.1$.
The left panel of Figure~\ref{fig:avBHAR} shows our estimates of  $\langle \mathrm{BHAR} \rangle$ in star-forming galaxies as a function of \sigone for our two broader redshift bins, compared to the correlations measured by \citet{ni_revealing_2021} at comparable redshifts (as well as a lower redshift bin at $z<0.8$). 
\redit{While our measurements do show an increase in  $\langle \mathrm{BHAR} \rangle$ with increasing \sigone, we find a significantly flatter slope compared to \citet{ni_revealing_2021}}. This is likely due to our correction for the AGN contribution to the optical/IR light, which can  assign centrally-concentrated blue light to an AGN and thus correctly reveal more extended underlying host emission, shifting some sources to lower \sigone (see top panels of Figure~\ref{fig:change_in_class}). 
Such corrections were not applied by \citet{ni_revealing_2021} which, as shown here (see Figure~ \ref{fig:fagn_vs_classl} and Appendix~\ref{sec:galonly_vs_best}), can lead to an overestimate of AGN activity in cSF galaxies.

\citet{ni_revealing_2021} also measured $\langle \mathrm{BHAR} \rangle$ as a function of \sigone in a sample of lower redshift ($z<0.8$) quiescent galaxies identified in the wide-area F814W-band \emph{HST} imaging of the COSMOS field.  They found a weak, positive trend over a limited dynamic range, consistent with no correlation within the uncertainties. 
In contrast, at higher redshifts  we find a \emph{negative} correlation between \sigone/\sigseq and the AGN fraction in quiescent galaxies,  i.e. a significantly higher incidence of AGN in more extended (lower \sigone) quiescent galaxies. 
In the right panel of Figure~\ref{fig:avBHAR} we compare our estimates of $\langle \mathrm{BHAR} \rangle$ for quiescent galaxies with the results from \citet{ni_revealing_2021}. 
While our measurements recover a negative correlation, the statistical errors on $\langle \mathrm{BHAR} \rangle$\footnote{We note that systematic uncertainties in $\langle \mathrm{BHAR} \rangle$ due to the assumed bolometric corrections and radiative efficiencies in Equation~\ref{eq:bhar} will affect all bins in a consistent manner and thus not impact any observed correlation.}, propagated from the uncertainties in our $p(\LX)$ distributions, are substantially larger than the errors on the AGN fractions and thus the negative trend is not  significant. 
Sample-averaged accretion rates such as those shown in Figure~\ref{fig:avBHAR} 
are substantially affected by both the incidence of high-luminosity sources (that can substantially skew a linear mean) and the precise shape of $p(\LX)$ at lower luminosities.  As both of these regimes are relatively poorly constrained by a given X-ray dataset, there are large statistical uncertainties on these average measurements. 
In contrast, AGN fractions tend to be much better constrained (see Section~\ref{sec:mainresults}), allowing us to reveal trends that are not apparent from a stacking analysis. 
The correlation here may be further obfuscated by using bins in absolute \sigone, rather than the relative bins we adopt in Section~\ref{sec:refined}. 
Nevertheless, the flat relationship that \citet{ni_revealing_2021} measured at $z<0.8$ may indicate that an enhancement of AGN activity in more extended quiescent galaxies is only present at higher redshifts, where  quiescent galaxies are typically younger in age, have quenched their star formation relatively recently, and thus may have stronger stellar winds that can provide sufficient fuel for radiatively efficient accretion onto their central black holes.

\subsection{Compactness versus SFR relative to the main sequence}
\label{sec:compactness_vs_SFRMS}

\citet{aird_x-rays_2019} measured AGN fractions in samples of galaxies based on their SFRs {\it relative} to the main sequence of star formation, where star-forming galaxies were split into three populations: above, on, and below the main sequence.
Out of these three populations, the AGN fraction was found to be \emph{lowest} within galaxies that lie on the main sequence, i.e. with SFRs within $\pm 0.4$~dex of the main sequence, although as such galaxies correspond to $\sim$70\% of star-forming galaxies they still host the majority of AGN.
Starburst galaxies, with SFRs $>0.4$~dex above the main sequence (at a given redshift),
were found to have an enhanced AGN fraction by a factor $\gtrsim$2, as expected given their enhanced SFRs.
Additionally, \citet{aird_x-rays_2019} found an enhancement of the AGN fraction for star-forming galaxies at $z > 0.5$ that lie below the main sequence, 
i.e.~galaxies with $-1.3<\log \mathrm{SFR/SFR_{MS}}(\mstel,z)<-0.4$ where $\mathrm{SFR_{MS}}(\mstel,z)$ is the SFR corresponding to the main sequence at \mstel and $z$, hereafter referred to as ``sub-MS'' galaxies.
The AGN fraction among these galaxies is enhanced by a factor $\sim$2--5 compared to galaxies on the star-forming main sequence. 

\begin{figure*}
\centering
\includegraphics[width=0.7\textwidth,trim=0 0.4cm 0 0]{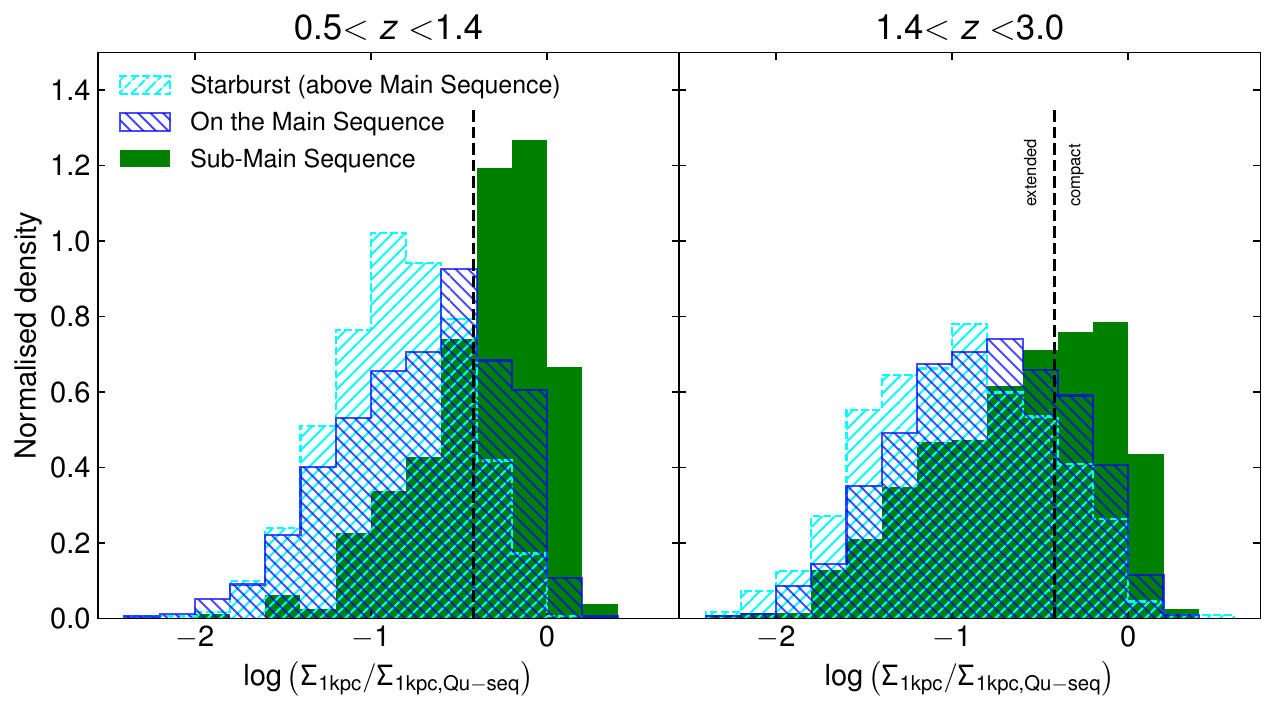}
\caption{
Normalised distributions of \sigone/\sigseq for star-forming galaxies, divided into three bins according to their SFRs relative to the star-forming main sequence \citep[following][]{aird_x-rays_2019}. 
Star-forming galaxies with SFRs that place them below the main sequence (solid green histograms) tend to be more compact, indicating an association between galaxy compaction and the onset of quenching. 
The enhancement of the AGN fraction in cSF galaxies that we measure in this paper is likely related to the enhancement in sub-main-sequence found by \citet{aird_x-rays_2019}.
In contrast, galaxies that lie on the main sequence (blue hatched) span the full range of \sigone/\sigseq, whereas starburst galaxies with SFRs placing them above the main sequence tend to be associated with exSF galaxies.
}
\label{fig:sigone_histograms_vs_ms}
\end{figure*}

\citet{aird_x-rays_2019} suggested this enhancement in sub-MS galaxies may be related to the build up of a central bulge component. 
This conclusion appears consistent with our findings here: that cSF galaxies also show an enhanced AGN fraction (relative to the more numerous exSF galaxies). 
To help consolidate these results, in Figure~\ref{fig:sigone_histograms_vs_ms} we show the normalised distributions of \sigone/\sigseq, which we use to measure compactness, for star-forming galaxies divided into these three populations based on their SFRs relative to the main sequence. 
The majority of star-forming galaxies, by definition, lie on the main sequence (dark blue hatched histograms in Figure~\ref{fig:sigone_histograms_vs_ms}) and are found to span the full range of \sigone/\sigseq, showing that such main-sequence galaxies have both compact and extended morphologies. 
In contrast, sub-MS galaxies tend to be compact (i.e. have $\log \sigone/\sigseq>-0.42$, above the black dashed line), especially in our lower redshift bin, indicating an association in star-forming galaxies between compactness (i.e. the central build-up of stellar mass), lower SFRs (possibly related to the the onset of quenching), and a high AGN fraction (indicating increased triggering of black hole accretion episodes). 
Thus, it appears that the processes that drive gas into the centres of galaxies---which forms stars, builds up their central bulges, and increase central mass density---also lead to an increase in AGN activity as well as an overall drop in SFR, i.e the onset of quenching \citep[as suggested by][]{kocevski_candels_2017,aird_x-rays_2019}.
However, it remains unclear whether the quenching of sub-MS, compact galaxies is caused by the increased presence of AGN (and thus associated with AGN feedback) given the mismatch between the short timescales of individual accretion episodes \citep[$\sim$0.1--1~Myr, e.g.][]{schawinski_active_2015,king_agn_2015} and the longer timescale for galaxy quenching \citep[$>$100~Myr, e.g.][]{wild_post-starburst_2009,barro_candels_2013}.

\subsection{Black hole mass growth during different galaxy evolution phases}
\label{sec:bhgrowth}

Our results indicate a complex relationship between the assembly of stellar mass in galaxies, their evolving properties, and their AGN content.
Here we use our measurements to quantify the amount of supermassive black hole growth that can occur in different galaxy populations, while in Section~\ref{sec:galpathways} below we consider the various evolutionary pathways that may be followed by individual galaxies to assess how and when the assembly of their central black holes takes place. 

\begin{figure*}
    \centering
    \includegraphics[width=0.78\textwidth,trim=0.15cm 0 -0.15cm 0.1cm]{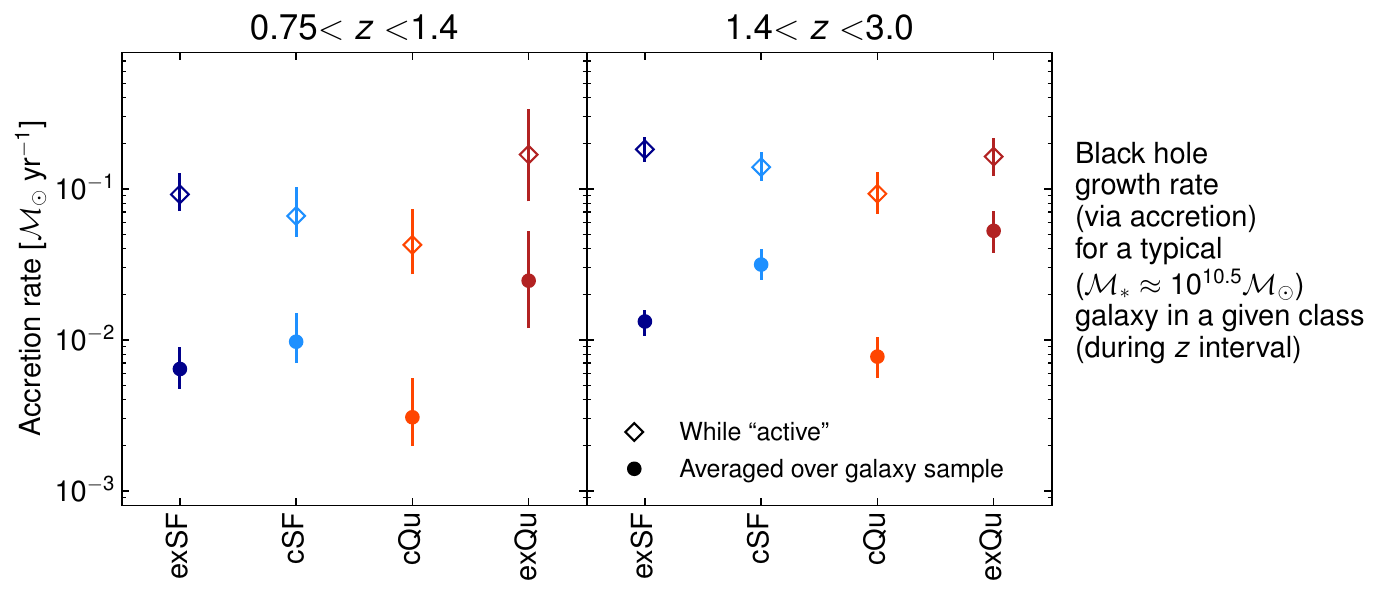}
    \includegraphics[width=0.78\textwidth,trim=0 0.2cm 0cm 0.2cm]{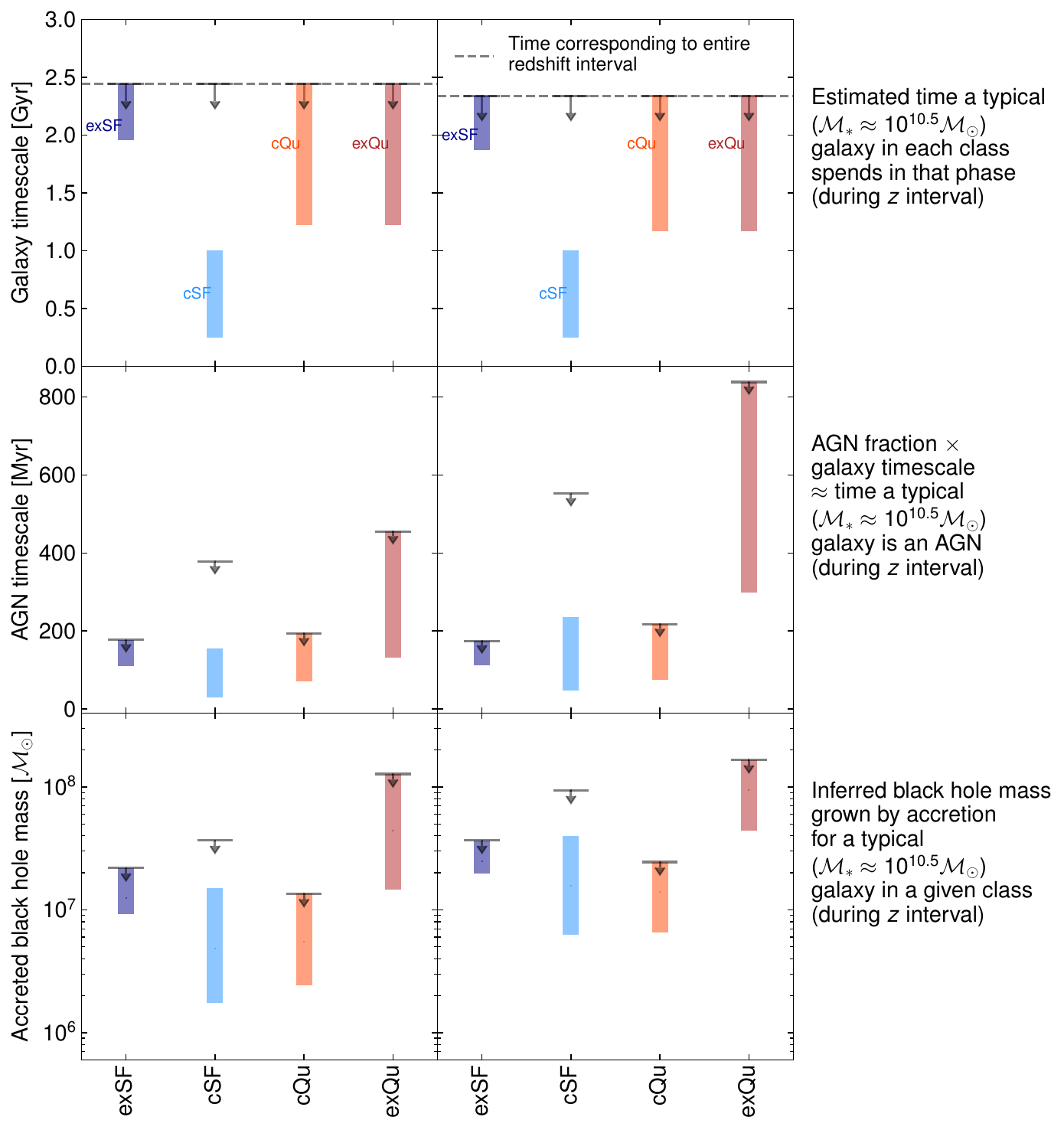}
    \caption{
   Estimating the typical black hole mass growth through accretion in the different galaxy populations.
   The top row shows our measurements of average accretion rates in exSF, cSF, cQu and exQu galaxies at $0.75<z<1.4$ and $1.4<z<3.0$ (as indicated), based on our full measurements of accretion rate probability distributions; solid circles show the accretion rate averaged over the galaxy sample whereas open diamonds correspond to the average during active periods only (see Equation~\ref{eq:av_active}). 
   The second row shows our assumed galaxy timescales, i.e.~how long a typical galaxy (with $\mstel\approx10^{10.5}\msun$ for our samples) spends in a given class within the redshift interval (see discussion in Section~\ref{sec:bhgrowth}). 
   The coloured bands indicate the range of our assumed values, and
   the grey arrows set a maximum corresponding to the cosmic time period covered by the entire redshift interval. 
   In the third row we estimate an AGN timescale, i.e. the length of time that galaxies of each class host an AGN and is given by the product of the galaxy timescale and our measured AGN fractions (N.B. this timescale is likely to be the sum of multiple, much shorter periods of AGN activity). 
   The bottom row combines our measurements of accretion rates and estimated timescales to calculate the total black hole mass grown at the centre of a typical galaxy of a given class during the epoch given by the redshift interval. 
    }
    \label{fig:BHgrowth}
\end{figure*}

The first step is to determine how much black mass assembly takes place in the typical galaxy within each of our four populations at different cosmic times, as a result of AGN activity. 
We can estimate the \emph{rate} of black hole growth directly from our measurements of accretion rate probability distributions. For each galaxy class we show two sets of measurements:
\begin{enumerate}
    \item 
    The accretion rate averaged over a galaxy sample (solid circles in the top panels of Figure~\ref{fig:BHgrowth}), which provides an estimate of the average black hole growth rate throughout the lifetime of a typical galaxy with a given population. We estimate this using our measurements of average X-ray luminosities using Equation~\ref{eq:bhar} above, assuming the same fixed bolometric correction, $k_\mathrm{bol}=20$, and radiative efficiency, $\epsilon=0.1$.
    
    \item
    The accretion rate while the black hole is ``active'', i.e. during periods that it would be identified as an AGN, with $\LX>10^{42}$~\ergs (open diamonds in Figure~\ref{fig:BHgrowth}). This accretion rate is calculated as above via Equation~\ref{eq:bhar} using the average \LX during active periods only, i.e.
    \begin{equation}
    \langle \LX^\mathrm{active} \rangle = 
    \frac{\int_{\log \LX = 42.0}^{\infty} \LX \; p(\LX) \; \mathrm{d}\log \LX}
         {\int_{\log \LX = 42.0}^{\infty}  p(\LX) \; \mathrm{d}\log \LX}
         \label{eq:av_active}
     \end{equation}
     where the denominator corresponds to the AGN fraction, $f(\log \LX>42)$, and ensures only active phases are considered.
\end{enumerate}

The accretion rate estimates shown in Figure~\ref{fig:BHgrowth} indicate that the rate of black hole growth is typically $\sim0.05-0.2$~\msun~yr$^{-1}$ during active periods, with relatively minor differences between the four galaxy populations within a given redshift range.\footnote{We have adjusted the limit of our lower redshift bin to $z=0.75$ in Figure~\ref{fig:BHgrowth} so that both redshift bins cover roughly equal periods of cosmic time.}  
Black holes in these galaxy samples, \emph{when active}, are typically accreting at similar rates, regardless of the host galaxy properties. 
However, as shown by our measurements of AGN fractions (e.g.~Figure~\ref{fig:fagn_vs_classl}), these active periods are \emph{less common} in certain galaxy populations and thus the accretion rates averaged over the entire galaxy lifetime (solid symbols in Figure~\ref{fig:BHgrowth}) show greater variation between galaxy populations, and are a factor $\sim$10 lower as they are averaged over substantial periods of inactivity. 

Our measurements provide estimates of the black hole growth \emph{rate}, though to convert these to estimates of the total black hole mass grown via accretion requires a timescale. 
The coloured bands in the second row of Figure~\ref{fig:BHgrowth} shows our estimates of the amount of a time a typical galaxy (for our samples this corresponds to a galaxy with $\mstel\approx10^{10.5}\msun$) in a given class spends in that putative evolutionary phase. 
For all classes, the maximum timescale is set by the cosmic time spanned by the redshift interval ($\sim $2.5~Gyr for the redshift bins used here). 
In general, we expect that exSF galaxies are long-lived: the majority of exSF will remain in this class throughout the redshift interval. However, a small fraction ($\sim$20\%) may undergo compaction that preferentially grows their central stellar mass, transforming them into cSF galaxies, or quench directly to form exQu galaxies. We thus adopt a range of timescales spanning $\sim2-2.5$~Gyr for the {\it typical} exSF galaxy. 
In contrast, the cSF phase is thought to be relatively short, $\sim$0.3--1~Gyr, before these galaxies quench and transform into cQu galaxies \citep[estimated from the evolving space densities of the two populations, see][]{barro_candels_2013,van_dokkum_forming_2015}, although we retain a strict upper limit corresponding to the entire redshift interval (grey arrow).

For quiescent galaxies the expected timescales are less clear. 
Some cQu galaxies will have formed within the redshift interval, following the quenching of a cSF galaxy.
Once a cQu galaxy has quenched, it may start to grow in size and thus transform into an exQu galaxy at a later epoch. 
However, with our evolving, relative definition of compactness, many of these galaxies will remain classed as cQu for the rest of cosmic time, even if they undergo some size growth.
In contrast, an exQu galaxy at high-redshift may already have undergone size growth or have quenched directly with an extended size. 
Such galaxies would only be expected to grow further in size, and thus remain as exQu for the remainder of cosmic time. 
However, the evolving \sigone--\mstel sequence could mean that some high-$z$ exQu galaxies would be classified as cQu at later epochs.
Assuming that these galaxies do not grow any further in size or mass, i.e.~they remain at the same \sigone and \mstel as observed for the remainder of cosmic time, we find that 14\% of our $z=1.4-3.0$ sample of exQu would be re-classified as cQu by $z=1.4$ due to our evolving definition (defined relative to \sigseq),
while 37\% would be classified as cQu by z=0.75 and 61\% by $z=0$. Such re-classification is consistent with the evolution of the compact quiescent sequence being driven---at least in part---by galaxies quenching at lower \sigone \redit{at later cosmic times (i.e. a progenitor bias effect)} and the exQu galaxies that we see at $z\sim2$ being early examples of such a population. 
To allow for this broad range of scenarios, we thus adopt conservative estimates of the galaxy timescales for the cQu and exQu galaxies, spanning a relatively wide range corresponding to 50--100\% of the redshift interval.

With an estimate of the galaxy timescale in hand and our measurements of AGN fractions, we can then estimate the AGN timescale, i.e.~the length of time (within a given $z$ interval) that a typical galaxy of a given class hosts a luminous AGN. 
We show our estimated AGN timescales in the third row of Figure~\ref{fig:BHgrowth},  given by the product of the galaxy timescale and the AGN fraction.
These timescales correspond to the {\it total} time a typical galaxy has an AGN within the $z$ interval shown.
However, individual episodes of AGN accretion are only expected to last $\sim0.1-10$~Myr \citep[see][and references therein]{hickox_black_2014,schawinski_active_2015,king_agn_2015} and thus these timescales will correspond to the sum of multiple, short-lived periods of activity as the AGN ``flicker'' (consistent with the broad distributions of accretion rates we measure in Section~\ref{sec:results}). 
For exSF, cSF and cQu galaxies, we estimate AGN timescales of $\sim$150~Myr. 
The much shorter galaxy timescale for the cSF phase is countered by the higher AGN fraction that we measure in this phase. 
In contrast, exQu galaxies are expected to be much longer lived \emph{and} have a high AGN fraction, resulting in a large amount of time when such galaxies could be growing their black holes, especially in our higher redshift interval ($z=1.4-3.0$), leading to estimates of $\sim$200--800~Myr for the AGN timescales in these galaxies.

Finally, in the bottom row of Figure~\ref{fig:BHgrowth} we  estimate the total black hole mass that is grown via accretion for a typical galaxy of a given class, during a given epoch of cosmic time.
\jaedit{We can take two approaches:
\begin{enumerate}
    \item use the overall galaxy timescale (second row of Figure~\ref{fig:BHgrowth}) combined with the accretion rate averaged throughout the galaxy lifetime (solid circles in the top row);     or
    \item
    use the AGN timescale (third row of Figure~\ref{fig:BHgrowth}) multiplied by the accretion rate while active (open diamonds in the top row).
\end{enumerate}
}
Both approaches produce consistent answers, showing that the bulk of black hole mass growth occurs during the multiple short-lived phases of luminous AGN activity.\footnote{We note that any black hole mass growth due to mergers is neglected in these estimates and thus formally they provide lower limits on the black hole mass growth. However, mergers remain relatively rare, even at high redshift, and would typically lead to growth by a factor $\lesssim2$ compared to the orders of magnitude in growth that can be achieved through accretion. 
We also note that mergers will increase both \mbh and \mstel at a roughly equal rate rather than preferentially growing one of these components.
}
We can also set an upper limit (grey arrows) by assuming the maximum possible galaxy lifetime (i.e., the entire $z$ interval) in combination with our measured AGN fractions and accretion rates.
We find that a typical ($\mstel\approx 10^{10.5}\msun$) exSF, cSF, or cQu galaxy in our sample is expected to grow its central black hole by $\sim2\times10^{7}\msun$ during the time period from $z=3$ to $z=1.4$, and by $\sim$10$^{7}$\msun between $z=1.4$ and $z=0.75$. 
We can set an upper limit on the black hole mass grown during cSF phases of $10^8$~\msun and $4\times10^7$~\msun in our higher and lower redshift intervals, respectively, although achieving such mass growth would require cSF galaxies to persist for $\gtrsim2$~Gyr which is much longer than our best estimate of 0.3--1~Gyr taken from \citet{barro_candels_2013} and \citet{van_dokkum_forming_2015}. 

In the exQu phase galaxies may be able to assemble up to $\sim$10$^{8}$\msun in black hole mass. 
For these rare galaxies that have already quenched at high redshift ($z\gtrsim2$) and possibly grown in size, the long periods of time spent as an exQu galaxy appear to be extremely important for assembling a large central black hole through repeated periods of AGN accretion (see also \citealt{georgakakis_investigating_2014} who found that a significant fraction of total black hole growth may be associated with quiescent galaxy phases).
At this point, star formation and AGN accretion appear disconnected: little additional stellar mass growth occurs due to star formation, whereas substantial black hole mass growth through accretion may continue.
\jaedit{We explore the impact of such growth on black hole mass--galaxy mass scaling relations for these galaxies in Section~\ref{sec:galpathways} below.}
Mass loss from the aging population of stars within the galaxy may provide a supply of hot, low angular momentum gas to the centre of the galaxy that can fuel regular episodes of AGN accretion
\citep[e.g.][]{ciotti_radiative_2007}.
The rate of stellar mass loss will be especially high in galaxies \redit{with a relatively young stellar population \citep{kauffmann_feast_2009}, explaining the high AGN fraction in high-$z$, exQu galaxies if they have quenched more recently than the typical cQu galaxy \citep[e.g.][]{van_der_wel_Size_2009, wu_fast_2018}.}
The same stellar mass loss process may also drive ongoing size growth of such galaxies \citep[e.g.][]{damjanov_red_2009}.
The repeated periods of AGN activity may also help maintain the hot gaseous atmospheres of such galaxies that keeps them quenched, via AGN feedback processes \citep[e.g.][]{yuan_active_2018}.

\subsection{Evolutionary pathways of galaxies as they assemble their stellar mass and central black holes}
\label{sec:galpathways}

Having estimated the amount of black hole mass growth in the different populations of galaxies above, we now assume that these populations correspond to distinct evolutionary phases and consider the pathways that individual galaxies may follow as they evolve and transition between populations. 
Specifically, we assess how and when black hole mass is assembled in high-$z$ galaxies that follow two possible quenching pathways---one that includes a significant compaction phase and one that does not---as well considering galaxies that continue to form stars and those that quench at much later epochs.

\begin{figure*}
    \centering
    \includegraphics[width=\textwidth,trim=0 0.5cm 0 0]{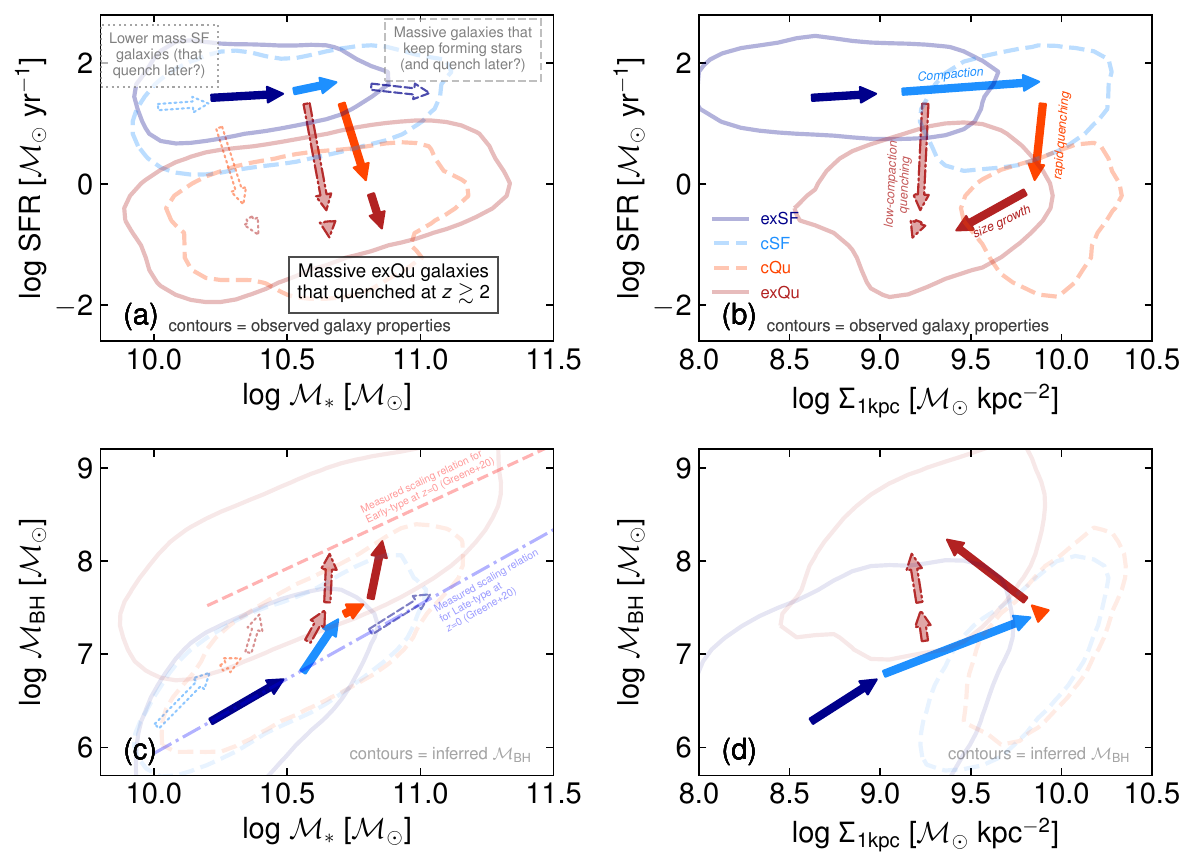}
    \caption{
    A schematic picture of the possible pathways followed by different galaxies as they evolve in stellar mass (\mstel), star formation rate (SFR), central stellar mass density (\sigone) and black hole mass (\mbh). 
    The contours in panels (a) and (b) enclose 80\% of exSF, cSF, cQu and exQu galaxies, based on the observed properties of our sample of galaxies at $z=1.4-3.0$.
    For panels (c) and (d) the faded contours instead show the inferred \mbh in exSF and exQu galaxies based on the scaling relations (including scatter) measured at $z=0$ for late-type and early-type galaxies (blue dot-dashed and red dashed lines in panel (c), respectively), measured using local galaxies with reliable dynamical black hole mass measurements \citep[]{greene_intermediate-mass_2020}. 
    For cSF and cQu galaxies, \mbh is inferred by applying an offset from the \citet{greene_intermediate-mass_2020} late-type relation based on the typical \mbh growth estimated in Figure~\ref{fig:BHgrowth}.
    The solid coloured arrows show a possible pathway for the formation of a massive exQu galaxy that quenches at $z\gtrsim2$ via the putative ``fast-track'' mechanism, i.e. starting as an exSF galaxy (dark blue arrow), undergoing compaction (light blue arrow), rapid quenching (light red arrow) and subsequent size growth (dark red arrow).
    The shaded red arrows show an alternative path
    \jaedit{that we call ``low-compaction quenching''}, 
     where a star-forming galaxy quenches at lower \sigone---without undergoing significant compaction---to directly form an exQu galaxy without requiring additional size growth.
    The arrows in panels (c) and (d) use our estimates from Figure~\ref{fig:BHgrowth} to estimate the assembly of black hole mass in such galaxies.
    An exQu galaxy that formed sufficiently early ($z\gtrsim2$) can undergo substantial growth in \mbh without associated \mstel growth (i.e. star formation), which can produce the offset between the early-type and late-type scaling relations at $z=0$ seen in panel (c).  
    The open arrows in panels (a) and (c) show two additional pathways:
    massive galaxies that continue to form stars, growing \mbh and \mstel in step and thus remaining on the late-type \mbh--\mstel relation (dashed dark-blue arrow); and lower mass star-forming galaxies that may quench much later (dotted arrows) and do not have sufficient time to grow substantial \mbh, remaining below the \citet{greene_intermediate-mass_2020} early-type scaling relation (which is poorly sampled at lower masses at $z=0$).
                                }
    \label{fig:pathways}
\end{figure*}

We consider the pathways that individual galaxies may follow as they evolve in terms of their total stellar mass (\mstel), star formation rate (SFR), central stellar mass density (\sigone), and central black hole mass (\mbh).
This multi-dimensional space is illustrated in Figure~\ref{fig:pathways}. 
In the top row (panels (a) and (b)), the contours show where our four populations of galaxies (at $z=1.4-3.0$) lie in terms of their observed \mstel, SFR and \sigone.
We use arrows to suggest potential evolutionary pathways followed by individual galaxies. The solid and shaded arrows consider the different paths to form a massive exQu galaxy, which must have quenched at a high redshift (i.e. $z\gtrsim2$).
We assume such a galaxy started off as a lower mass, exSF galaxy that lay on the star-forming main sequence, gradually building up its stellar mass (dark blue arrow). 
The galaxy may then have undergone compaction (light blue arrow), where gas inflow or secular processes lead to a rapid build up of stellar mass in the central kpc, shifting the galaxy rapidly to the right in panel (b). Such a galaxy would then quench to become a cQu galaxy (solid light red arrow), dropping rapidly in SFR, and then grow in size and reduce in \sigone to form an exQu galaxy (solid dark red arrow). 
We also consider an alternative pathway to form an exQu galaxy at $z\sim2$ from a star-forming galaxy that quenches at lower \sigone---without undergoing significant compaction---to directly form an exQu galaxy without requiring significant size growth (shaded red arrows, \jaedit{marked as ``low-compaction quenching''}). 
\redit{Such a pathway may in fact be the primary route for forming exQu galaxies and become more dominant at later cosmic times, leading to a progenitor effect that drives the evolution of the quiescent galaxy size--mass sequence.}

In panel (a) of Figure~\ref{fig:pathways} we show two further pathways with the open arrows: i)~lower mass star-forming galaxies that may follow a similar evolution to described above but quench at a somewhat later epoch at lower redshift (dotted arrows); and ii)~massive star-forming galaxies that continue to grow in stellar mass, never undergo compaction, and may or may not quench at a much later epoch (dashed dark blue arrow).

In panels (c) and (d) of Figure~\ref{fig:pathways} we consider the build up of \mbh. 
As we do not have measurements of \mbh for our galaxy samples, we instead infer \mbh based on the scaling relations between \mbh and \mstel measured at $z=0$ and our estimates of \mbh growth in the different populations (faded contours in panels (c) and (d), cf.~panels (a) and (b) that use observed data).
We assume that exSF and exQu galaxies follow the observed  correlations (with scatter) between \mbh and \mstel at $z=0$ for late-type galaxies (blue dot-dashed line) and early-type galaxies (red dotted line), respectively, based on the recent compilation of reliable dynamical black hole mass measurements for nearby galaxies from \citet{greene_intermediate-mass_2020}.
To infer \mbh for the contours for cSF and cQu galaxies we then apply a systematic shift of 0.2 and 0.25~dex from the \citet{greene_intermediate-mass_2020} late-type relation corresponding to the typical fractional growth in \mbh from our estimates in Section~\ref{sec:bhgrowth}. 

We now consider what the measurements in this paper, via our analysis from Section~\ref{sec:bhgrowth} and Figure~\ref{fig:BHgrowth} above, tell us about the typical black hole mass growth for galaxies following the pathways illustrated in panels (a) and (b) and how the assembled black hole mass compares to that inferred based on the observed $z=0$ scaling relations.
The arrows in panels (c) and (d)  of Figure~\ref{fig:pathways} follow the same positions in \mstel and \sigone as in panels (a) and (b), 
but we now estimate the black hole mass that is grown through accretion for the typical galaxy following these pathways, based on our measurements.
The dark blue arrow tracks the exSF phase where the galaxy begins with  $\mbh\approx1.5\times10^6\msun$, placing it on the \citet{greene_intermediate-mass_2020} late-type relation, although the precise starting point has a minimal impact on the subsequent \mbh values, which are dominated by the mass grown through accretion.
We assume that this galaxy is part of the $\sim$20\% of the exSF population that transform into another class and thus adopt a short timescale during which $\sim4\times10^6 \msun$ in \mbh growth occurs, keeping the galaxy on the \citet{greene_intermediate-mass_2020} late-type relation. 
The galaxy then undergoes a compaction and a short ($\sim$0.3--1~Gyr) phase as a cSF galaxy, associated with a high incidence of AGN activity that grows \mbh by $\sim 2\times10^7\msun$ (light blue arrrow). 
The galaxy quenches and a further $\sim10^7\msun$ growth in \mbh may take place during a cQu phase (light red arrow), although this appears as relatively little increase in the logarithmic axes of Figure~\ref{fig:pathways}. 
We assume that these phases have taken place early in cosmic time and that the galaxy has subsequently grown in size and transformed into an exQu by $z\sim3$. It is then able to grow \mbh by $\sim10^8\msun$ as an exQu at $z=1.4-3.0$, followed by a further $\sim3\times10^7\msun$ at $z=0.75-1.4$.
We assume, conservatively, that black holes in exQu galaxies continue to grow at a factor $\sim5$ lower accretion rate from $z=0.75$ to $z=0$ (extrapolating from our measurements of $\langle \LX \rangle$ versus redshift), which covers a further $\sim$6~Gyr of cosmic time, and thus an additional $\sim1.5\times 10^7\msun$ growth in \mbh can occur. The overall growth during the exQu phase is shown by the dark red arrow.
These phases of black hole growth, while the galaxy is extended and quiescent, enhance the black hole mass without substantially increasing \mstel and can thus explain the observed offset of the \mbh--\mstel scaling relation for early-type (quiescent) galaxies compared to late-type (star-forming) galaxies at $z=0$ \citep[see also \citealt{reines_relations_2015,terrazas_quiescence_2016}]{greene_intermediate-mass_2020}.

The shaded red arrows illustrate the alternative ``low-compaction quenching'' path for a galaxy that quenches at lower \sigone and thus transforms directly into an exQu galaxy;
again, assuming such a galaxy quenched sufficiently early, it can assemble substantial black hole mass \emph{after} star formation has ended to reach the early-type scaling relation at $z=0$. 
These periods as an exQu galaxy appear to be a crucial stage in the assembly of the massive black holes at the centres of local ellipticals. 
Achieving their high \mbh values requires that such galaxies quenched early, at $z\gtrsim2$, and were thus able to spend sufficient time as an exQu galaxy at high $z$ where we measure a high AGN fraction.
Indeed, it is {\it only} through substantial growth as an exQu galaxy that galaxies can grow their \mbh sufficiently, without also growing \mstel due to star formation, to reach the higher normalisation of the elliptical galaxy scaling relation observed at $z=0$.

Our conclusion that substantial \mbh growth occurs in exQu galaxy phases differs from the model developed by \citet{chen_quenching_2020}, who assume that significant black hole growth occurs during galaxy compaction and is thus associated with the cSF phase.
While we find that AGN are common in cSF galaxies, they are \emph{not} ubiquitous: we measure an AGN fraction in this population of $\sim$20\%. 
Our measurements of the accretion rates of AGN in cSF galaxies also place limits on how much \mbh can be assembled during this phase. 
For substantial \mbh growth to occur in cSF galaxies would require a substantially higher AGN fraction or enhanced accretion rates that is not consistent with our X-ray observations 
\emph{and} a significantly longer lifetime for the cSF phase that is inconsistent with observational constraints on galaxy evolution \citep{van_dokkum_forming_2015,barro_structural_2017}.
Nonetheless, the onset of galaxy-wide quenching may still occur when the radiative energy released by black hole growth in the compact phase reaches a threshold (set by the halo binding energy), as proposed by \citet{chen_quenching_2020}.
However, this threshold must be lower than assumed by \citet{chen_quenching_2020} so that it can be achieved by the levels of AGN activity that we measure in cSF galaxies and is consistent with the relatively modest amounts of black hole growth that occur in this short evolutionary phase. 
The bulk of black hole growth, required to reach the $z=0$ scaling relation for early-type galaxies, must then take place later in exQu phases.

It is important to understand that these proposed 
evolutionary paths, with substantial \mbh assembly as a high-$z$ exQu galaxy, are \emph{not} followed by most galaxies. 
Most moderate-mass ($\mstel \sim10^{10-10.5}\msun$) galaxies remain as star-forming galaxies throughout cosmic time, building both \mstel and \mbh roughly in step and thus remaining on the lower, late-type scaling relation. 
The majority of the X-ray AGN population at any epoch are hosted by star-forming galaxies \citep[e.g.][]{rosario_nuclear_2013,georgakakis_investigating_2014,stanley_remarkably_2015}.
Thus, in most galaxies black hole and galaxy growth proceed together at a ratio of roughly 1:$10^4$, consistent with direct estimates of the average growth ratio \citep[e.g.~][]{yang_linking_2018,delvecchio_galaxys_2019}, similar to the scaling required to match the total black hole accretion rate density and star formation rate density over cosmic time \citep[e.g.][]{madau_cosmic_2014,aird_evolution_2015},
and consistent with direct measurements of the \mbh--\mstel scaling relation using AGN samples \citep[e.g.][]{reines_relations_2015,bentz_black_2018,shankar_black_2019}.
Our findings for exQu galaxies correspond to the subset of galaxies that quench early in cosmic time and where black hole growth continues despite relatively little stellar mass growth, allowing them to grow more massive black holes and placing them on a higher scaling relation, where the \mbh to \mstel ratio is $\sim$1:$10^3$. 

In panel (c) of Figure~\ref{fig:BHgrowth} we also use open arrows to show the \mbh assembly for galaxies following two other possible evolutionary paths, as shown in panel (a). 
Massive galaxies that continue to form stars will evolve along the \citet{greene_intermediate-mass_2020} late-type relation (dark-blue dashed arrow), growing both \mstel and \mbh in step.
Lower mass star-forming galaxies that quench at later epochs (light-blue, red and dark-red dotted arrows showing compaction, quenching and size growth phases, respectively) may not have sufficient time as an exQu galaxy to grow substantial, additional \mbh, and thus remain under-massive compared to the \citet{greene_intermediate-mass_2020} early-type scaling relation (dashed red line). 
We note that the measurement of the slope and normalisation of this relation is dominated by samples of early-type galaxies at much higher \mstel, where dynamical \mbh measurements are possible. 
Obtaining accurate measurements of \mbh in an unbiased sample of lower mass early-type galaxies in the nearby universe is vital to establish whether such galaxies continue to have enhanced \mbh compared to their equally massive, late-type counterparts.
Indeed, \citet{bentz_black_2018} suggest that at lower stellar masses the offset between the late-type and early-type galaxy \mbh--\mstel scaling relations is substantially reduced, consistent with the downsizing picture of galaxy evolution whereby lower mass galaxies quench later and consistent with our suggestion that such galaxies will have less time to grow enhanced \mbh.

\section{Summary and Conclusions}
\label{sec:conclusions}

This paper provides detailed measurements of how the incidence of AGN and their distributions of accretion rates varies as a function of the compactness of the galaxy population (quantified based on central stellar mass density, \sigone) from $z\approx0.5$  to $z\approx3$.  In this section, we first summarise our methodology and observational results and then list our major conclusions. 

We select stellar-mass-limited ($\mstel>10^{10}\msun$) samples of galaxies based on the deep optical-to-NIR imaging of the five CANDELS survey fields and quantify the AGN content using deep \textit{Chandra} imaging. 
An important feature of our work is that we carefully assess and correct for the potential impact of AGN light on our measurements of the stellar population properties and---most crucially---the structural properties of our galaxy sample. 
Corrections are required for 57\% of the X-ray detected sources and for those sources we typically assign $\sim$10\% of the light to the AGN, although more extreme corrections are required for a small fraction of individual sources. 
As a result, 142 of the 678 X-ray detected sources in our mass-limited sample (21\%) change from a compact to an extended classification.

With these robust measurements of stellar and structural properties in hand, we classify our galaxy samples into four different populations---extended star forming (exSF), compact star forming (cSF), compact quiescent (cQu) and extended quiescent (exQu) galaxies---based on their SFRs (relative to the evolving main sequence of star formation) and where they lie in the \sigone--\mstel plane (relative to the evolving quiescent galaxy sequence in this parameter space).
We extract hard (2--7~keV) X-ray data for all galaxies and use the Bayesian methodology of \citet{aird_x-rays_2017,aird_x-rays_2018} to measure the accretion rate probability distribution function within the four galaxy populations at different redshifts.
From these probability distributions we derive robust measurements of AGN fractions and sample-averaged accretion rates, adopting three different tracers of accretion rate: \LX, \LX/\mstel, and \LX/\sigone.

We find that $\sim$10--25\% of cSF galaxies host an AGN with $\LX>10^{42}$~\ergs, depending on redshift, which is a factor $\sim2-3$ higher than within the more numerous exSF population. 
A similar enhancement is found based on \LX/\mstel limits but is much weaker (or non-existent) based on \LX/\sigone, indicating that cSF galaxies may have already assembled a substantial central black hole and are not growing their black holes further at an enhanced rate relative to their central stellar density. 
The AGN fraction in cQu is $\sim$5--8\%, lower than in the cSF population but comparable to within exSF galaxies. 
Most notably, we measure the highest AGN fractions at a given redshift ($\sim$10--30\%) within the relatively rare population of exQu galaxies. 
We also demonstrate that these trends---an increase in AGN fraction with increasing compactness in star-forming galaxies and the \emph{opposite} trend in quiescent galaxies---are also found as a function of relative \sigone \emph{within} each of the exSF, cSF, cQu and exQu galaxy populations (see Figure~\ref{fig:fagn_vs_Sigma1relative} and Section~\ref{sec:refined}). 

We also carry out a number of tests to establish the robustness of our results and assess the importance of elements of our methodology. 
We find that accounting for the impact of the AGN light on the galaxy structural and star formation properties has a significant impact on our estimates of the AGN fraction in star-forming galaxies: neglecting these corrections leads to an over-estimate of the AGN fraction in cSF galaxies by a factor up to $\sim2$, which may explain the higher AGN fractions found in prior studies.
In contrast, there is a negligible impact on our measurements of AGN fractions in cQu galaxies and a small systematic effect for exQu that is not significant. 
Adopting hard (2--7~keV) X-ray selection---provided appropriate corrections for the varying sensitivity limits are applied---is also important and results in a factor $\sim2$ higher AGN fraction compared to soft (0.5--2~keV) selection across all four galaxy populations  due to the missed fraction of obscured AGN.
\redit{
We also show that there are significant field-to-field variations,  thus combining multiple fields is required when studying rare but important galaxy evolution phases such as the exQu galaxies.}

Finally, we use our measurements to explore how the incidence of AGN varies during different possible phases of galaxy evolution, compare to prior studies, and assess how and when galaxies assemble their central black holes. Our main conclusions are as follows:
\begin{enumerate}
 \setlength{\itemsep}{4pt}    \item 
    AGN activity is enhanced during the cSF phase, \redit{with $\sim$10--25\% of compact star-forming galaxies found to host an AGN}, although the enhancement is only by a factor $\sim$2 compared to within the bulk of the star-forming population (i.e. exSF galaxies). However, AGN are \emph{not} ubiquitous in cSF galaxies.
    In addition, while they may have an enhanced triggering rate, they have a broad range of accretion rates that are not substantially enhanced compared to  the exSF population. 
    
    \item
    The enhanced AGN fraction in cSF galaxies may explain the higher AGN fractions in star-forming galaxies with SFRs that place them below the main sequence (i.e.~in sub-MS galaxies), 
    given the anti-correlation that we find between compactness and SFR within the star-forming galaxy population.
    
    \item
    Prior studies     have over-estimated the AGN fraction in cSF galaxies as they did not correct for the impact of the AGN light. 
    \redit{Our method to account for the impact of AGN light changes $\sim$20\% of the X-ray sources from compact to extended galaxy classifications and reduces the AGN fraction in cSF galaxies by a factor $\sim$2.}
    In addition, obtaining accurate estimates of AGN fractions requires careful consideration of X-ray sensitivity limits and should use hard X-ray selection to avoid selection biases against obscured AGN. 

    \item
    Even though the AGN fraction is higher in the cSF phase, the time that galaxies spend in this phase is relatively short, thus this is \emph{not} the phase where the bulk of black hole mass growth takes place for moderately massive galaxies ($\mstel\sim10^{10.5}\msun$) 
    at $z<3$.
    
    \item
    For massive galaxies that quench at early cosmic epochs, substantial 
    black hole growth 
        takes place in the exQu phase, \redit{where we measure high AGN fractions of $\sim$10--30\%. Such growth may be fuelled by strong stellar winds from a recently quenched stellar population.
        This additional black hole growth \emph{after} star formation (and thus stellar mass growth) has ceased         enables 
    massive quiescent galaxies to assemble substantial additional black hole mass, building the most massive black holes in the local Universe and creating the enhanced \mbh--\mstel scaling relation at $z=0$ that is measured for massive, early-type galaxies.}
    
    \item
    In contrast, in the \emph{majority} of galaxies (that remain extended and star-forming throughout cosmic time), stellar and black hole growth proceed broadly in step, explaining why the \mbh-\mstel scaling relation for such (late-type) galaxies at $z=0$ is offset to lower \mbh at a given \mstel  compared to the relation for early-type galaxies (which dominate  samples of dynamic black hole mass measurements). 

\end{enumerate}

Our work shows that while the compaction process may be an important phase in the evolution and  quenching of some massive galaxies at cosmic noon and does lead to an associated increase in AGN activity, it is vital to 
consider the timescales of different phases and the evolutionary pathways followed by different galaxies when analysing the assembly of supermassive black holes.
In particular, we find that the extended quiescent galaxy evolution phase is \emph{crucial} to build the most massive black holes in the local Universe.

\section*{Acknowledgements}

\redit{We thank the referee for helpful comments that improved the clarity of this paper.}
JA acknowledges support from a UKRI Future Leaders Fellowship (grant code: MR/T020989/1).
ALC acknowledges support from the Ingrid and Joseph W.\ Hibben endowed chair at UC San Diego.
DK acknowledges support from NASA Chandra grant GO5-16150B.
This work is based in part on observations taken by the CANDELS Multi-Cycle Treasury Program and the 3D-HST Treasury Program (GO 12177 and 12328) with the NASA/ESA HST, which is operated by the Association of Universities for Research in Astronomy, Inc., under NASA contract NAS5-26555.
The scientific results reported in this article are based to a significant degree on observations made by the \textit{Chandra} X-ray Observatory.
\redit{For the purpose of open access, the authors have applied a Creative Commons Attribution (CC BY) licence to any Author Accepted Manuscript version arising from this submission.}

\section*{Data Availability}

The HST CANDELS data used in this publication, including the original source catalogues, are publicly available at \url{https://archive.stsci.edu/prepds/candels/}, while the \textit{Chandra} X-ray data are available at \url{https://heasarc.gsfc.nasa.gov}.
The derived data generated for this paper will be shared on reasonable request to the corresponding author.

\appendix

\section{Testing the robustness of our results}
\label{appendix:robustness}
\label{sec:robustness}

The results presented in this paper rely on the accuracy of our measurements of galaxy structural and star formation properties (in particular when an AGN is present), as well as our ability to identify AGN using our \textit{Chandra} X-ray data (including correcting for incompleteness in the inhomogenous X-ray imaging of the five CANDELS fields). 
In this \redit{appendix, we describe a number of tests} to establish the robustness of our results and investigate potential effects that could alter our conclusions. 
In each case, we alter the underlying measurements of galaxy or AGN properties and repeat our analysis, comparing to our fiducial results. 
For clarity, we show the impact on our results in terms of \LX/\mstel only, our accretion rate tracer that accounts for stellar-mass-dependent selection effects. 
We show the impact for each of the original four galaxy populations, presenting our measurements of $f(\log \LX/\mstel>32)$ across all of our redshift bins in Figure~\ref{fig:fagn_checks} 
\redit{and the impact on measurements of the full probability distributions in Figure~\ref{fig:pagn_checks} (shown at $1.4<z<2.2$ only)}.  
Unless otherwise stated, the overall conclusions are consistent for the other summary statistics or accretion rate tracers.

\subsection{Accounting for the impact of AGN light on galaxy properties}
\label{sec:galonly_vs_best}

\begin{figure*}
    \centering
    \includegraphics[width=0.72\textwidth,trim=0 0.3cm 0 0]{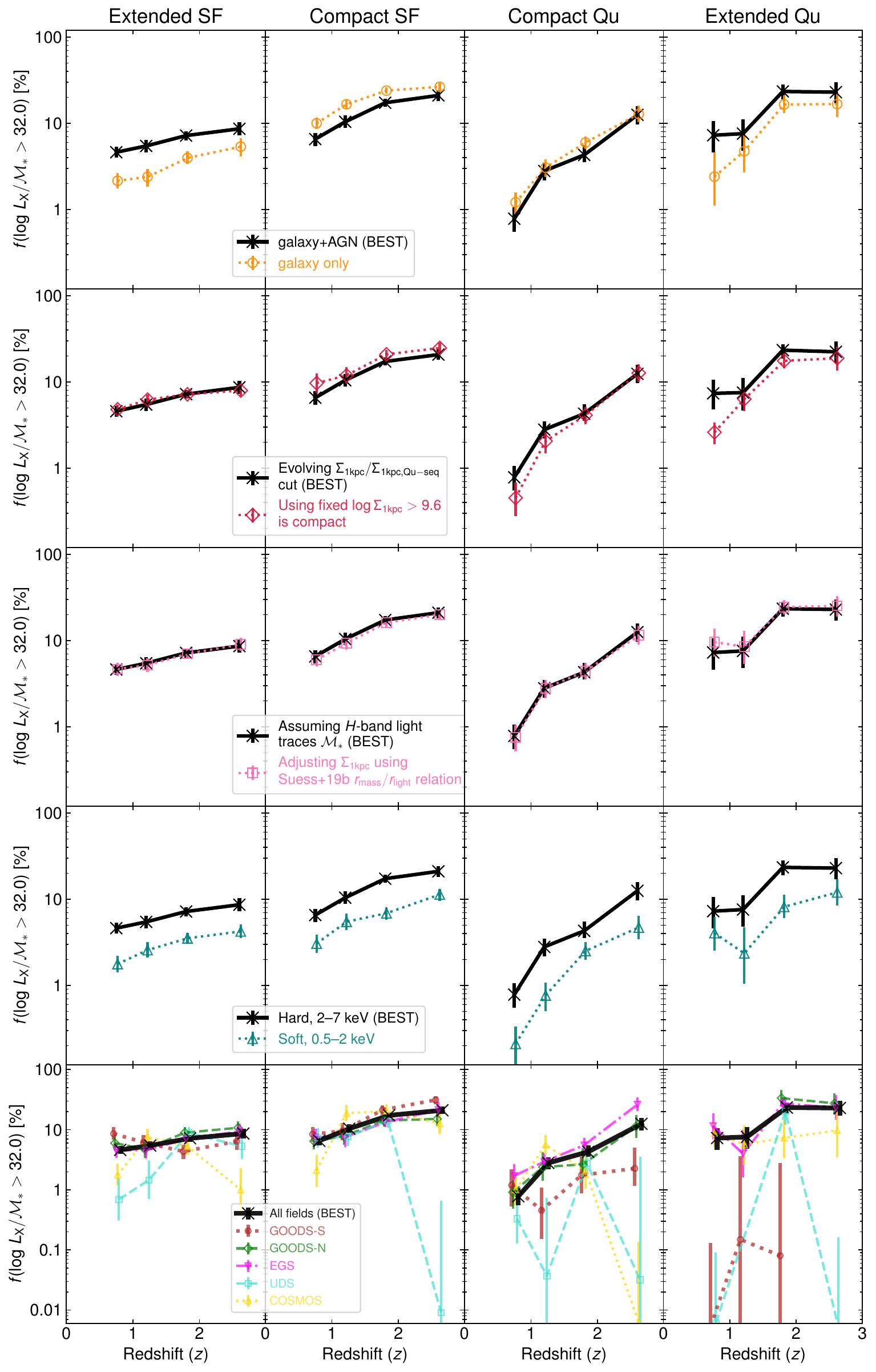}
    \caption{The impact on our measurements of AGN fractions, $f(\log \LX/\mstel>32.0)$, of our various robustness checks as detailed in Appendix~\ref{sec:robustness}, shown as a function of redshift for our four galaxy populations (exSF, cSF, cQu and exQu) as indicated. 
    In each panel the black crosses correspond to our best analysis.
    In the top row, we compare our best results that allow for both galaxy and AGN light in the SED and 2D light profile to galaxy-only measurements (orange circles), where the AGN component is neglected. 
    The second row compares our best measurements that adopt an evolving, relative definition of compactness to results based on using a single threshold of $\log \sigone>9.6$ (red diamond). 
    \redit{The third row shows the impact of accounting for differences in the stellar mass profile compared to the $H$-band light profile when measuring \sigone and classifying galaxies, which has a negligible impact on our AGN fractions.}
    The fourth row compares our best measurements that use the hard (2--7~keV) \emph{Chandra} X-ray data to measurements based on the soft (0.5--2~keV) band data instead (green triangles).
     While the soft band data is typically deeper, X-ray selection at these energies remains biased against moderately obscured X-ray sources and results in an underestimate of the AGN fraction by a factor $\sim$2 at all redshifts and in all four galaxy populations. 
     The final row compares out best measurements to estimates based on only a single field (coloured points, as labelled). While there are large field-to-field variations (note the expanded y-axis range compared to the other panels), the measurements in the individual fields are consistent---given their larger uncertainties---with our best estimates.
    }
    \label{fig:fagn_checks}
\end{figure*}

\begin{figure*}
    \centering
    \includegraphics[width=0.72\textwidth,trim=0 0.3cm 0 0]{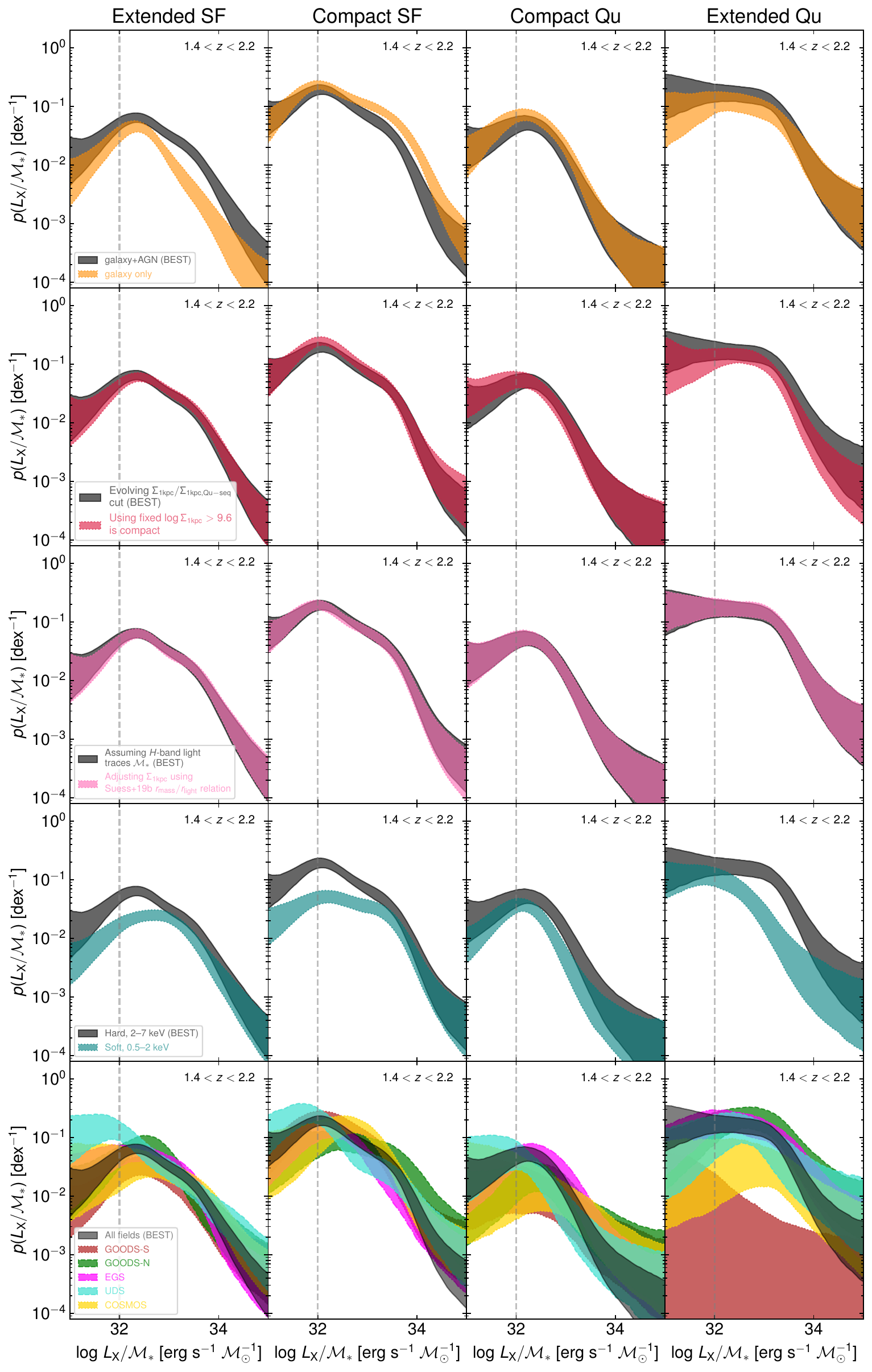}
    \caption{
    The impact of our various robustness checks on our measurements of $p(\LX/\mstel>32.0)$, shown for just one of our redshift bins ($1.4<z<2.2$ for the four galaxy populations (exSF, cSF, cQu and exQu), as indicated. 
    The black shaded regions in each panel correspond to our best analysis, whereas the coloured regions indicate (\emph{orange/top row}) allowing only for the galaxy component in our SFR and structural measurements and neglecting the AGN light, (\emph{red/second row}) using a single threshold of $\log \sigone>9.6$ rather than our evolving definition of compactness,
    \redit{(pink/third row) accounting for the difference of the mass profile versus the light profile when measuring \sigone and classifying galaxies,}
    (\emph{green/fourth row}) using soft (0.5--2~keV) X-ray data rather than the hard (2--7~keV) that is less affected by absorption, and (\emph{multiple colours/bottom row}) measurements made in each individual field compared to the combination of all five. 
    These measurements of the full probability distributions (in each redshift bin) are used to estimate the AGN fractions shown in Figure~\ref{fig:fagn_checks}.
    }
    \label{fig:pagn_checks}
\end{figure*}

Firstly, we consider the impact of the corrections we apply to our measurements of galaxy structural and star formation properties for X-ray detected sources to account for the effects of any AGN light:~allowing for an AGN component in the SED fitting (see Section~\ref{sec:sedfit}) and allowing for a central point source contribution in the 2D light profile (see Section~\ref{sec:galfit}). 
In the top row of Figure~\ref{fig:fagn_checks}, we compare our best measurements (black crosses) to the results obtained without any allowance for an AGN contribution, i.e. associating all of the observed light with the host galaxy (orange circles). 

It is clear that the improvements to account for the presence of an AGN in our best analysis make a significant difference to the results obtained in star-forming galaxies. 
As previously seen in Figure~\ref{fig:change_in_class}, sources with significant central AGN light may be mis-classified as cSF galaxies. 
Correctly accounting for the AGN contamination moves many sources to systematically lower values of \sigone and reveals the underlying  exSF hosts of many of these AGN, leading to a significant (factor $\sim2$) increase in our measured AGN fraction within the exSF galaxy population and a corresponding decrease in the AGN fraction in cSF galaxies.
Applying such corrections is clearly crucial when considering the dependence of AGN activity on the structural properties of star-forming galaxies. 
We also note that applying these improvements reduces the difference between the AGN fraction in cSF and exSF galaxies from a factor $\sim$5 to a factor $\sim$2. 

In contrast, there is little impact on our measurements for cQu galaxies.
In exQu galaxies, we find a small systematic shift in our AGN fractions but this does not significantly alter our results (in all cases the differences are $<2\sigma$). 
While we do identify a number of galaxies with bright central AGN that we subsequently classify as exQu (identifying an extended, red host galaxy once the AGN contribution is removed, see Figures~\ref{fig:change_in_class} and \ref{fig:color_thumbnails}), it is re-assuring that these extreme cases are \emph{not} driving our results and---most crucially---are not responsible for the relatively high AGN fraction that we measure in exQu galaxies.

We also note that the trends as a function of \sigone/\sigseq identified \emph{within} each of the four galaxy populations in Section~\ref{sec:refined} are still found when using the galaxy-only analysis. Thus, the existence of these trends within a given population---in particular the negative correlation between \sigone/\sigseq and AGN fraction for quiescent galaxies---is not a systematic effect due to the corrections for central AGN light for the X-ray detected sources, although the trend is slightly weaker in star-forming galaxies with our best analysis.

\subsection{Adopting a strict cut in \sigone to define compact galaxies}
\label{sec:strictsigone}

Our analysis uses a relative definition of compactness, based on where galaxies lie compared to the clear \sigone--\mstel sequence that is found for quiescent galaxies at a given redshift (see Figure~\ref{fig:sfrandsigma1_vs_mstel} and Section~\ref{sec:galsample}).
As such, our definition of a compact galaxy changes with redshift, in that a higher redshift galaxy requires a higher \sigone\ to be defined as compact, as well as with stellar mass, in that higher stellar mass galaxies require higher \sigone (at a given redshift) to meet this definition of compactness.
While an evolving, relative definition is a more robust way of defining compactness as it accounts for the overall distribution of galaxy properties at a given redshift, it is instructive to explore how our results would change if absolute values of \sigone are adopted instead, as used in some previous studies to trace compactness \citep[e.g.][]{kocevski_candels_2017,ni_does_2019}. 

The second row of Figure~\ref{fig:fagn_checks} compares our best estimates of AGN fractions using our evolving \sigone/\sigseq compactness definition (black crosses) with estimates that use an absolute threshold of $\log \sigone[\msun\;\mathrm{kpc}^{-2}]>9.6$ at all redshifts \citep[red diamonds, cf. appendix of][]{kocevski_candels_2017}.
In general this change has a minor impact on our results.
We find no impact on the AGN fraction in exSF galaxies.
The bulk of this population are distributed broadly toward lower \sigone (see Figure~\ref{fig:refined_sigone_bins}) and thus changes in the exact position at which the rarer population of cSF galaxies are removed has a minimal impact on the overall sample or our measured AGN fractions. 
Our measurements in cSF galaxies rise slightly. 
A strict cut in \sigone will preferentially select higher \mstel\ star-forming galaxies and the known dependence of AGN fractions on \mstel \citep[e.g.][]{aird_x-rays_2018} likely drives this small increase. 
We note that a more substantial effect is seen for our measurements to \LX limits (a factor $\sim$1.5--2 increase, not shown here), which are more significantly affected by the change in stellar mass. 
Adopting measurements in terms of \LX/\mstel accounts for the broad selection bias that makes it easier to detect AGN in higher \mstel galaxies, due to weakly accreting but typically higher mass black holes producing a higher absolute luminosity, although an underlying stellar mass dependence does remain.

The AGN fraction in cQu galaxies is only marginally impacted by the change to a strict \sigone cut. 
The bulk of this population (that dominate the overall quiescent galaxy number density in our considered \mstel range) remain classified as compact with the strict cut for the redshift ranges considered here.
Adopting a strict cut in \sigone\ has a slightly larger impact on our measurements of the AGN fraction in exQu galaxies, which are intrinsically rarer and thus more susceptible to changes in thresholds. Considering Figure~\ref{fig:refined_sigone_bins}, it is clear that a strict (i.e. horizontal) cut will have the tendency to classify lower mass quiescent galaxies as ``extended'', even though they clearly lie on the well-defined \sigone--\mstel sequence. 
Our best measurements show that such galaxies (that we originally classify as cQu) tend to have a low AGN fraction, so mixing them in with the truly exQu galaxies naturally reduces the overall AGN fraction. 
In addition, high-mass quiescent galaxies with $\log \sigone>9.6$ but that still lie below the \sigone--\mstel sequence (i.e. are classified as exQu for our best estimates) are removed from the exQu population when the strict cut is adopted, again lowering the AGN fraction. 
The impact due to exclusion of these galaxies has an even larger impact on the AGN fractions to \LX limits due to the stellar-mass-dependent effects discussed above, as a high proportion of these rare, high-mass but \emph{relatively} extended quiescent galaxies are found to host X-ray AGN.

While adopting a strict threshold in \sigone to measure compactness generally has a minimal impact on our results, at least when adopting \LX/\mstel as an accretion rate tracer to mitigate stellar-mass-dependent effects, we retain our relative compactness definition elsewhere in this paper. 
A relative definitions ensures we consider compactness compared to the bulk of the (quiescent) galaxy population at a given redshift, although how individual galaxies evolve---and if and when they transition between populations---needs careful consideration when interpreting our results (see Section~\ref{sec:discuss}).

\subsection{\redit{Accounting for differences in the mass versus the light profile of galaxies}}
\label{sec:mass_vs_light}

\redit{Our estimates of \sigone\ are based on the assumption that the stellar mass follows the \Hband-band light profile, as described in Section~\ref{sec:galfit}, and do not allow for variations in the stellar population (and thus mass-to-light ratios) across the extent of the galaxy \citep[see e.g.][]{wuyts_sizes_2010,szomoru_sizes_2012,mosleh_connection_2017}.
Measuring variations in stellar populations across the physical extent of a galaxy to obtain accurate \emph{mass}-profiles and improve the estimates of \sigone requires complex and involved methodologies \citep[e.g.][]{szomoru_stellar_2013,chan_sizes_2016,suess_half-mass_2019} that have not been tested in the presence of central AGN light and thus remain beyond the scope of this paper. 
However, \citet{suess_half-mass_2019-1} find that moderately massive galaxies at $0<z<2.5$ (without AGN) have a median ratio of half-mass to half-light radii of $\sim0.65-1$, depending on redshift, total \mstel and galaxy type (star-forming versus quiescent) but with a weak dependence on other galaxy properties.
Here, we use these measurements to adjust our estimates of \sigone and assess the impact of spatial variations in the mass-to-light ratios on our classification of different galaxy populations and our ultimate measurements of AGN fractions.}

\redit{First, we approximate the \citet{suess_half-mass_2019-1} findings for the median ratio of the half-mass to half-light radii, $r_\mathrm{mass}/r_\mathrm{light}$ of galaxies as}
\begin{equation}
    \log \frac{r_\mathrm{mass}}{r_\mathrm{light}} = -0.1(\log \mstel - 10.5) + 0.11z + c
    \label{eq:mtol}
\end{equation}
\redit{where $c=-0.28$ for star-forming galaxies and -0.23 for quiescent galaxies, reflecting the fact that the light profile of a quiescent galaxies tends to more closely track the mass (i.e. $r_\mathrm{mass}/r_\mathrm{light}$ is closer to 1) than for star-forming galaxies of equivalent \mstel and $z$. 
We also apply limits of $0.5\leq r_\mathrm{mass}/r_\mathrm{light} \leq 1$ to avoid applying large and unphysical corrections to galaxies at the extremes of our mass or redshift range.
Combining values from Equation~\ref{eq:mtol} with Equation~\ref{eq:sigone} provides an estimate of the change in \sigone to account for the difference of the mass profile versus light profile for each galaxy in our sample,
\begin{equation}
   \frac{\Sigma_\mathrm{1kpc,mass}}{\Sigma_\mathrm{1kpc,light}} = 
   \frac{P(2n, b_n r_\mathrm{mass}^{-1/n})}{P(2n, b_n r_\mathrm{light}^{-1/n})}
\end{equation}
where $P(s,x)$ is the regularised lower incomplete gamma function, $n$ is the S\'ersic index of the galaxy measured with GALFIT, and $b_n$ is given by Equation~\ref{eq:bn}.}

\redit{Using this method, our \sigone estimates increase by up to $\sim$0.2~dex for star-forming galaxies and up to $\sim$0.1~dex for quiescent galaxies, although this change depends systematically on \mstel and $z$ as well as the measured light profile. 
We thus need to update our definition of the quiescent galaxy \sigone sequence which is used to classify galaxies as compact or extended. We update Equation~\ref{eq:sigone_qu_seq} to
\begin{equation}
    \log \sigseq = 0.75(\log \mstel - 10.5) + 0.6(1+z) + 9.6
\end{equation}
and, as before, identify galaxies with $\log (\sigone/\sigseq)>-0.4$ as compact and galaxies below this cut as extended.
By construction, redefining \sigseq means that the size of our cQu and exQu galaxy samples do not change (although the classification of individual sources can change).  
In contrast, the size of the cSF galaxy sample increases by $\sim$10\%, with a corresponding reduction in the exSF sample, as star-forming galaxies tend to have both flatter light profiles (lower $n$) and lower $r_\mathrm{mass}/r_\mathrm{light}$ which both lead to a greater increase in our \sigone estimates, on average. 
Crucially, however, we see the same changes in the number of X-ray sources classified as cSF and exSF.
Thus, making these adjustments to \sigone has a negligible impact on our measurements of AGN fractions or probability distributions, as shown in the third rows of Figures~\ref{fig:fagn_checks} and \ref{fig:pagn_checks}.}

\redit{Ultimately, differences in the mass versus light profile may affect the absolute values of \sigone and are important to consider when determining the size evolution of populations over cosmic time \citep[e.g.][]{suess_dissecting_2021}.
However, our \emph{relative} definition of compact versus extended galaxies remains robust.
Furthermore, a shift in \sigone measurements will affect both AGN and non-AGN galaxies equally and thus has a minimal effect on our measurements of the AGN incidence in different galaxy populations.
Given that robustly measuring stellar mass profiles for \emph{individual} galaxies---especially in the presence of AGN light---is beyond the scope of the current paper, we choose to retain our current ``BEST'' estimates from Section~\ref{sec:galfit} throughout our main analysis.
}

\subsection{Hard X-ray selection and absorption effects}
\label{sec:hard_vs_soft}

Next we consider our choice of hard X-ray selection in identifying AGN. 
In our fiducial results, we use the observed 2--7~keV \textit{Chandra} data when measuring rest-frame 2--10~keV X-ray luminosities and determining the incidence of AGN within galaxies.
The 2--7~keV band, especially at $z\gtrsim1$, probes relatively hard rest-frame energies, mitigating the effects of any absorption of the X-ray light and ensuring that we have a reliable tracer of the intrinsic X-ray luminosity. 
However, \textit{Chandra} has greater sensitivity at soft X-ray energies. 
Thus, in the third row of Figure~\ref{fig:fagn_checks} we compare our best (hard band) estimates (black crosses) to measurements repeated using the soft (0.5--2~keV) observed energy band to infer the rest-frame 2--10~keV luminosities (green triangles).\footnote{We retain our assumption of a $\Gamma=1.9$ X-ray spectrum with Galactic absorption only, as used in our hard band measurements, and make no additional corrections for intrinsic absorption.}

In all four galaxy populations, we find a factor $\sim2$ lower AGN fraction when adopting a soft X-ray selection \emph{at all redshifts}. 
These results show that soft X-ray selection is severely biased against the dominant, moderately absorbed AGN populations, even at $z\gtrsim1$.\footnote{See also figure~7 of \citet{aird_evolution_2015} that shows a similar discrepancy between hard and soft X-ray measurements of the X-ray luminosity function for moderate luminosities at all redshifts.}
The third row of Figure~\ref{fig:pagn_checks} shows the impact of soft selection on $p(\LX/\mstel)$. 
In the star-forming galaxy populations, the lower accretion rate sources are clearly under-represented in the soft-band measurements, consistent with the well-established rise in the fraction of absorbed sources at lower luminosities \citep[e.g.][]{ueda_cosmological_2003,aird_evolution_2015}. 
The pattern is less clear in the quiescent galaxy populations, although the measured AGN fractions remain a factor $\sim$2 lower using soft selection, as in the star-forming galaxies. We thus conclude that using hard X-ray data is essential to obtain accurate measurements of AGN fractions, including at $z\sim1-3$, even though such data are unavoidably shallower than at soft X-ray energies (and thus the raw numbers of X-ray \emph{detections} will in fact be lower). 

It is also worth noting that our hard (2--7~keV) X-ray selection will still be biased against the most heavily obscured, Compton-thick sources (with equivalent hydrogen column densities \hbox{$N_\mathrm{H}\gtrsim10^{24}$~cm$^{-2}$}), even at $z\sim2$ where we are probing rest-frame energies $\sim$6--21~keV. 
The luminosities of such sources are likely to be severely underestimated using our methodology and thus their contribution will not be included in our measured AGN fractions. 
As discussed by \citet{aird_agn-galaxy-halo_2021}, if it is assumed that the incidence of Compton-thick AGN simply traces the incidence of Compton-thin populations, with little  dependence on \LX or \LX/\mstel \citep[e.g.][]{buchner_obscuration-dependent_2015,ricci_close_2017} or host galaxy properties, then our AGN fractions and accretion rate probability distributions could be corrected simply by applying a scale factor $\sim1.25-2.0$ (allowing for remaining uncertainties in the Compton-thick fraction). 
Accurately determining whether Compton-thick AGN are preferentially found in certain galaxy populations---and thus whether this assumption is valid---is beyond the scope of this paper and is deferred to future work.

\subsection{Field-to-field variations}
\label{sec:field-to-field}

Finally in the bottom row of Figure~\ref{fig:fagn_checks} we demonstrate the differences in our measurements between the five CANDELS fields.
Our galaxy sample is selected to a uniform, conservative magnitude limit of $\Hband<25.5$ across all fields, with a further cut on stellar mass $\mstel>10^{10}$~\msun, and thus variations in the depth of the optical/infrared imaging will have a negligible impact. 
However, the depth of the \textit{Chandra} X-ray imaging varies significantly across the five fields (from $\sim$160~ks in COSMOS to $\sim$4~Ms in GOODS-S).
Furthermore, sample variance, cosmic large-scale structure within an individual field, and small number statistics can have a significant impact on results derived from a single field alone.

For the exSF and cSF populations we do not find significant differences in the measurements of the AGN fraction between our five fields, although the uncertainties are larger for measurements based on a single field (in particular within UDS where the measurements show larger discrpeancies as well as larger uncertainties). 
The overall consistency within the uncertainties, however, is reassuring given the widely varying X-ray depths and indicates that our methodology is correctly accounting for the varying X-ray sensitivity. 

For the cQu and exQu galaxy populations, 
we find greater field-to-field variation. 
For cQu galaxies this is likely due to the intrinsically lower AGN fraction, meaning there are relatively few X-ray detections within any single field, whereas for exQu galaxies it is the much lower overall size of the parent galaxy samples that leads to the greater field-to-field scatter.
However, the measurements based on individual fields are consistent with our best estimates in all cases, given the large statistical uncertainties. 
In particular, we note that our measurement of $f(\log \LX/\mstel>32)$ in exQu galaxies in the GOODS-S field (with the deepest X-ray data) is highly uncertain but lies systematically below our best estimate at all redshifts.\footnote{A measurement is not shown in the highest redshift bin for GOODS-S as the size of the exQu galaxy sample is insufficient to obtain a reliable estimate.} 
In contrast, we measure a much higher AGN fraction in GOODS-N, with the next deepest X-ray imaging. 
Combining the power of all five CANDELS fields---and allowing for the varying X-ray depths---is clearly vital to obtain an accurate measurement of the incidence of AGN within relatively rare galaxy populations, such as the exQu galaxies at high redshift.

\bsp	\label{lastpage}
\end{document}